\documentclass{aa}

\usepackage{txfonts,graphicx,natbib,rotating}
\usepackage{longtable}
\bibpunct{(}{)}{;}{a}{}{,}

\begin{document}

\title{ESC\thanks{European Supernova Collaboration \newline http://www.mpa-garching.mpg.de/$\sim$rtn} Supernova spectroscopy of non-ESC targets}


\author{A. H. Harutyunyan \inst{1, 2}
        \and P. Pfahler \inst{3}
        \and A. Pastorello \inst{4}
	\and S. Taubenberger \inst{3}
	\and M. Turatto \inst{1}
	\and E. Cappellaro \inst{1}
	\and S. Benetti \inst{1}
	\and N. Elias-Rosa \inst{3}
	\and H. Navasardyan \inst{1}
	\and S. Valenti \inst{5, 6}
	\and V. Stanishev \inst{7}
	\and F. Patat \inst{5}
	\and M. Riello \inst{8}
	\and G. Pignata \inst{9}
	\and W. Hillebrandt \inst{3}
}
\institute{INAF - Osservatorio Astronomico di Padova, vicolo dell'Osservatorio 5, 35122 Padova, Italy \newline e-mail: avet.harutyunyan@oapd.inaf.it
           \and Dipartimento di Astronomia, Universit\`{a} degli Studi di Padova, vicolo dell'Osservatorio 3, 35122 Padova, Italy
	   \and Max-Planck-Institut f\"{u}r Astrophysik, Karl-Schwarzschild-Str. 1, D-85741 Garching bei M\"{u}nchen, Germany
	   \and Astrophysics Research Centre, School of Mathematics and Physics, Queen's University Belfast, Belfast BT7 1NN
	   \and European Southern Observatory (ESO), Karl-Schwarzschild-Str. 2, D-85748 Garching bei M\"{u}nchen, Germany
	   \and Dipartimento di Fisica, Universit\`{a} di Ferrara, via Saragat 1, 44100 Ferrara, Italy
           \and Department of Physics, Stockholm University, AlbaNova University Center, 10692 Stockholm, Sweden
	   \and Institute of Astronomy, Madingley Rd., Cambridge CB3 0HA
	   \and Departamento de Astronom\'{i}a, Universidad de Chile, Casilla 36-D, Santiago, Chile
}
\abstract{}
	 {We present the spectra of 36 Supernovae (SNe) of various types, obtained
	 by the European Supernova Collaboration. Because of the spectral
	 classification and the phase determination at their discovery the
	 SNe did not warrant further study, and the spectra we present are the only available
	 for the respective objects. In this paper we present and discuss this material
	 using a new software for the automated classification of SNe spectra.}
	 {As a validation of the software, we verify the classification and phase estimate
	 reported for these objects in their discovery / classification circulars.
	 For the comparison, the software uses the library of template spectra of Padova-Asiago
	 Supernova Archive (ASA).}
	 {For each spectrum of our sample we present a brief, individual discussion,
	 highlighting the main characteristics and possible peculiarities. The comparison
	 with ASA spectra confirms the previous classification of all
	 objects and refines the age estimates. For our software we determine numerical
	 limits of ``safe'' spectral classification and the uncertainties of the
	 phase determination.}
	 {}
\maketitle

\section{Introduction}
\label{intro}
Supernovae (SNe) are the catastrophic events terminating the evolution of different
kinds of stars.
Two main explosion mechanisms are commonly considered: the core
collapse of a massive star and the thermonuclear runaway of a white dwarf accreting
material from a companion. Core-Collapse (CC) SNe include several types (SNe II, Ib, Ic,
IIn, etc) while the Thermonuclear SNe are observationally called SNe Ia
\citep[see][for reviews]{turatto03a, turatto07}.

CC SNe are important tools in understanding the latest epochs of
the life of massive stars, the chemical enrichment of galaxies
and the star formation history of the universe \citep[e.g.][]{botticella07},
while type-Ia Supernovae (SNe Ia), on the account of their high luminosity and
homogeneous properties, are considered
one of the most accurate distance indicators. Indeed, the measurement of SNe Ia at
high redshifts gives the best evidence that we are living in an accelerating
Universe \citep[see][and references therein]{perlmutterschmidt03}. However, the fact that
progenitor systems of SNe Ia and their explosion
mechanisms are still debated leaves room for a possible luminosity
evolution with redshift, which would undermine the results reported so
far. Studies of larger and larger datasets confirmed the existence of rather vast
variety of SN Ia properties \citep[e.g.][]{benetti04, benetti05}.

The European Supernova Collaboration (ESC) was a
European Research Training Network (RTN), founded in 2002 
with the goal
to improve our understanding of SN Ia physics through a detailed
study of nearby (v$_{rec} <$ 6000 km s$^{-1}$) SNe Ia. During the
following years detailed spectroscopic and photometric monitoring
of 15 nearby SNe Ia (plus 1 SN Ic) was carried
out. The results for 10 of these objects are already published in
dedicated papers \citep{benetti04, eliasrosa06, eliasrosa08, pignata04b, kotak05, stanishev07,
taubenberger06, taubenberger08, altavilla07, pastorello07a, pastorello07c, garavini07}.
A number of papers with results on the other objects are in preparation (Pignata et al.,
Stanishev et al., Kotak et al., Salvo et al., Kerzendorf et al., Elias-Rosa et al.,
in preparation).
Using also SNe Ia from this sample, analyses of systematic
properties of SNe Ia have been presented in \citet{benetti05}, \citet{mazzali05},
\citet{hachinger06} and \citet{mazzali07}.

\begin{table*}
\caption{SN sample and observations}
\label{table1}
\begin{center}
\begin{tabular}{l c c c c c c}
\hline
SN & Discovery & Acquisition & Instrumentation      & Spectral range & Resolution$^{1}$ & Reference\\
&&&&&&\\
   & DD/MM/YY  & DD/MM/YY    & Telescope+Instrument & \AA            & \AA              & IAUC / CBET\\
\hline
&&&&&&\\
2002an & 22/01/02 & 05/02/02 & 1.82m Mt.Ekar+AFOSC & 3700 - 7620 & 25   & 7805, 7808, 7818, 7828\\
2002bh & 24/02/02 & 05/03/02 & 1.82m Mt.Ekar+AFOSC & 4000 - 7600 & 26   & 7837, 7840, 7844\\
2002cs & 05/05/02 & 07/05/02 & TNG+DOLORES         & 3500 - 7930 & 15   & 7891, 7894\\
2002dg & 31/05/02 & 15/06/02 & ESO NTT+EMMI        & 3800 - 9330 & 10   & 7915, 7922\\
2002dm & 04/05/02 & 15/06/02 & ESO VLT U4+FORS2    & 4290 - 10250$^{2}$ & 12  & 7921, 7923\\
2002ej & 09/08/02 & 30/08/02 & 1.82m Mt.Ekar+AFOSC & 3800 - 7630 & 25   & 7951, 7963\\
2002hg & 28/10/02 & 02/11/02 & CA 2.2m+CAFOS       & 3600 - 8690 & 12   & 8004, 8007\\
2002hm & 05/11/02 & 06/11/02 & CA 2.2m+CAFOS       & 3500 - 8700 & 13   & 8009\\
2002hy & 12/11/02 & 15/11/02 & ESO 3.6m+EFOSC2     & 3500 - 9820 & 13   & 8016, 8019\\
2003hg & 18/08/03 & 22/08/03 & 1.82m Mt.Ekar+AFOSC & 3800 - 7360 & 24   & 8184, 8187, 40\\
2003hn & 25/08/03 & 26/08/03 & ESO 3.6m+EFOSC2     & 3600 - 9960 & 13   & 8186, 8187\\
2003ie & 19/09/03 & 22/09/03 & 1.82m Mt.Ekar+AFOSC & 4000 - 7440 & 25   & 8205, 8207\\
2004G  & 19/01/04 & 21/01/04 & CA 2.2m+CAFOS       & 3500 - 8090 & 14   & 8272, 8273\\
2004aq & 02/03/04 & 10/03/04 & NOT+ALFOSC          & 3900 - 8910 & 19   & 8301\\
2004bs & 16/05/04 & 19/05/04 & CA 2.2m+CAFOS       & 3800 - 8650 & 12   & 8341, 66, 8344, 8348\\
2004cc & 10/06/04 & 14/06/04 & NOT+ALFOSC          & 3500 - 8990 & 19   & 8350, 8353\\
2004dg & 19/07/04 & 21/07/04 & 1.82m Mt.Ekar+AFOSC & 3800 - 7780 & 25   & 8375, 8376, 8383\\
2004dk & 30/07/04 & 03/08/04 & CA 2.2m+CAFOS       & 3500 - 8750 & 12   & 8377, 8379, 8404, 75\\
2004dn & 29/07/04 & 05/08/04 & CA 2.2m+CAFOS       & 3800 - 8690 & 12   & 8381\\
2004fe & 30/10/04 & 02/11/04 & NOT+ALFOSC          & 3500 - 8900 & 19   & 8425, 8426\\
2004go & 18/11/04 & 07/12/04 & 1.82m Mt.Ekar+AFOSC & 3800 - 7580 & 24   & 8448, 8450, 8454 \\
2005G  & 14/01/05 & 18/01/05 & 1.82m Mt.Ekar+AFOSC & 3600 - 7300 & 24   & 8465, 8568\\
2005H  & 15/01/05 & 17/01/05 & CA 2.2m+CAFOS       & 4000 - 8700 & 10   & 8467\\
2005I  & 15/01/05 & 18/01/05 & CA 2.2m+CAFOS       & 3800 - 8610 & 12   & 8467\\
2005N  & 19/01/05 & 22/01/05 & CA 2.2m+CAFOS       & 3700 - 8600 & 12   & 8470\\
2005V  & 30/01/05 & 31/01/05 & CA 2.2m+CAFOS       & 3500 - 8780 & 14   & 8474, 8572\\
2005ab & 05/02/05 & 09/02/05 & 1.82m Mt.Ekar+AFOSC & 4250 - 8070 & 25   & 8478, 8479, 8480\\
2005ai & 12/02/05 & 14/02/05 & CA 2.2m+CAFOS       & 3800 - 8700 & 12   & 8486, 8487\\
2005br & 28/03/05 & 25/05/05 & ESO VLT U1+FORS2    & 4000 - 9710$^{2}$ & 12   & 8516, 156, 8538\\
2005bs & 19/04/05 & 25/05/05 & ESO VLT U1+FORS2    & 3800 - 9290$^{2}$ & 12   & 8517, 143, 156, 8538\\
2005cb & 13/05/05 & 25/05/05 & ESO VLT U1+FORS2    & 3700 - 9720$^{2}$ & 12   & 8530, 156, 8538\\
2005ce & 28/05/05 & 29/05/05 & NOT+ALFOSC          & 3400 - 8800 & 19   & 158, 159\\
2005de & 02/08/05 & 06/08/05 & CA 2.2m+CAFOS       & 3500 - 8640 & 15   & 8580, 191, 8581, 193\\
2005dv & 04/09/05 & 09/09/05 & CA 2.2m+CAFOS       & 3500 - 8710 & 12   & 8598, 217, 218\\
2005dz & 10/09/05 & 12/09/05 & NOT+ALFOSC          & 3400 - 8850 & 19   & 222, 225\\
2005kl & 22/11/05 & 24/11/05 & CA 2.2m+CAFOS       & 4200 - 8750 & 12   & 8634, 300, 305\\

\hline
\end{tabular}
\end{center}
\footnotesize
Notes:\\
$^{1}$ measured on the FWHM of the night sky lines when available,
otherwise the typical FWHM for the respective telescope-instrument combination is written\\
$^{2}$ obtained by merging two spectra with different spectral coverage\\
1.82m Mt.Ekar + AFOSC - 1.82m Copernico Telescope + Asiago Faint Object Spectrograph and Camera, Asiago, Italy\\
TNG + DOLORES - 3.5m Telescopio Nazionale Galileo + Device Optimized for the LOw RESolution, La Palma, Spain\\
ESO NTT + EMMI - 3.5m New Technology Telescope + ESO Multi-Mode Instrument, ESO La Silla, Chile\\
ESO VLT U1, U4 + FORS2 - 8m Very Large Telescope, Unit 1, 4 + visual and near UV FOcal Reducer and low dispersion Spectrograph 2, ESO Paranal, Chile\\
CA 2.2m + CAFOS - 2.2m Calar Alto Telescope + Calar Alto Faint Object Spectrograph, Almer\'ia, Spain\\
ESO 3.6m + EFOSC2 - 3.6m Telescope + ESO Faint Object Spectrograph and Camera (v.2), ESO La Silla, Chile\\
NOT + ALFOSC - Nordic Optical Telescope + AndaLucia Faint Object Spectrograph and Camera, La Palma, Spain
\normalsize
\end{table*}

In order to select RTN target candidates, prompt classification
of newly discovered SNe was required. Thereafter some
objects became targets of extensive monitoring,
many others did not pass the RTN selection criteria and 
no follow-up observations were triggered. For these SNe
only classification spectra are therefore available.
In this paper we present spectra of 36 SNe,
obtained during the ESC--related programs, for which 
follow-up observations were not activated.
These data have not been studied or published so far although
in some cases, they contain interesting information. In particular,
some of these spectra (4) were obtained at or before maximum light
when unique information about the physical conditions and chemical
structure of the progenitor star can be retrieved.

The spectra have been classified by means of a new automated tool,
specifically developed for this purpose, which compares an input spectrum
with the spectra of the Padova-Asiago Supernova Archive (ASA), and identifies
the best matching template spectrum. In this way it is possible to determine
the spectral classification and the age of the new objects.
A set of type-Ia SN spectra of ESC objects with good temporal
coverage was also used to test and calibrate the software tool.

The structure of the paper is the following: in Sect. \ref{snsample}
we present our SN sample, observations and data reduction techniques.
In Sect. \ref{specclassif} the issue of SN spectral classification is
addressed, and in Sect. \ref{gelato} the software is presented.
A discussion of individual spectra is done in Sect.
\ref{spectra} and finally the Sect. \ref{summ} gives a short summary.

\section{SN sample, observations and data reduction}
\label{snsample}
The sample of 36 SNe is presented in Table \ref{table1}, where
also the information on instrumental configurations and observational
details is listed. Seven different
telescope+instrument combinations have been used for the observations.
The last column of the table contains the references for the
discovery and the classification circulars (IAUC, CBET) for each object.

All two-dimensional images were pre-reduced (trimmed, overscan,
bias and flatfield corrected) with standard
IRAF\footnote{Image Reduction and Analysis Facility, http://iraf.noao.edu}
subroutines. Further data processing was performed using the
procedures from the IRAF CTIOSLIT package. In particular, after the optimised
extractions performed with the APALL task, the one-dimensional
spectra were wavelength-calibrated by comparison with spectra
of arc lamps obtained during the same night and with the same
instrumental configuration. The wavelength calibration was
checked against bright night-sky emission lines. The SN spectra
were then flux-calibrated using response curves derived from
the spectra of standard stars preferably observed during the
same night\footnote{For a
detailed discussion on data reduction techniques see,
for instance, \citet{pastorello07a}.}.\\
In some cases two spectra of the same object observed
during the same night in different spectral ranges
were merged into a single spectrum to gain a wider
wavelength coverage and/or higher signal-to-noise ratio (SNR).

\section{SN classification}
\label{specclassif}
Prompt SN type determination is important
in the study of SNe. In particular, the success of extensive and
time consuming observational campaigns depends crucially
on the proper planning of observations, typically obtained in
Target of Opportunity mode. A rapid SN classification is usually
performed on the basis of one optical spectrum (usually obtained
near maximum light) and in general is quite reliable. 
In a few cases a definitive classification requires the analysis
of the spectral evolution.

In the earliest phase when the SN is optically thick, the
photosphere emits a continuum radiation field, while line
formation occurs above. The ejecta is expanding and the lines
are characterised by P-Cygni (P-Cyg) profiles,
an emission peak near the rest wavelength of the line and
a blueshifted absorption feature. The emission peak is formed by line
scattering into the line of sight of photons emitted by the photosphere
and would be symmetrical to the line center wavelength if there was not
the absorption. The absorption is formed by scattering out the line of
sight of photospheric photons emitted toward the observer. Since this
occurs in front of the photosphere, the absorption is blueshifted
\citep[see][for a review on spectra formation]{jeffery90}.

Type-Ia SNe are classified by the presence of lines
of intermediate mass elements such as Si, Ca and S
during the maximum light phase and by the absence of H at any time.
Type-Ib SNe are spectroscopically classified by the absence of
H Balmer and Si II lines and by the presence of He I features,
though, as \citet{branch02} showed, weak, broad H$_\alpha$ may
also be present.
Type-Ic SNe are similar, but with the absence of He lines,
though some weak contamination of He seems to be common to
several type-Ic SNe \citep{filippenko95}. 
Type-II SNe are characterised by strong H Balmer lines.

Most of the SNe fall in one of the
above mentioned 4 classes. However, there is an increasing number
of peculiar objects, with unprecedented properties
and evolution
\citep[e.g. SN 2006jc, for which type-Ibn label was coined,][]{pastorello07b, pastorello08},
that do not fit within the scheme above. The current taxonomy of SNe
is therefore incomplete, ambiguous and not fully satisfactory
\citep[see][and references therein]{turatto07}. In a situation in which
the classification criteria are not exhaustive, 
a conservative approach for SN classification is via
comparison with other SNe.
With this motivation, our group has developed a
tool that compares a given input spectrum with the set of
spectra from ASA, a large
archive of SN spectra, collected by the members of
the group during the last decades.
%

\section{SN spectra comparison tool}
\label{gelato}
A few  groups \citep[see][for a discussion]{blondin07} have developed
software tools for SN spectra analysis. Using
different technical approaches from $\chi^{2}$ minimisation 
to cross-correlation, these tools aim at different immediate results,
from rapid automatic SN spectra classification to quantification
of spectral differences. The goal of our software is the quantitative classification of SN spectra
by comparison with a large set of template spectra of various SN types 
at different phase.
  

The PAdova Supernova Spectra comPARison TOOl (passpartoo) is
a collection of software procedures performing automatic
comparison of SN spectra. Designed for different purposes
and working with slightly different algorithms, all the
procedures carry out an automatic comparison of a given
(input) spectrum with a set of well-studied SN spectra
(templates), in order to find the template spectrum that
is most similar to the given one. The first version of the
tool was presented in \citet{harutyunyan05}.
In this paper we used the GEneric cLAssification TOol (GELATO),
which is a software for objective classification of SN spectra.

One of the major issues to solve is the treatment of reddening,
due both to the Galaxy and to the SN host. We wanted our tool to minimise
the impact of the reddening on the comparison result with no
assumption on the reddening law, which can be very unusual
\citep[cfr.][]{eliasrosa06, eliasrosa08}. To this aim,
GELATO divides the input and every template spectrum in a number
of separate spectral bins.
This approach for spectra comparison was discussed in
\citet{riess97} and \citet{rizzi98}
and has the advantage of giving priority to the presence and strength
of spectral features. The bins are selected to contain the spectral features most
significant for SN classification and dating. The same set of bins is adopted for
comparison of spectra of all types with satisfactory results. The current version of GELATO uses 11 bins,
8 of which are from \citet{riess97}, slightly modified to meet few technical
requirements, while 3 have been added to include other spectral features and to enlarge
the working spectral range. Table \ref{gelato_bins} lists the GELATO bins and
the corresponding main spectral features.
\begin{table}
\caption{GELATO bins with wavelength ranges and corresponding main spectral features.}
\label{gelato_bins}
\begin{center}
\begin{tabular}{c c c c c}
\hline
Bins & Range & \multicolumn{3}{c}{SN spectra features} \\
\cline{3-5}
     & \AA              & Ia             & Ib/c            & II   \\
\hline
&&&&\\
1    & 3504 - 3792 & Ca II          & Ca II           & Ca II\\
2    & 3800 - 4192 & Si II, Ca II   & Ca II           & Ca II, H$_{\delta}$\\
3    & 4200 - 4576 & Mg II, Fe II   & Fe II           & Mg II, Fe II, H$_{\gamma
}$\\
4    & 4584 - 4936 & Fe II          & Fe II           & Fe II, H$_{\beta}$\\
5    & 4944 - 5192 & Fe II          & Fe II           & Fe II\\
6    & 5200 - 5592 & S II           & S II, OI        & S II\\
7    & 5600 - 5896 & Si II, Na I    & Na I, He I      & Si II, Na I\\
8    & 5904 - 6296 & Si II          & He I            & Si II\\
9    & 6304 - 6800 & Fe II          & Si II, He I     & O I, H$_{\alpha}$\\
10   & 6808 - 7904 & O I            & O I             & O I\\
11   & 7912 - 9000 & Ca II          & Ca II           & Ca II\\
\hline
\end{tabular}
\end{center}
\end{table}

\begin{figure}
\includegraphics[width=9cm]{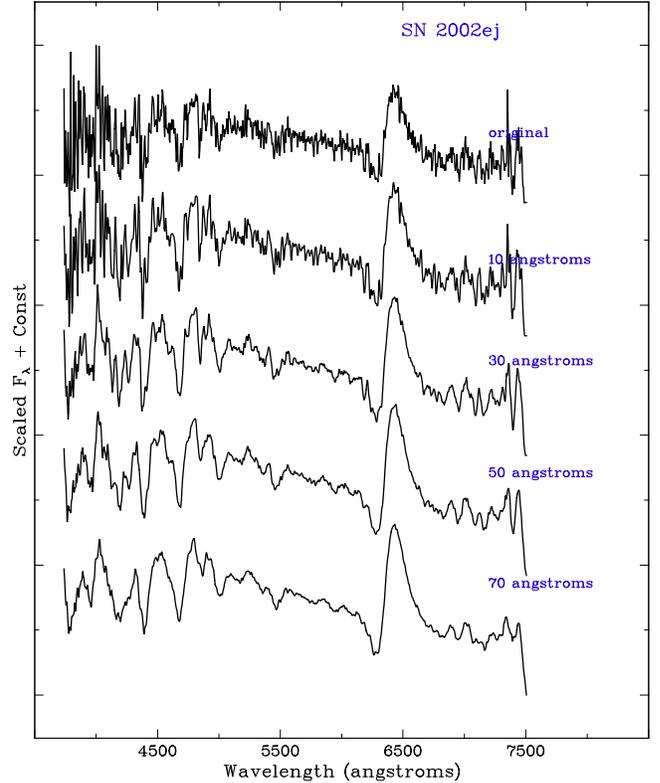}
\caption{The original and smoothed spectra of SN 2002ej (Section \ref{spectra}).
The spectra are in the parent galaxy restframe. The number near smoothed spectra
are the corresponding box sizes.}
\label{smooth_results}
\end{figure}

The GELATO code works as follows: for all the $N$ bins of an input
spectrum it computes $\delta_j$, the mean relative distance
between the $j$-th bin of the input spectrum and the corresponding
bin of the template, scaled in flux to match the input one.
This operation is iterated through all the template
spectra and the average of $\delta_j$ values for each template
spectrum is calculated. 
If $f_i$ is the input spectrum at wavelength
$\lambda_i$ and $F_i$ the template spectrum, we define
$\delta_j$ as follows:
\begin{equation}
\label{eq_delta_small}
\delta_j = \frac{1}{n \cdot <f>_j} \sum_{i=1}^{n}{|f_i - F_i^{norm}|},
\end{equation}
where $n$ is the number of spectral elements in the bin,
$F_i^{norm}$ is the $F_i$ flux scaled to the
input spectrum flux within the given bin, and $<f>_j$ is the
mean value of the flux in the bin. Then, the average of
$\delta_j$ values is computed as:
\begin{equation}
\label{eq_delta_big}
\Delta = \frac{1}{N} \sum_{j=1}^{N}{\delta_j},
\end{equation}
where N is the number of bins. The most similar template
to the input spectrum is the one that minimises the value of $\Delta$.
Before the comparison a boxcar smoothing is performed on all spectra
to reduce high-frequency noise components.
After tests with artificial noise addition, a box size corresponding
to $\sim$40-70$\AA$ is adopted to perform the smoothing. This smoothing
is sufficient to remove high-frequency noise components and
at the same time preserves the spectral features. Figure \ref{smooth_results}
shows the original and smoothed spectra of SN 2002ej (cfr. Section \ref{spectra}).

\begin{figure*}
\includegraphics[height=9cm, angle=270]{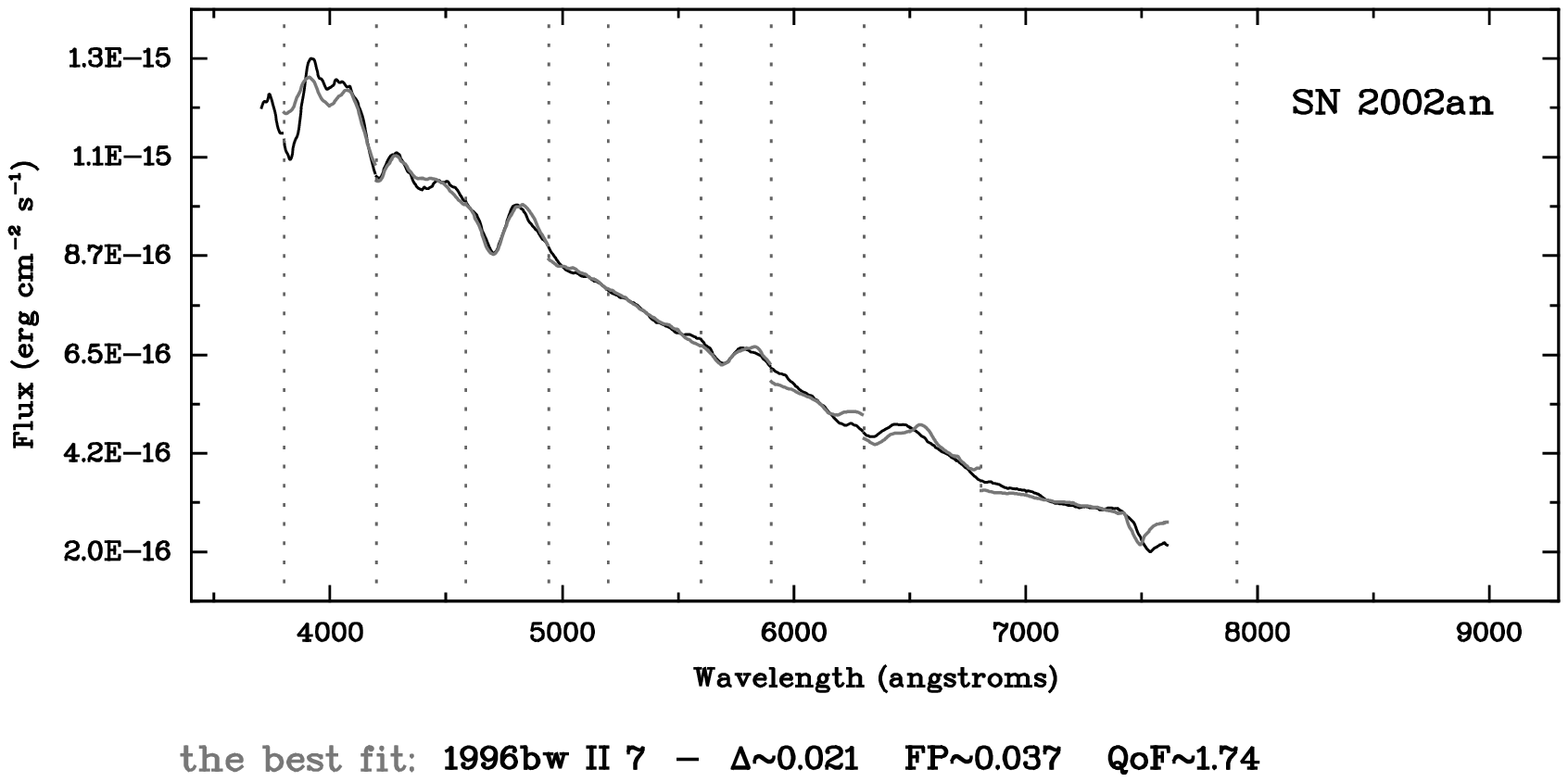}
\includegraphics[height=9cm, angle=270]{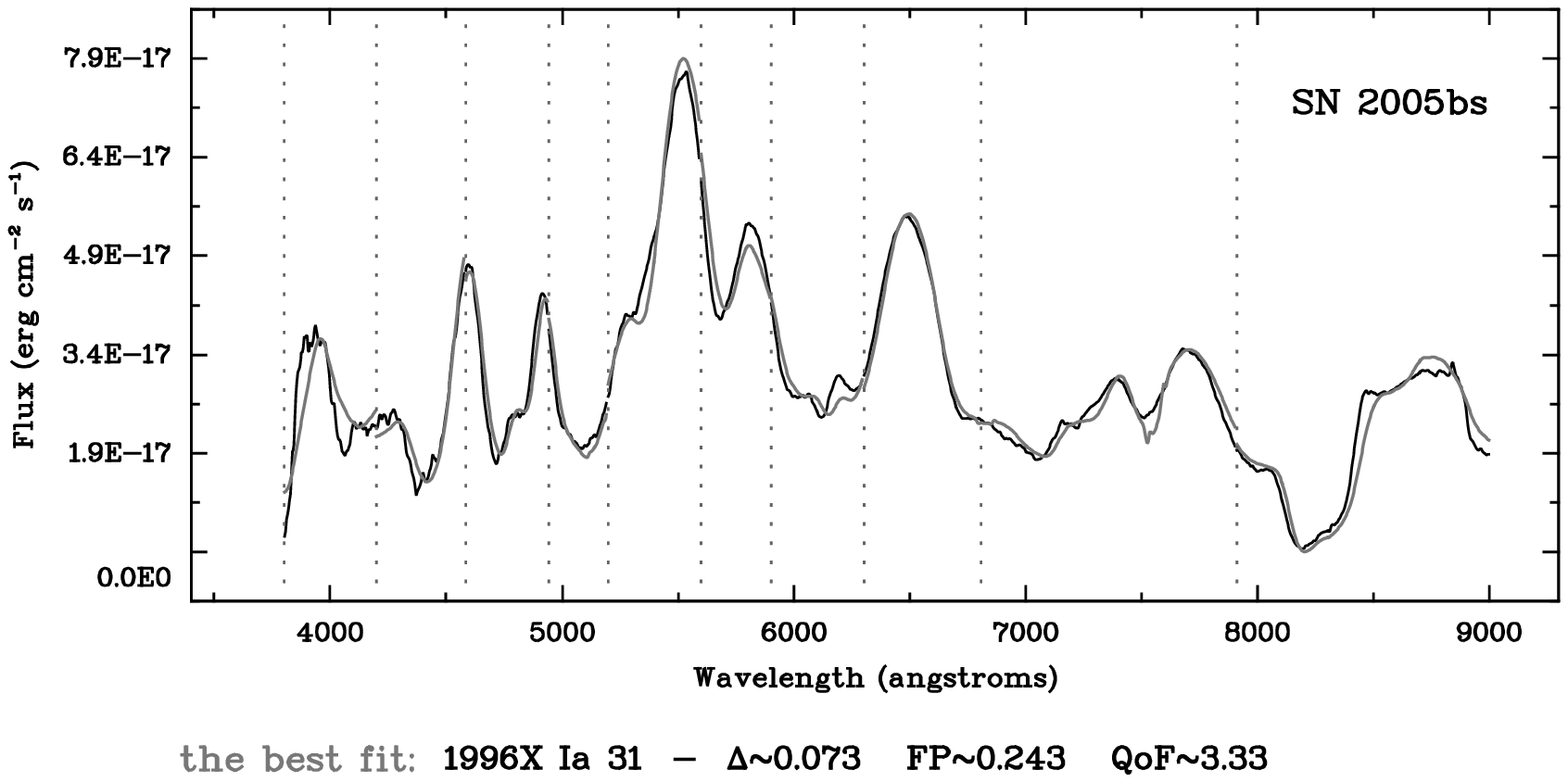}\\
\caption{The graphical output of GELATO for the cases of
SN 2002an and SN 2005bs from our sample. The spectra are in 
the parent galaxy rest frame. The black lines are
the input spectra and the gray ones the best fitting template
spectra, divided in bins and scaled in flux to match the input
spectra in the respective bins.}
\label{pl_fit1}
\end{figure*}

The software achieves the goal of finding the best
fitting archive spectrum to a given input spectrum, through the
minimisation of $\Delta$ values. However, $\Delta$ alone does not
describe the quality of the fit, which is necessary for a quantitative
comparison of the fits. Thus, we tried to define a quantitative
value providing a measure of quality of fits. In general,
we noted that the values
of $\Delta$ computed for the cases of spectra
with dominating continuum and weak spectral features are
systematically lower than those of spectra having many strong
spectral features which can vary in relative intensities and position.
To account for this, we weighted the
$\Delta$ values using coefficients that indicate whether the
spectra are more or less feature-rich. We defined the
feature-parameter (FP) as follows:
\begin{equation}
\label{eq_fp}
FP = \frac{1}{N}\sum_{j=1}^{N}\frac{1}{n}\sum_{i=1}^{n}\frac{|f_i - F_i^{flat}|}{<sp>_j},
\end{equation}
where $F_i^{flat}$ is the best fitting straight line
to the spectrum in the given bin. Then, for each fit we
define the quality of fit (QoF) value by the following expression:
\begin{equation}
\label{eq_qof}
QoF = \left(\frac{\Delta}{FP}\right)^{-1}.
\end{equation}

The QoF allows a numerical estimate of the quality of the fit,
which can be used to compare the fits to different SN spectra.
Table \ref{table2} reports the QoF values of the best fits 
to our set of spectra. In Figure \ref{pl_fit1} the
graphical outputs of GELATO for the cases of SN 2002an and 2005bs are
presented. In the figures the smoothed versions of the input spectrum (black line)
and best fitting template (gray line) together with bin boundaries
(dotted lines) are displayed. The $\Delta$
value of the fit in the SN 2002an case is smaller than that of the SN 2005bs
case ($\sim$0.02 and 0.07, respectively), but SN 2005bs has a greater FP
(i.e. has more prominent features) than SN 2002an. The resulting QoF is greater for
SN 2005bs ($\sim$3.3) than for SN 2002an ($\sim$1.7), thus according to
GELATO the former should be regarded as a better fit than the latter.
Further discussion for these spectra can be found in  Section \ref{spectra}.

We tested the QoF values comparing spectra to different sets of
templates.
The results combined
with visual inspections of the fits showed that
QoF $\geq$ 1.4 means a high / satisfactory quality of the fit 
and safe SN type determination, while QoF $<$ 1
occurs when the input has no match in the archive.
Intermediate values 1 $\leq$ QoF $<$ 1.4 may result both for
fair/good fits with correct type detection or poor
fits with incorrect type detection. Further tests will refine
the QoF and establish the confidence levels.

We stress that thanks to the above mentioned subdivision in
spectral bins, the best fit procedure depends very little on
the slope of the continuum, hence reddening. This can be seen,
for example, in the SN 2004aq spectrum fit by SN 1991al
(Fig. \ref{plt2}d), for which the QoF = 2.95, despite the
significant difference in SED. To verify this
we used the original spectrum of SN 2005bs (see Section \ref{spectra}
and Fig. \ref{plt3}j) and progressively reddened it up with
two reddening laws \citep[R$_{V}$=3.1 and 1.8,][]{eliasrosa06}.
Then we searched the best fitting templates with GELATO.
Table \ref{reddening_test} summarizes the results. The best
fitting template for all the cases was found to be SN 1996X 31
day after maximum, though with slightly decreasing QoF.
The test confirms that GELATO is little sensible to the
amount of reddening and to the reddening law.

\begin{table}
\caption{Comparison results for reddened spectra of SN 2005bs.}
\label{reddening_test}
\begin{center}
\begin{tabular}{c c c c c c}
\hline
E(B-V) & \multicolumn{2}{c}{R$_{V}$=3.1} & \multicolumn{2}{c}{R$_{V}$=1.8} & Best template \\
\cline{2-3} \cline{4-5}
 & A$_{V}$ & QoF & A$_{V}$ & QoF & \\
\hline
&&&&&\\
0.0  & 0   & 3.3 & 0   & 3.3 & SN 1996X\\
0.16 & 0.5 & 3.1 & 0.3 & 3.1 & SN 1996X\\
0.32 & 1   & 2.8 & 0.6 & 2.9 & SN 1996X\\
0.48 & 1.5 & 2.6 & 0.9 & 2.7 & SN 1996X\\
0.64 & 2   & 2.3 & 1.2 & 2.4 & SN 1996X\\
\hline
\end{tabular}
\end{center}
\end{table}

\begin{figure}[]
\includegraphics[height=9cm, angle=270]{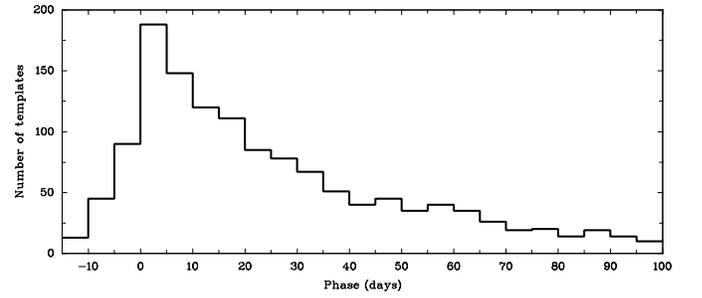}
\caption{The phase distribution of the template SN spectra. Typically,
the phases are relative to the B-maximum for SNe Ia and Ib/c and to the
explosion epoch for SNe II.}
\label{hist1}
\end{figure}

The second crucial ingredient for the automated spectra
classification, besides the software tool, is the availability of an
extended archive of spectra.
Spectra of all SN types from ASA
are used as templates for the comparison procedure. These spectra
have been collected by the Padova SN group in the course of
several long-term projects devoted to the study of the physical
properties of SNe including type-Ia SN spectra obtained by the ESC.
The archive has also been enriched with publicly
available material from the literature. Table \ref{archive_table}
lists the archive content on December 2007. It provides the type, redshift, number
spectra and phase range of the spectra for each SN. The information
about the types and redshifts of SNe are obtained from the up-to-date
version of the Asiago SN catalogue \citep{barbon99}.
Currently there are spectra of 155 type-Ia (including objects like
SN 1991T, 1991bg, 2000cx), 168 type-II (including SNe IIL, IIn, IIb)
and 65 type-Ib/c SNe (including hypernovae and objects like 1997dq, 2006jc),
for a total database of over 2500 spectra.
Figure \ref{hist1} shows the distribution of
the templates on the basis of their phase. As one can see from the
figure, the template temporal coverage is good at all phases though,
as expected, the number of the template spectra is greater
near maximum light.

\begin{figure}
\includegraphics[height=9cm, angle=270]{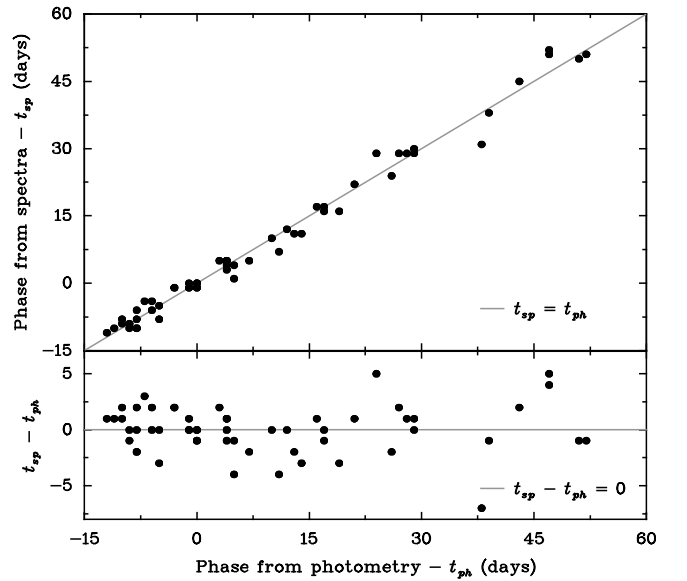}
\caption{The phases determined with GELATO compared to the phases
from photometry for a sample of type-Ia SNe. In the bottom panel
are the residuals.}
\label{pl_ph1}
\end{figure}
\begin{figure}
\includegraphics[height=9cm, angle=270]{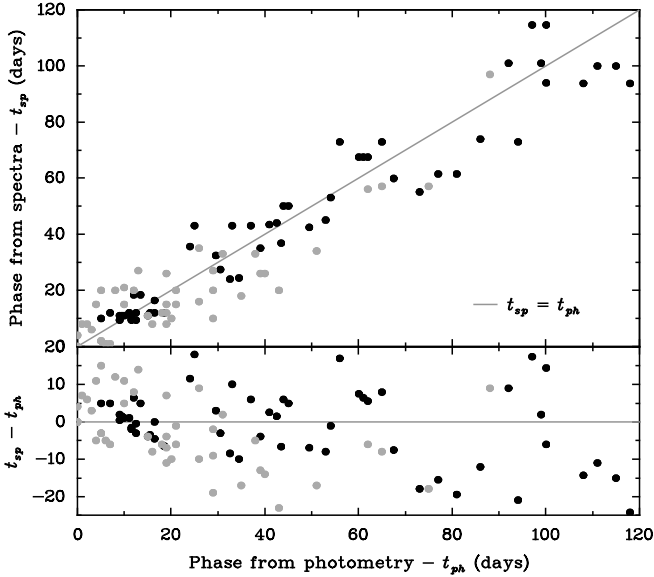}
\caption{The phases derived with GELATO compared to the phases
obtained from photometry for a sample of type-II (black dots) and
type-Ib/c (gray dots) SNe. In the bottom panel are the residuals.}
\label{pl_ph2}
\end{figure}

In general, we adopt as an estimate of the input spectrum age the
epoch of the corresponding best fitting template found by GELATO.
We carried out several tests to check the ability of GELATO, in
combination with ASA to determine the spectral age. To this aim,
we used the spectra of  a set of well-studied type-Ia SNe 2002bo,
2003cg, 2003du, 2004dt, 2004eo, 2005cf observed by the ESC,
for which detailed light curves are available. Each spectrum of these
objects was compared with those of a sample of a dozen of
template SNe Ia, which include objects spanning the
entire range of SN Ia properties (e.g. including SN 1991bg and 1991T).
Then, we compared the phases of the best fitting template spectra
obtained from the spectral comparison with the phases derived from
photometry. The results are plotted in Figure \ref{pl_ph1}. The plot
shows that the two phases are in a good agreement. The difference
of the phases has a RMS of about 3.1 days in the phase range from 15
to 60 days, while in the range -15 to 15 days it is as small as 1.9
days. The increase of the error with age is expected, because the
spectra of SNe Ia change less at late times.

Similar tests done with a set of type-II and type-Ib/c SNe give
RMS of about 10 days for both samples (see Figure \ref{pl_ph2}).
The errors are larger compared to those of type-Ia SNe due to the
heterogeneous behaviour of CC SNe (compared to SNe Ia),
to the lack of templates with very good temporal coverage and to
the uncertainty in the determination of the explosion epoch
for SN II.

\section{Individual spectra}
\label{spectra}

\begin{table*}
\caption{GELATO's best match templates.}
\label{table2}
\begin{center}
\begin{tabular}{c c c c c c c}
\hline
SN & Type & Best match  & QoF & Template phase & Template spectrum & Phase reference\\
   &      & SN template &     & (days)         & reference         &                \\
\hline
&&&&&&\\
2002an & II  & 1996bw & 1.7  &  7  $^{*}$  & ASA, \citet{benetti96} & 1995ad, 12d - \citet{pastorello03}\\
2002bh & II  & 1995V  & 1.4  &  8          & ASA                    & \citet{fassia98}\\
2002cs & Ia  & 2004dt & 1.6  & -7          & \citet{altavilla07}    & \\
2002dg & Ib  & 1998dt & 1.7  &  15 $^{**}$ & ASA                    & \citet{matheson01}\\
2002dm & Ia  & 1994ae & 9.1  &  92         & ASA                    & \\
2002ej & II  & 1995ad & 2.9  &  24         & \citet{pastorello03}   & \\
2002hg & II  & 1999em & 2.4  &  41         & \citet{elmhamdi03}     & \\
2002hm & II  & 1998ce & 3.8  &  10 $^{*}$  & ASA, \citet{patat98}   & 1995ad, 12d - \citet{pastorello03}\\
2002hy & IIb & 2001gh & 1.0  &  11 $^{*}$  & ASA, \citet{altavilla01}         & 1993J, 4d - \citet{barbon95}\\
2003hg & II  & 1995V  & 2.1  &  8          & ASA                    & \citet{fassia98}\\
2003hn & II  & 1995ad & 2.4  &  10         & \citet{pastorello03}   & \\
2003ie & II  & 1998A  & 1.2  &  37         & \citet{pastorello05a}  & \\
2004G  & II  & 1993S  & 2.4  &  90 $^{*}$  & ASA                    & 1995ad, 100d - \citet{pastorello03}\\
2004aq & II  & 1991al & 2.9  &  25 $^{*}$  & ASA                    & 2001du, 18d - \citet{smartt03}\\
2004bs & Ib  & 1998dt & 2.0  &  17 $^{**}$ & ASA                    & \citet{matheson01}\\
2004cc & Ic  & 1994I  & 1.4  &  10         & \citet{filippenko95}   & \\
2004dg & II  & 2001du & 4.8  &  18         & \citet{smartt03}       & \\
2004dk & Ic  & 2004aw & 1.4  &  4          & \citet{taubenberger06} & \\
2004dn & Ic  & 2004aw & 1.4  &  4          & \citet{taubenberger06} & \\
2004fe & Ic  & 1994I  & 1.9  & -3          & \citet{filippenko95}   & \\
2004go & Ia  & 1996X  & 2.2  &  24         & \citet{salvo01}        & \\
2005G  & Ia  & 1994D  & 3.5  &  11         & \citet{patat96}        & \\
2005H  & II  & 2002gd & 1.0  &  6          & \citet{pastorello03}   & \\
2005I  & II  & 2003gd & 3.3  &  101        & ASA                    & \citet{hendry05}\\
2005N  & Ib  & 1990I  & 2.1  &  88         & \citet{elmhamdi04}     & \\
2005V  & Ic  & 2004aw & 1.3  &  22         & \citet{taubenberger06} & \\
2005ab & II  & 1997du & 1.5  &  26 $^{*}$  & ASA, \citet{patat97}   & \\
2005ai & Ia  & 1994D  & 2.7  &  24         & \citet{patat96}        & \\
2005br & Ib  & 1997X  & 1.9  &  40         & ASA                    & 1990U, 48d - \citet{piemonte96}\\
2005bs & Ia  & 1996X  & 3.3  &  31         & \citet{salvo01}        & \\
2005cb & Ic  & 1994I  & 2.5  &  1          & \citet{filippenko95}   & \\
2005ce & Ib/c& 1996aq & 2.0  &  5 $^{*}$   & ASA                    & 1994I, 10d - \citet{filippenko95}\\
2005de & Ia  & 2005cf & 2.8  & -5          & \citet{garavini07}     & \\
2005dv & Ia  & 2002bo & 2.9  &  0          & \citet{benetti04}      & \\
2005dz & II  & 2007T  & 1.4  &  4 $^{*}$   & ASA, \citet{benetti07} & 2002gd, 6d - \citet{pastorello03}\\
2005kl & Ic  & 2004aw & 1.2  &  4          & \citet{taubenberger06} & \\
\hline
\end{tabular}
\end{center}
\footnotesize
Notes:\\
Phases are relative to the B-maximum epoch for type-Ia and Ib/c SNe and to explosion epoch for
type-II SNe.\\
$^{*}$ spectral epoch relative to the discovery date.
In these cases no reference of the explosion epoch is found and,
when available, the second best fitting template spectrum with
phase relative to the explosion epoch is reported in the
``Phase reference'' column.\\
$^{**}$ relative to the R-maximum\\
\normalsize
\end{table*}

\begin{figure*}
\includegraphics[height=9cm, angle=270]{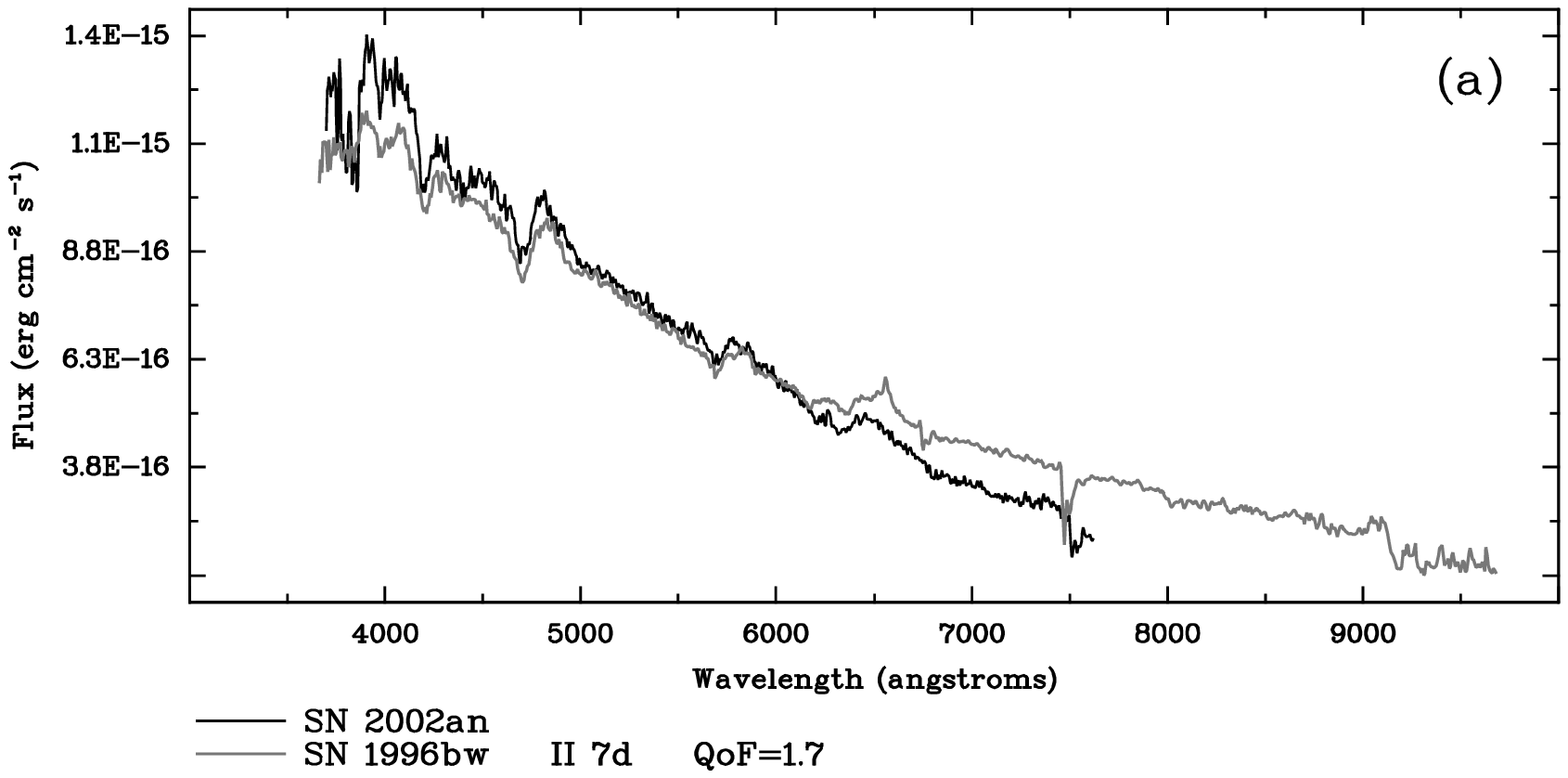}
\includegraphics[height=9cm, angle=270]{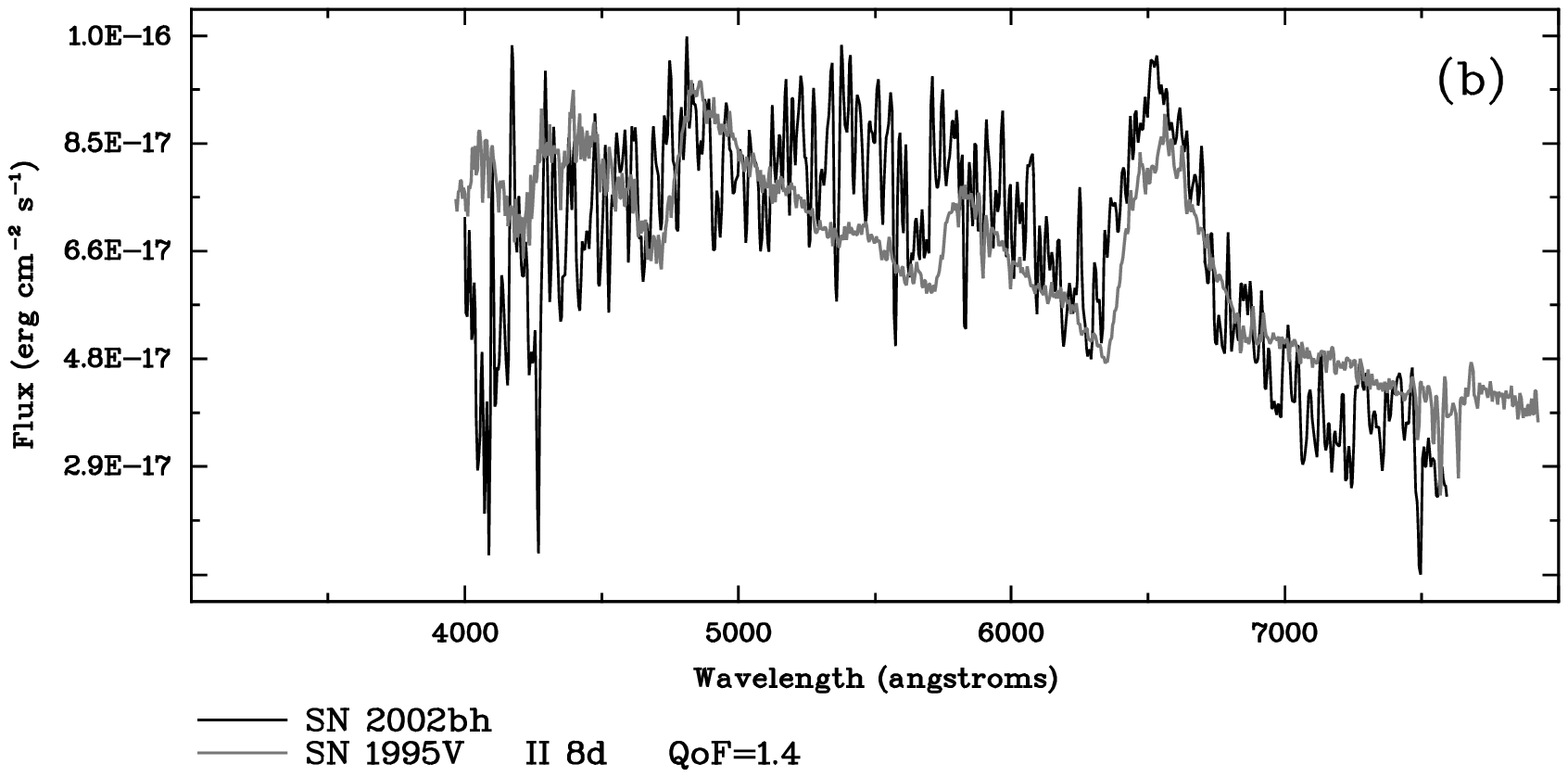}\\
\includegraphics[height=9cm, angle=270]{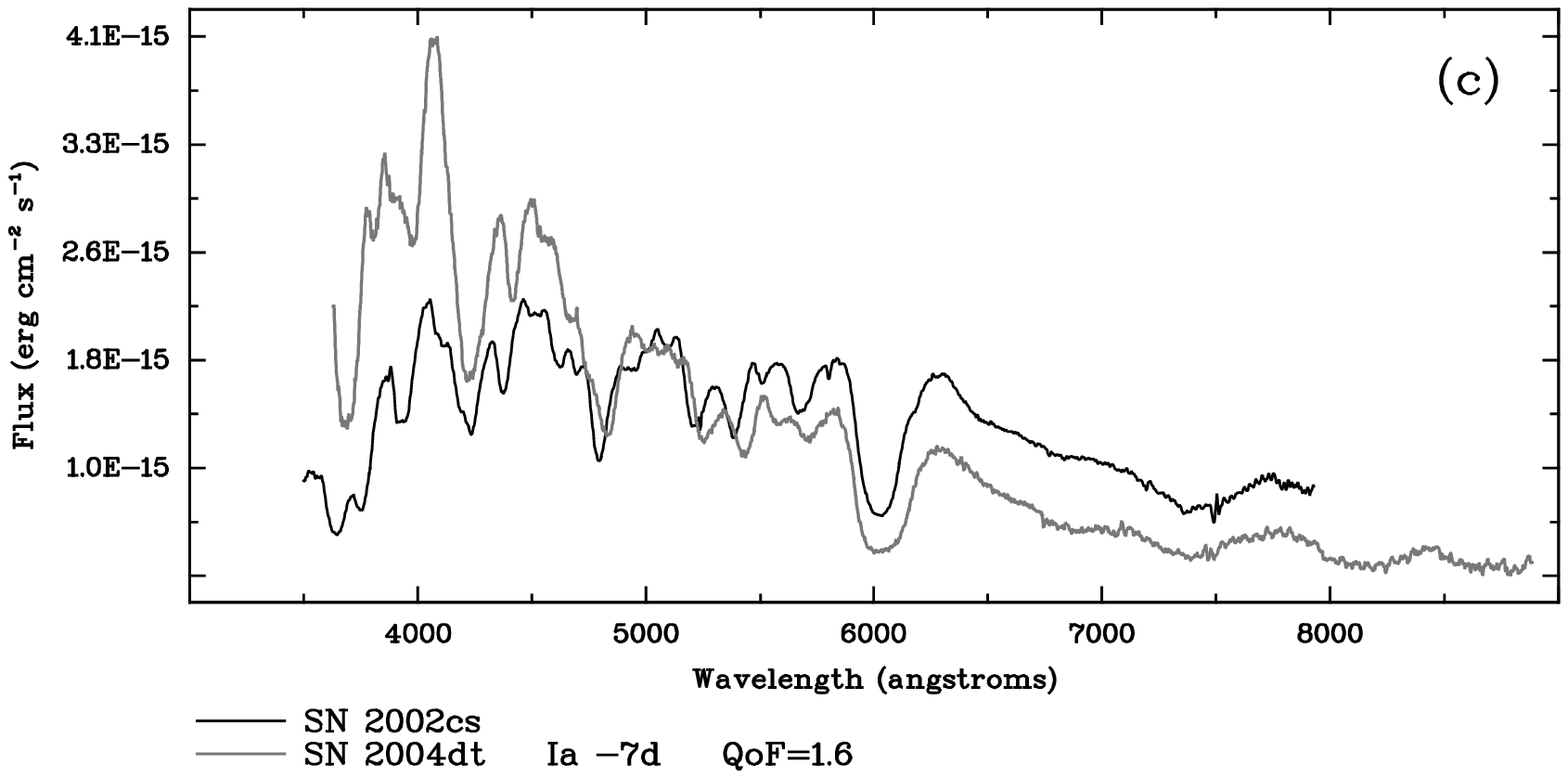}
\includegraphics[height=9cm, angle=270]{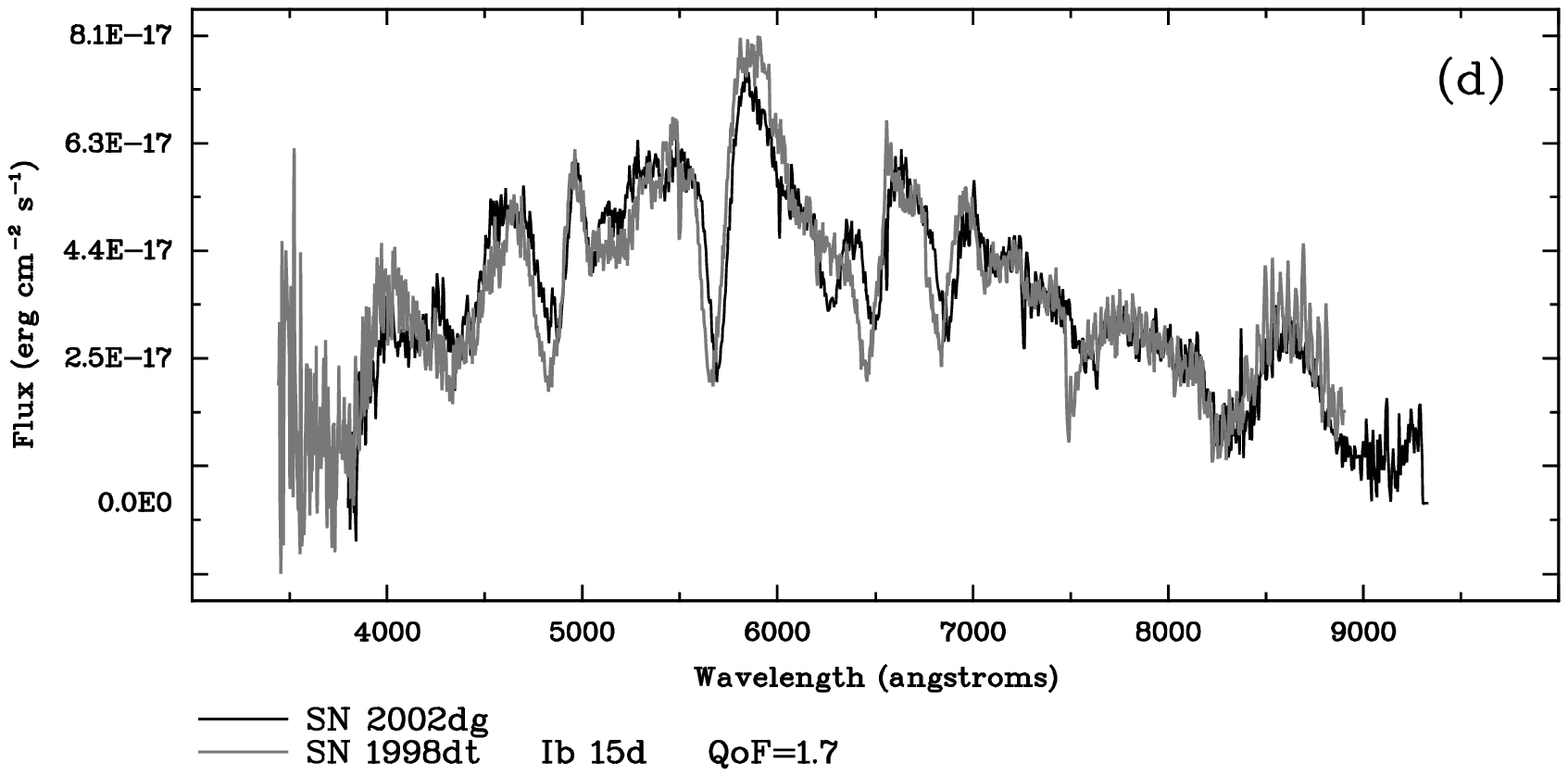}\\
\includegraphics[height=9cm, angle=270]{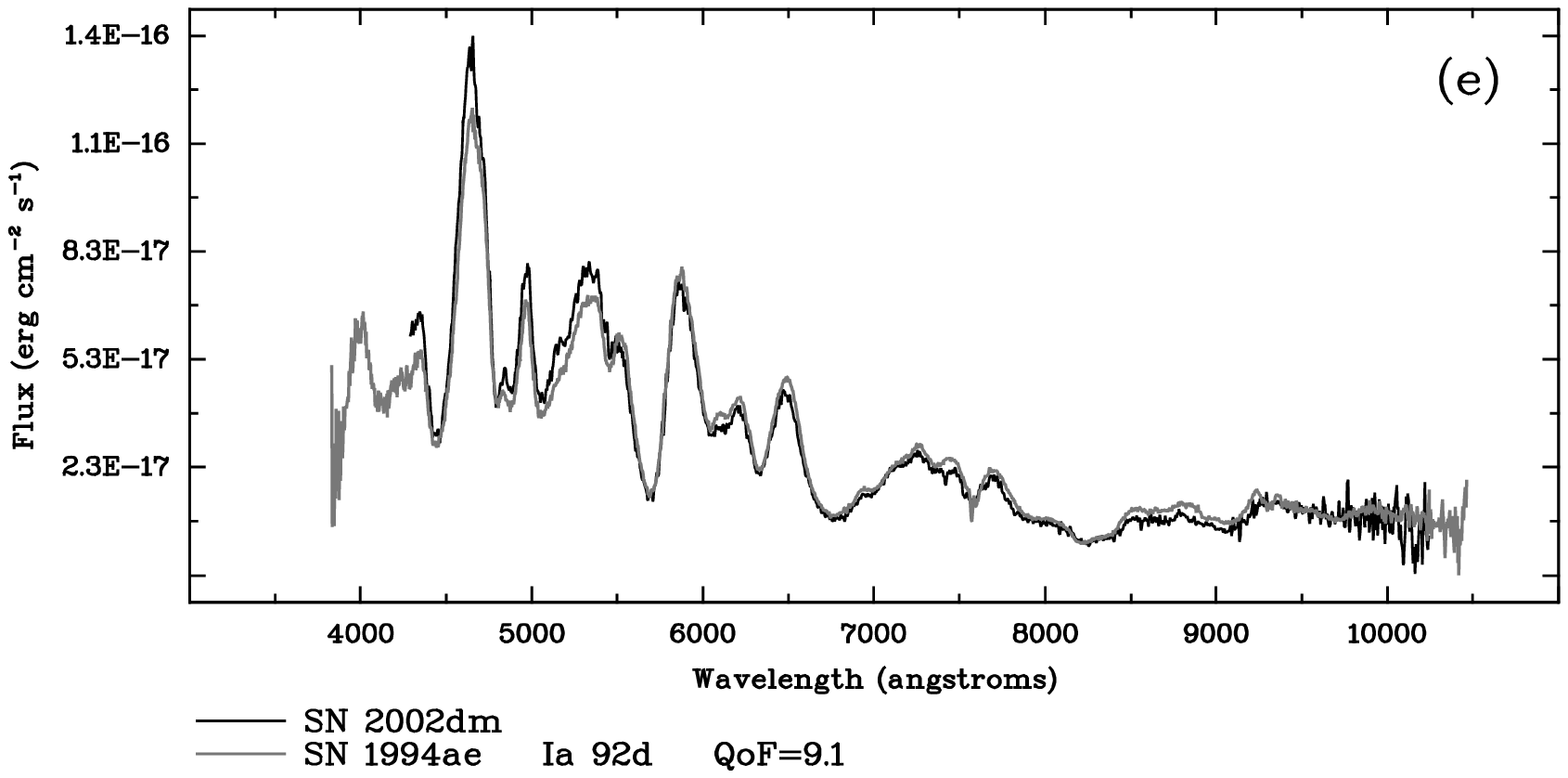}
\includegraphics[height=9cm, angle=270]{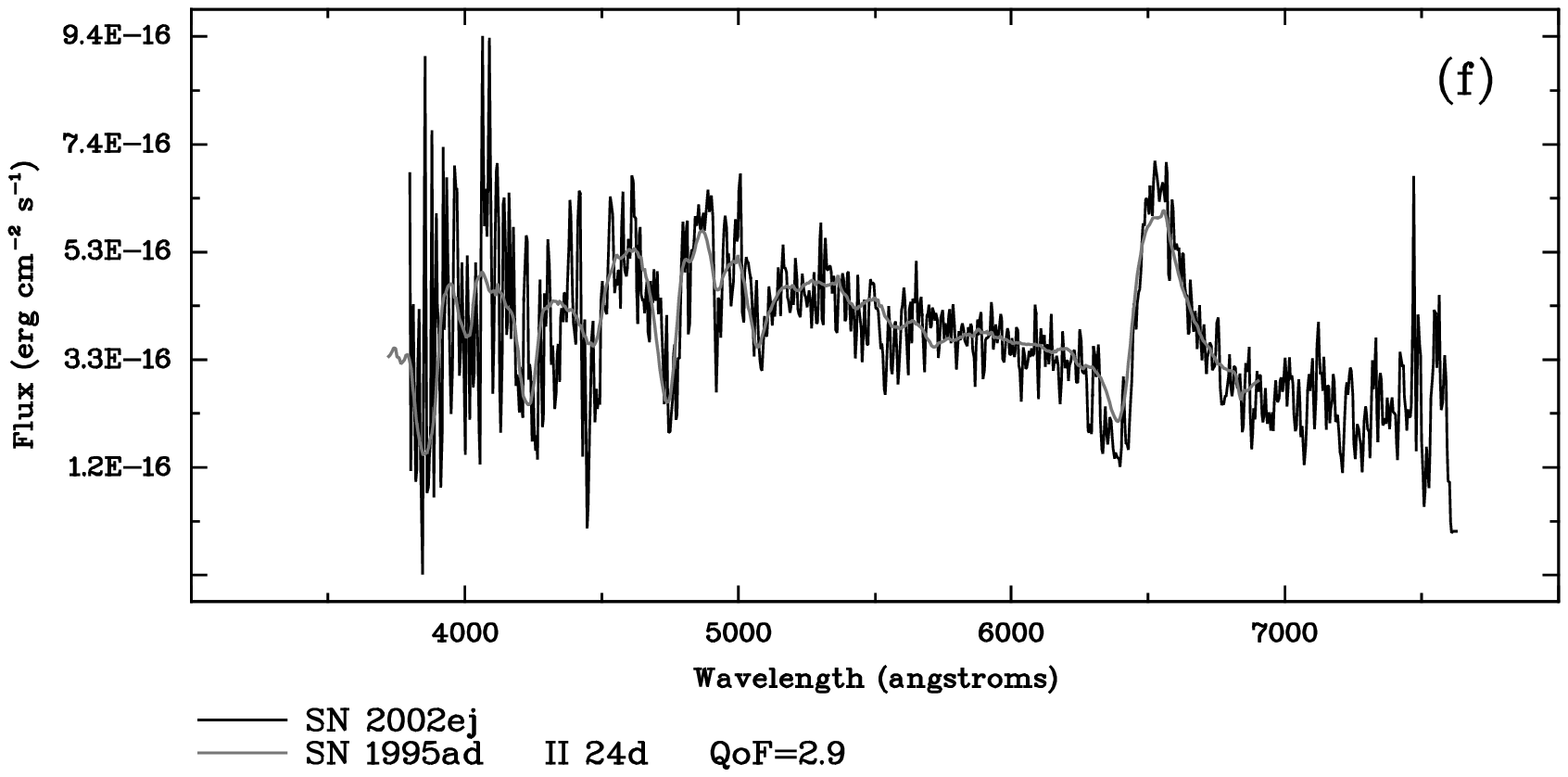}\\
\includegraphics[height=9cm, angle=270]{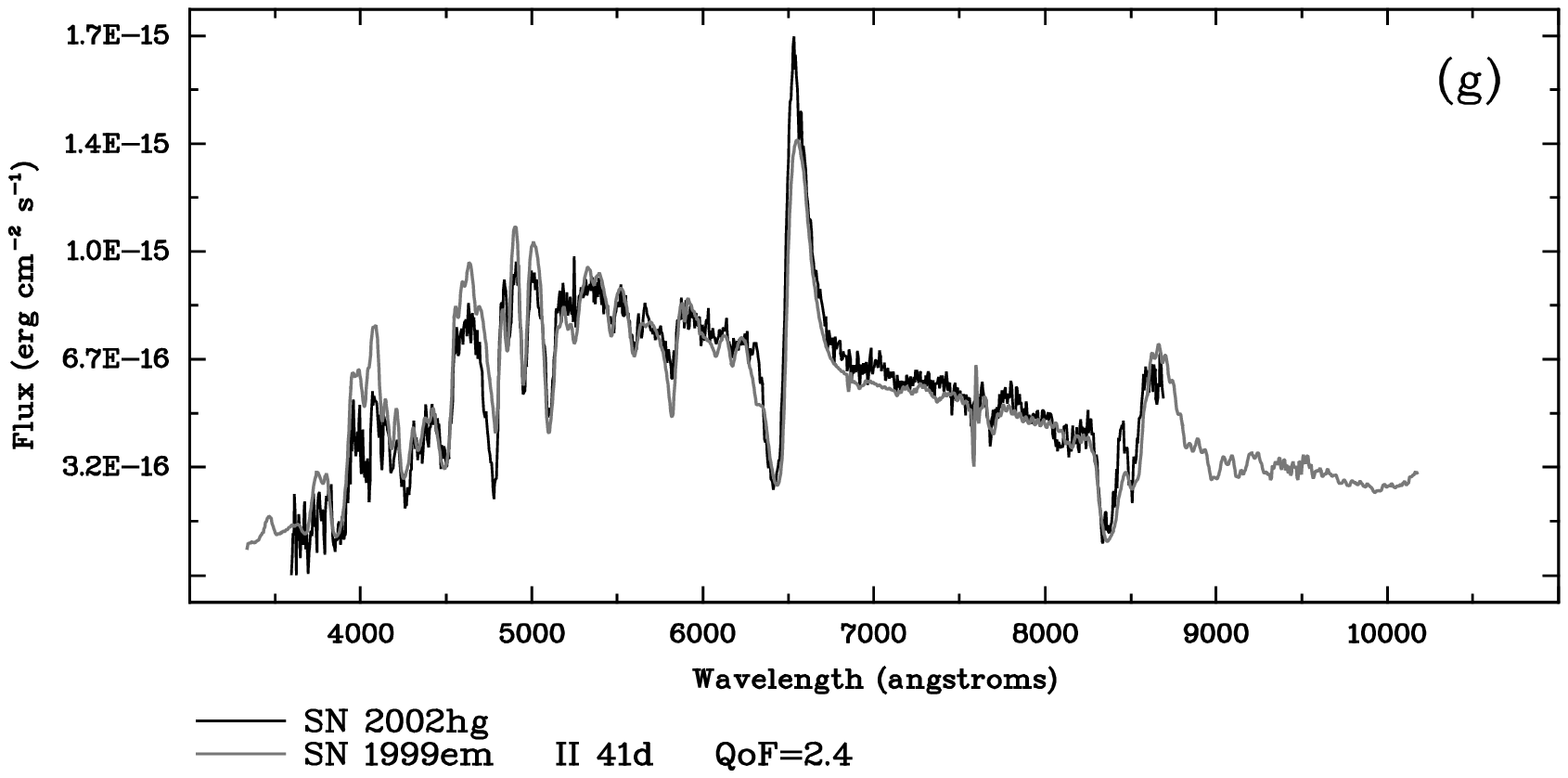}
\includegraphics[height=9cm, angle=270]{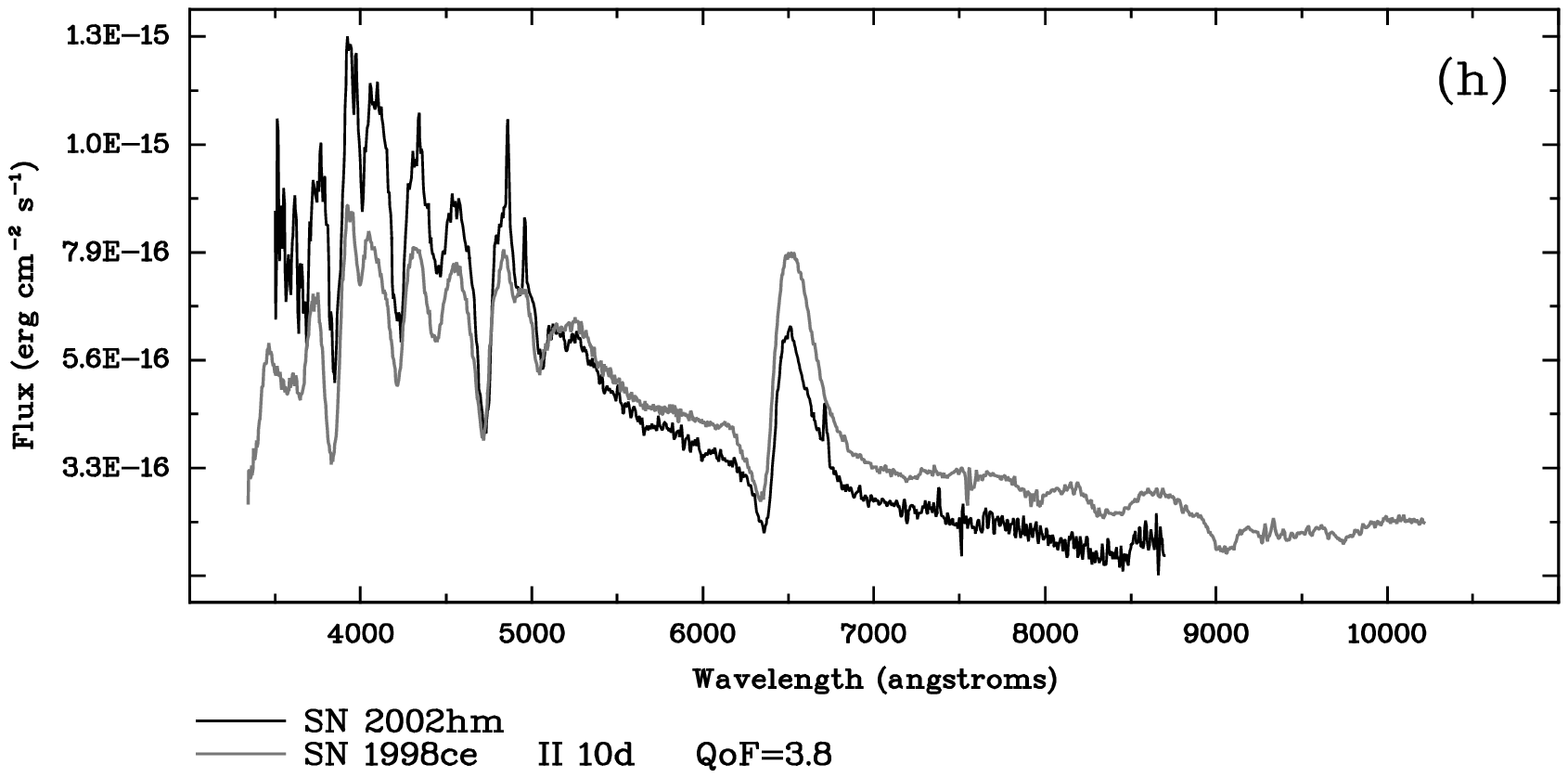}\\
\includegraphics[height=9cm, angle=270]{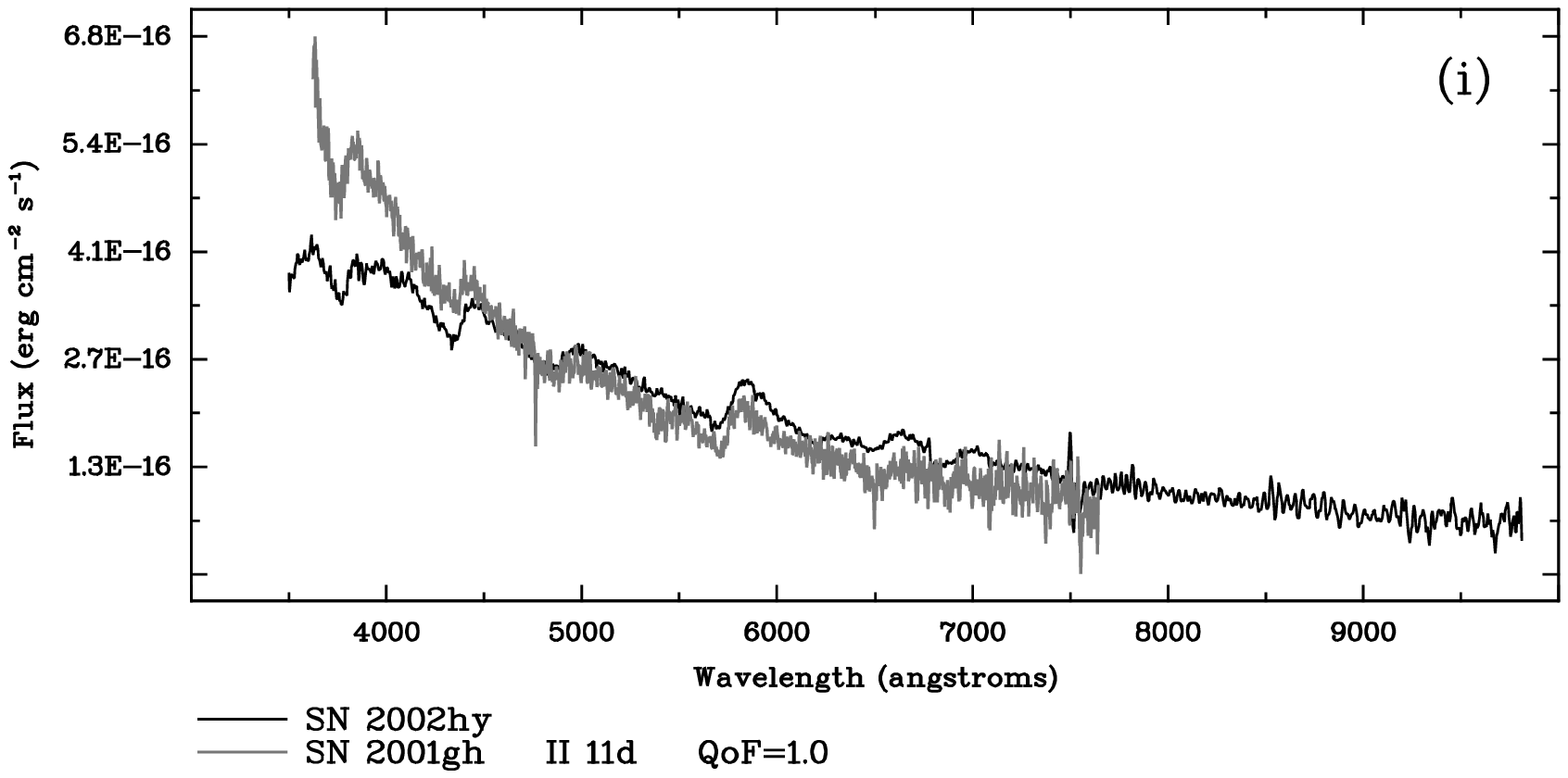}
\includegraphics[height=9cm, angle=270]{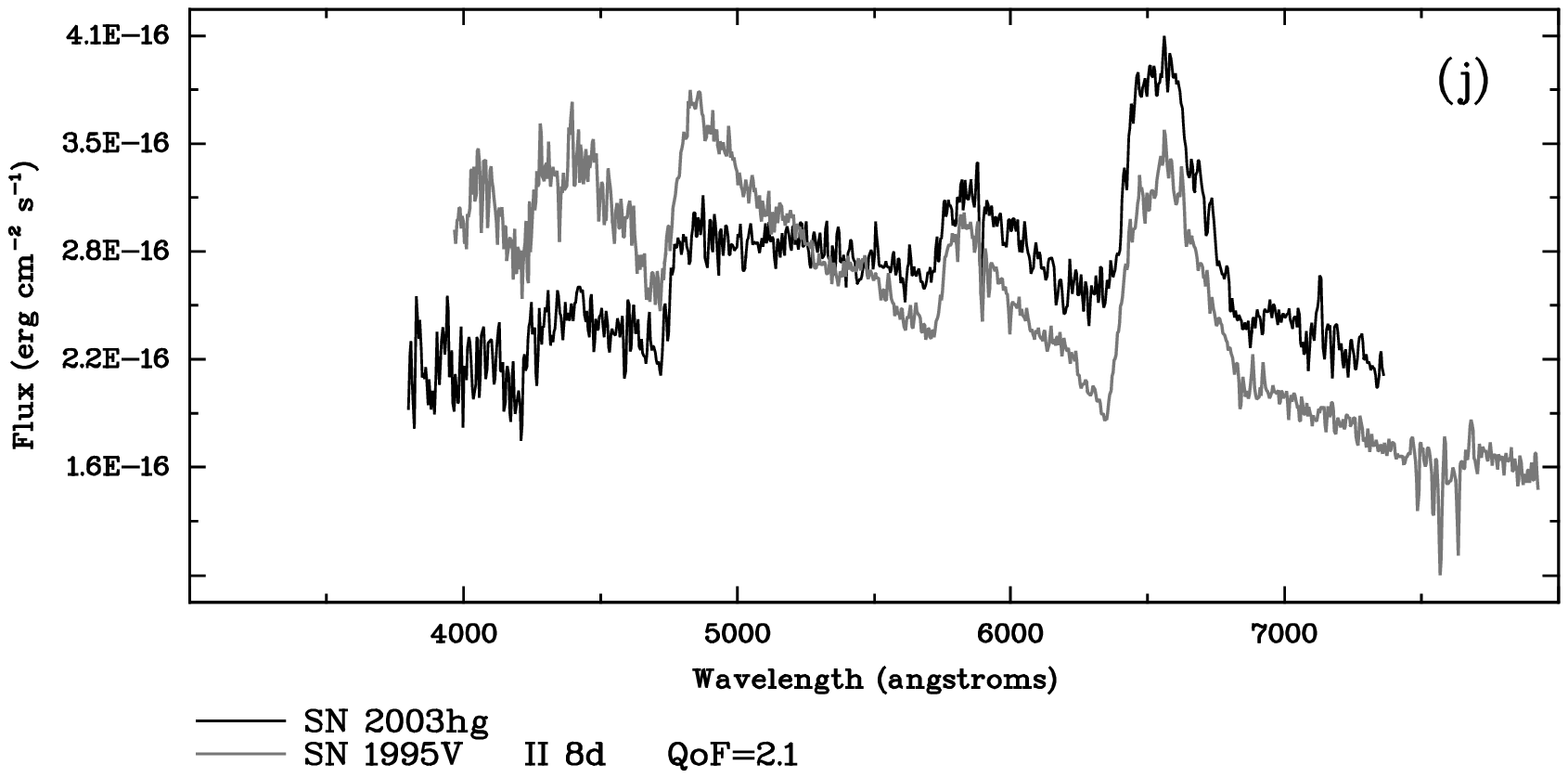}\\
\caption{The comparison of the ESC SN spectra of non-ESC
targets with their best fitting templates. The spectra
are in the parent galaxy restframe and not corrected for
extinction. The black lines are the ESC spectra, while
the gray ones display template spectra.}
\label{plt1}
\end{figure*}

\begin{figure*}
\includegraphics[height=9cm, angle=270]{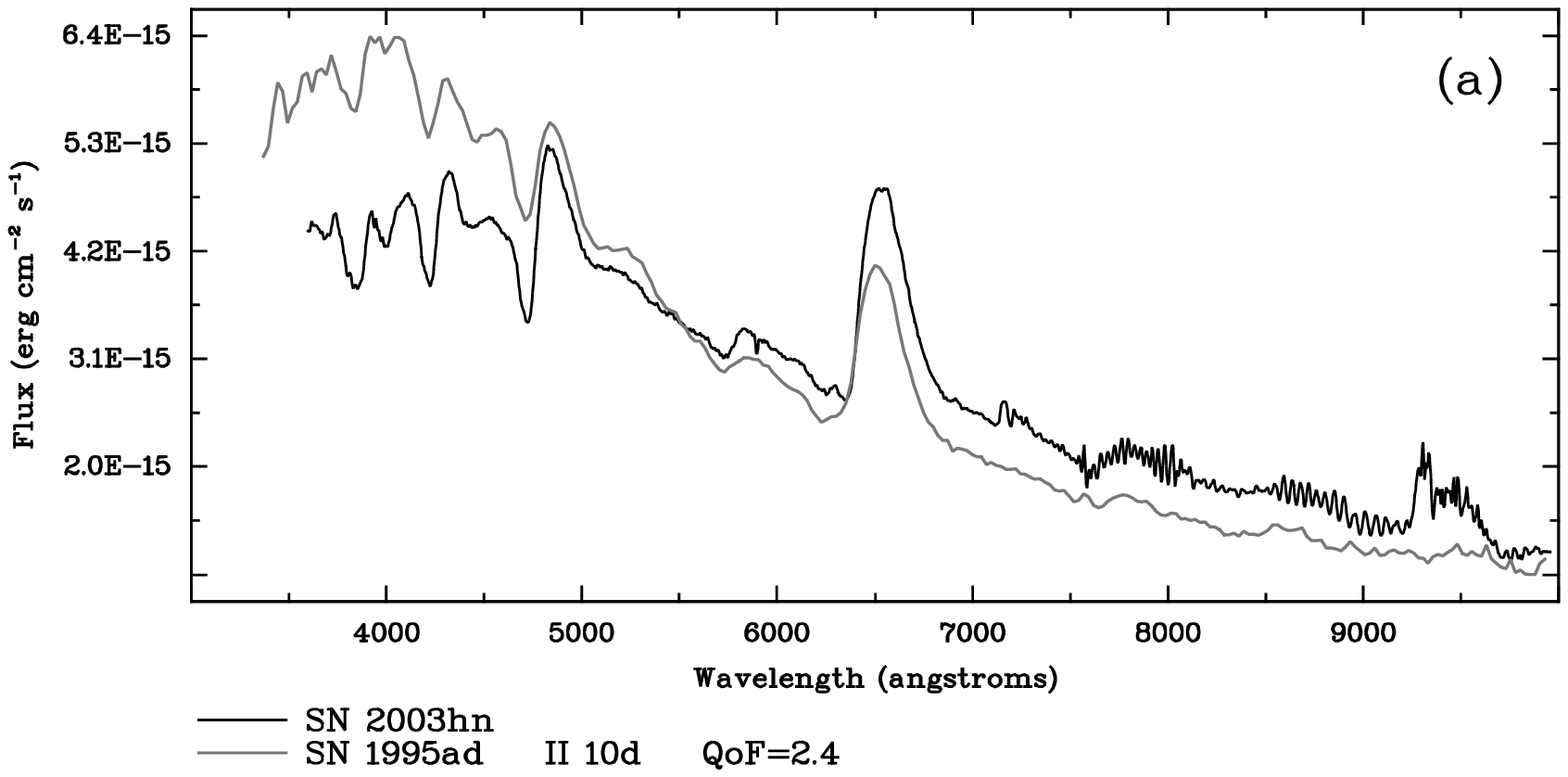}
\includegraphics[height=9cm, angle=270]{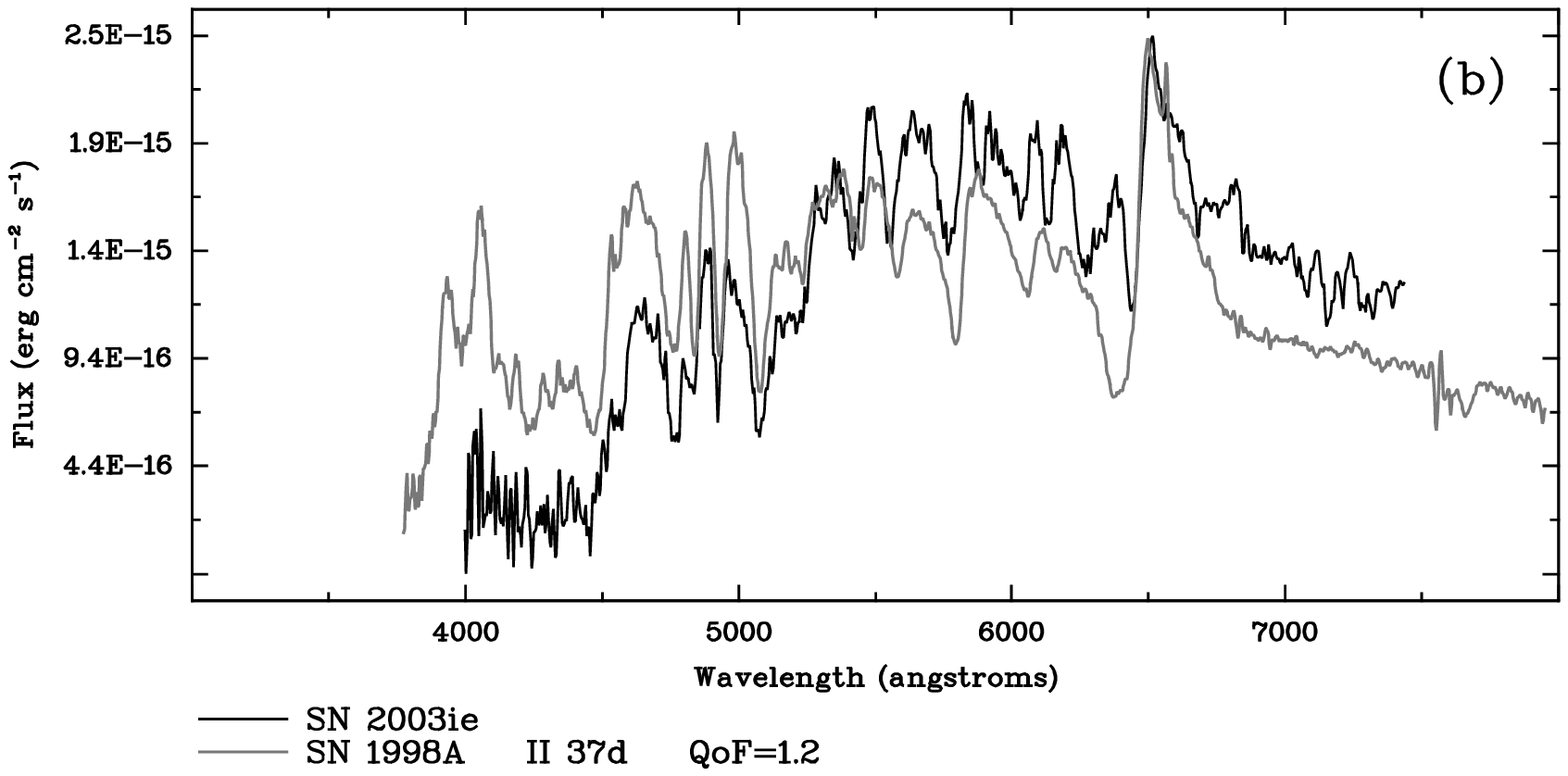}\\
\includegraphics[height=9cm, angle=270]{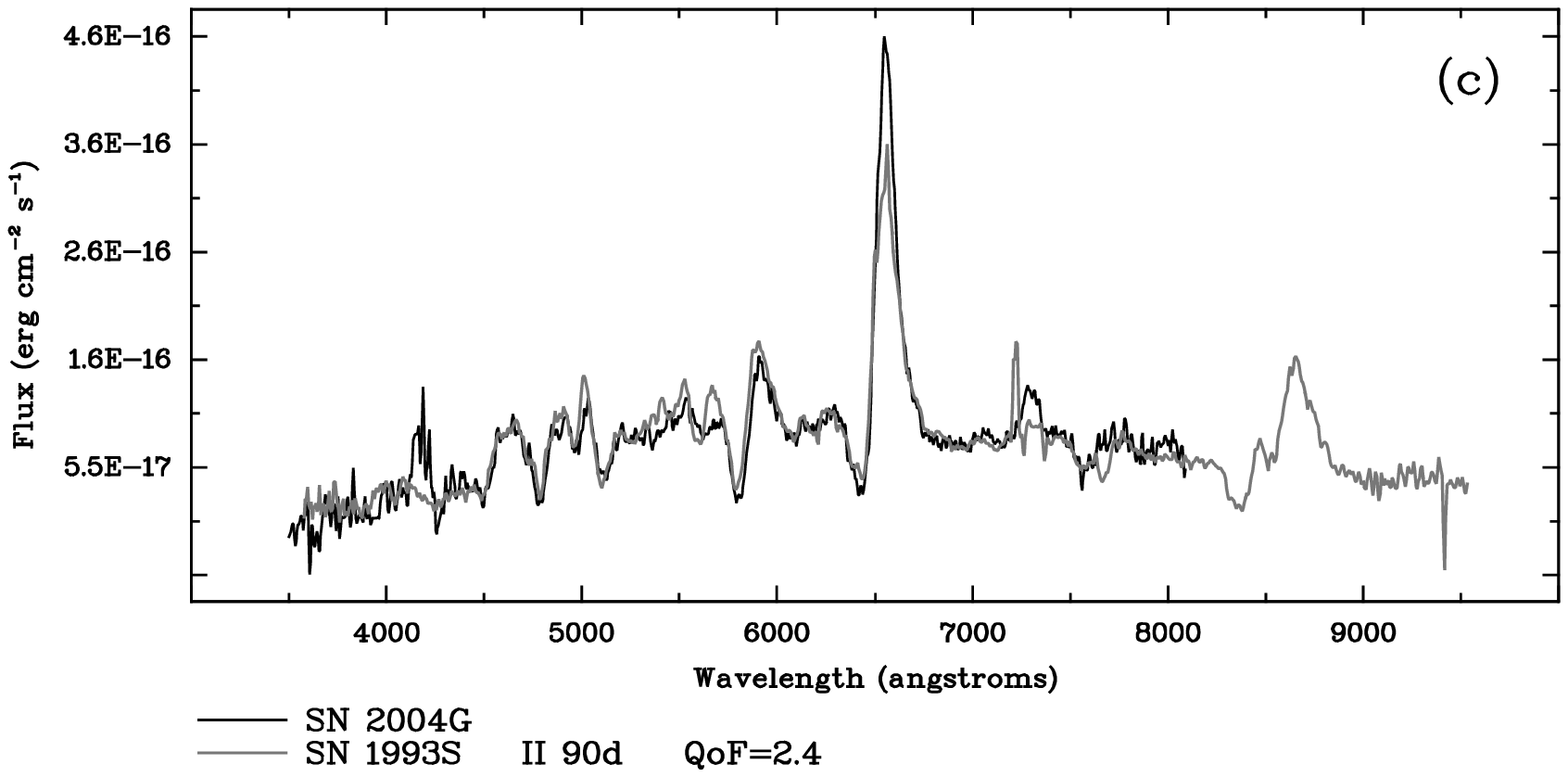}
\includegraphics[height=9cm, angle=270]{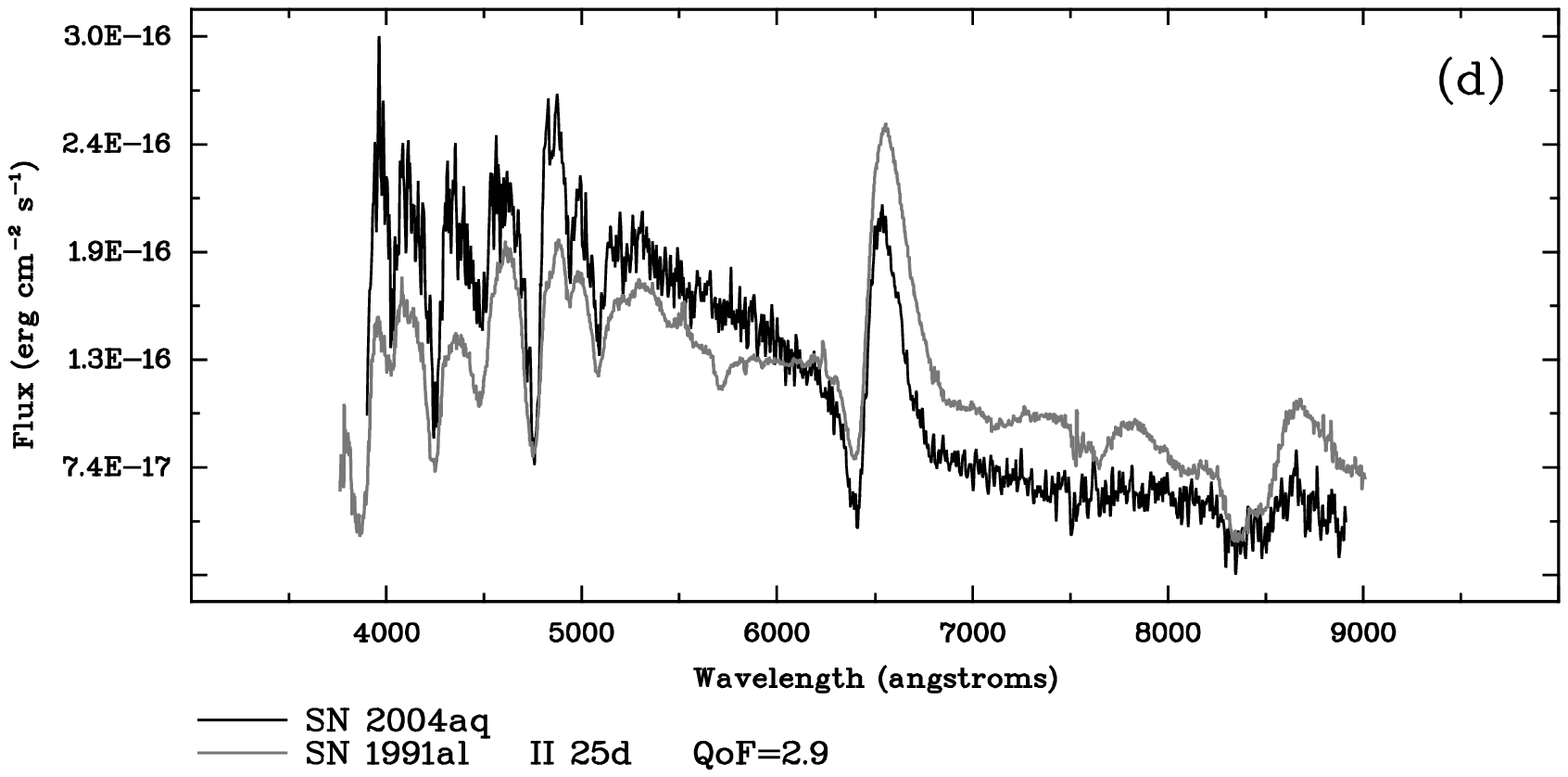}\\
\includegraphics[height=9cm, angle=270]{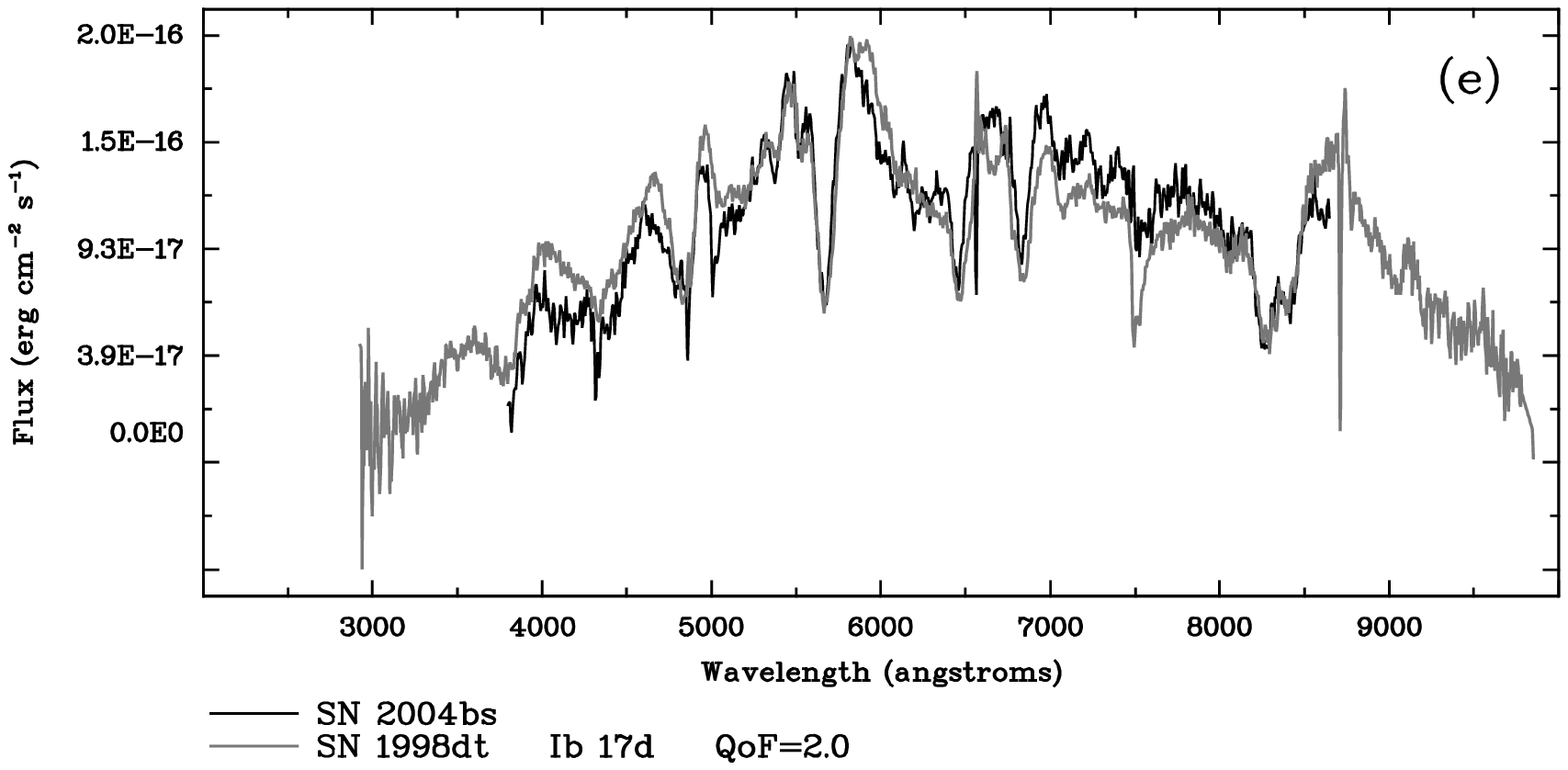}
\includegraphics[height=9cm, angle=270]{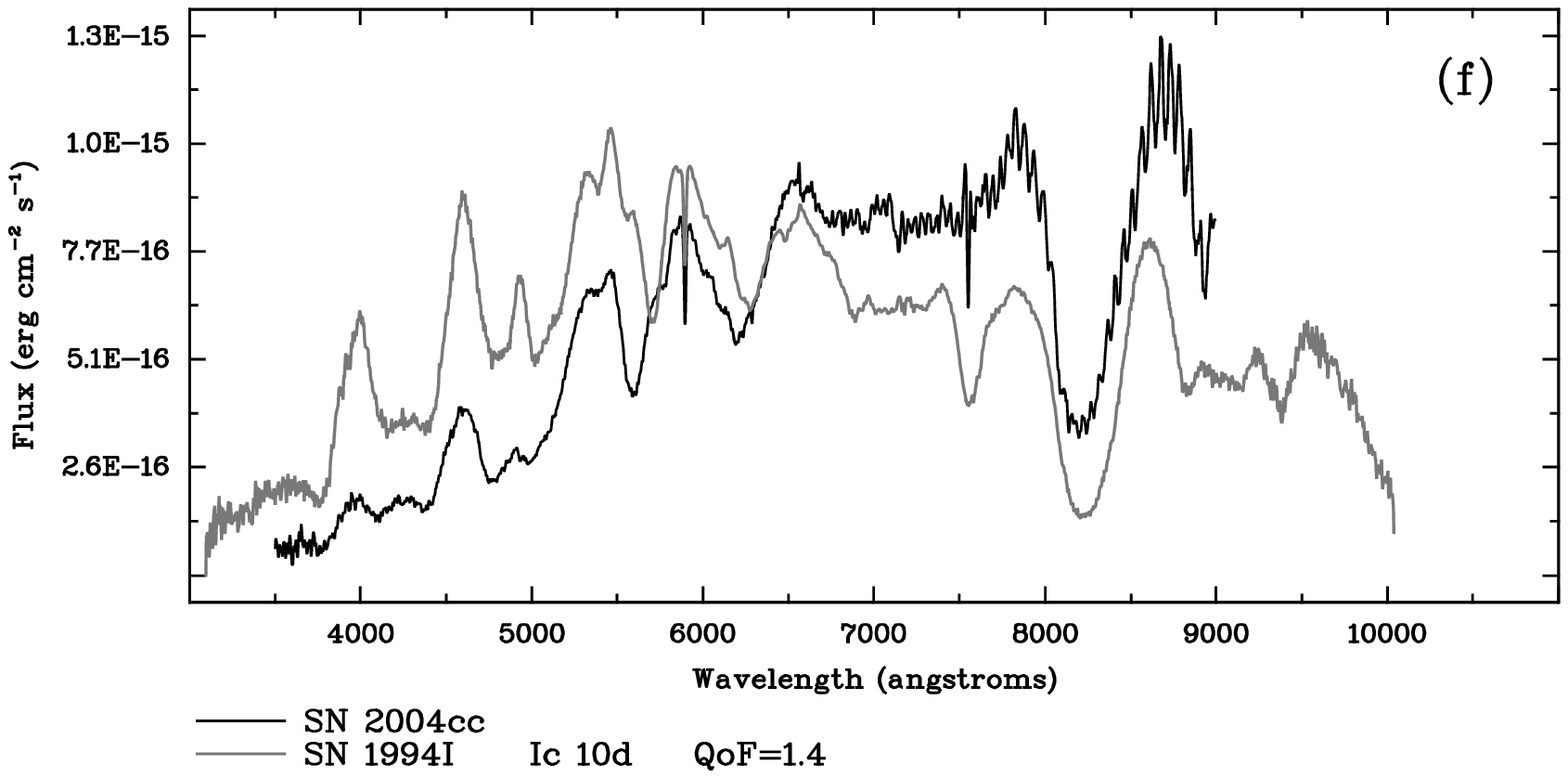}\\
\includegraphics[height=9cm, angle=270]{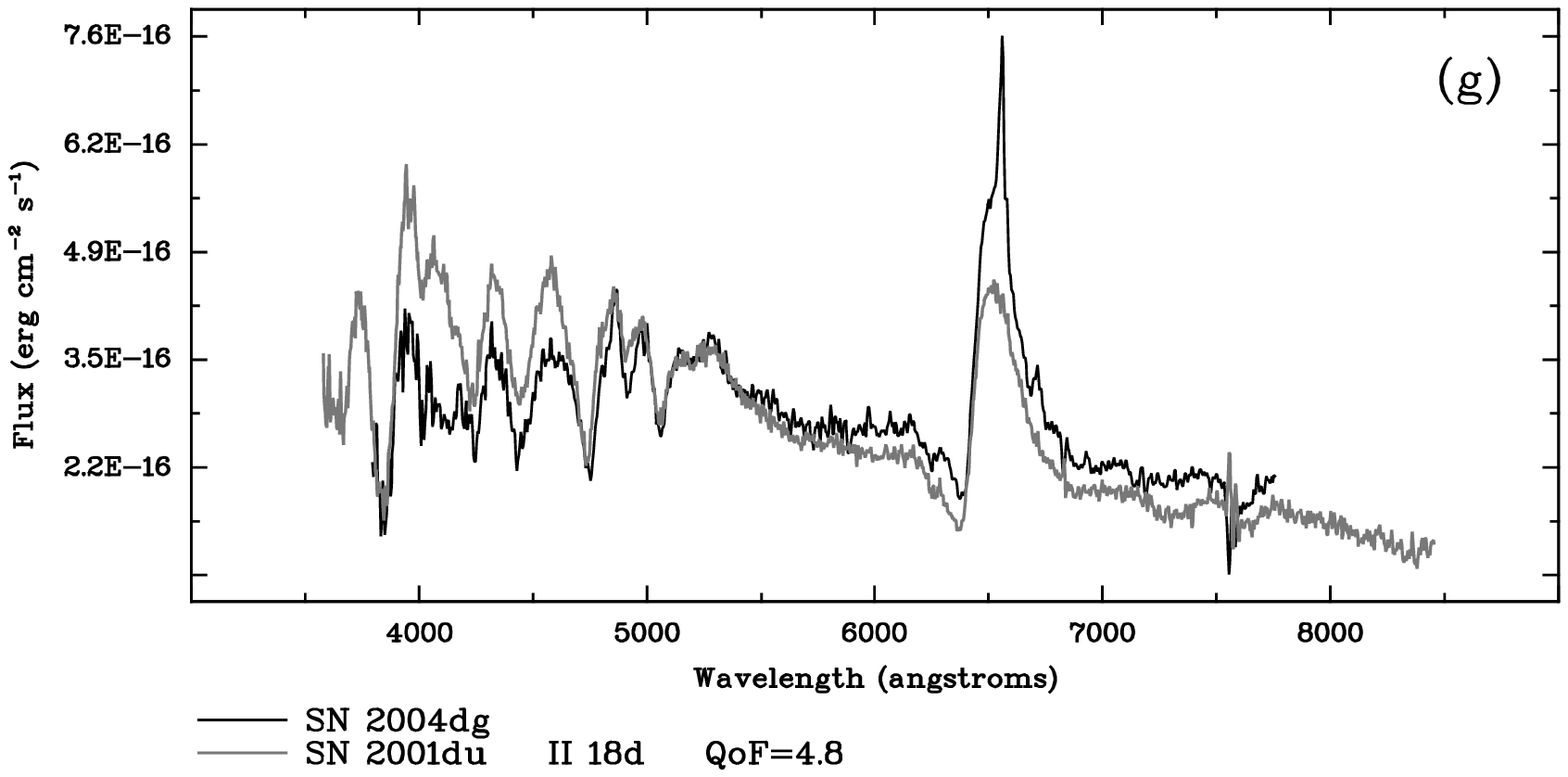}
\includegraphics[height=9cm, angle=270]{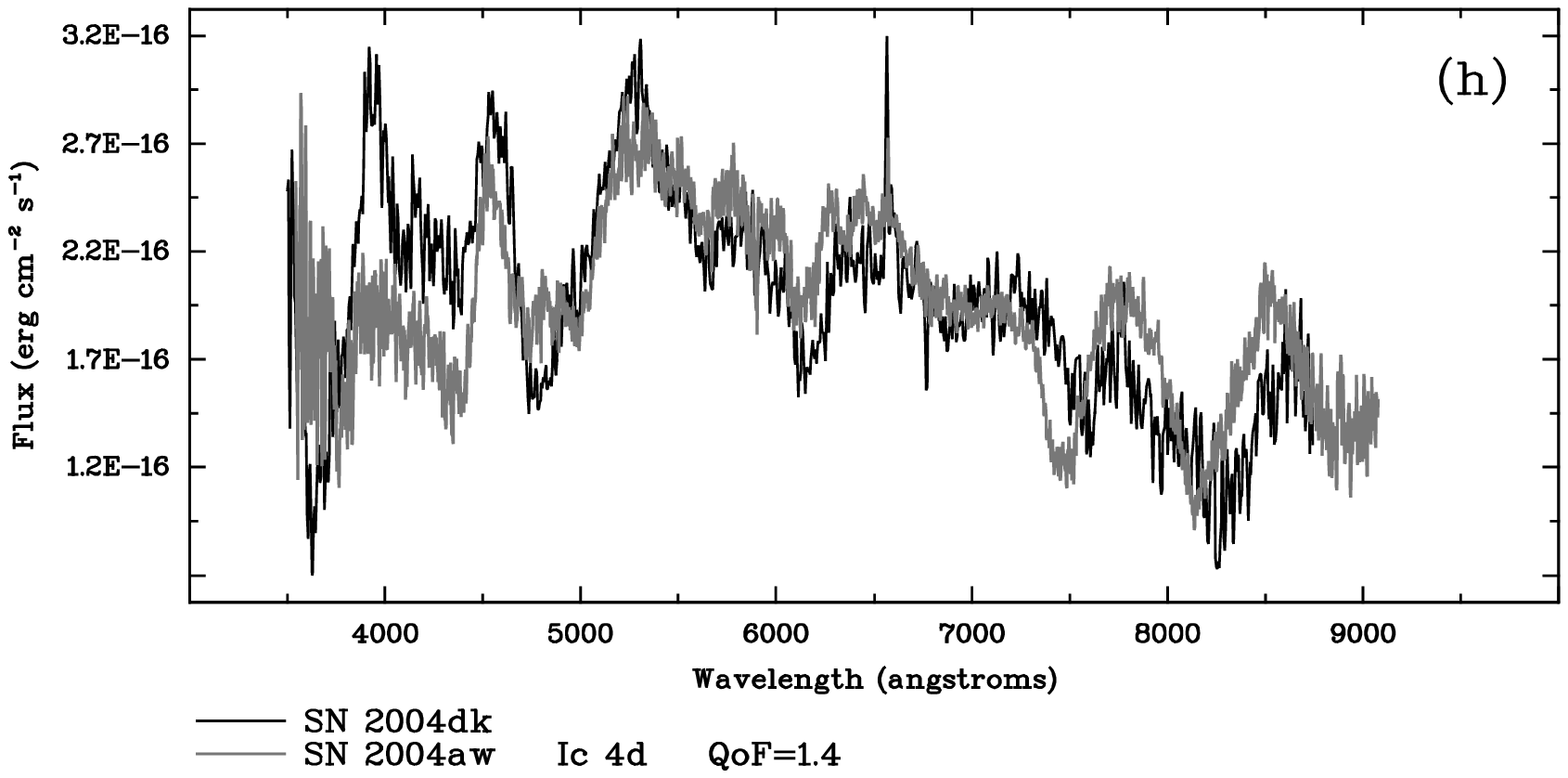}\\
\includegraphics[height=9cm, angle=270]{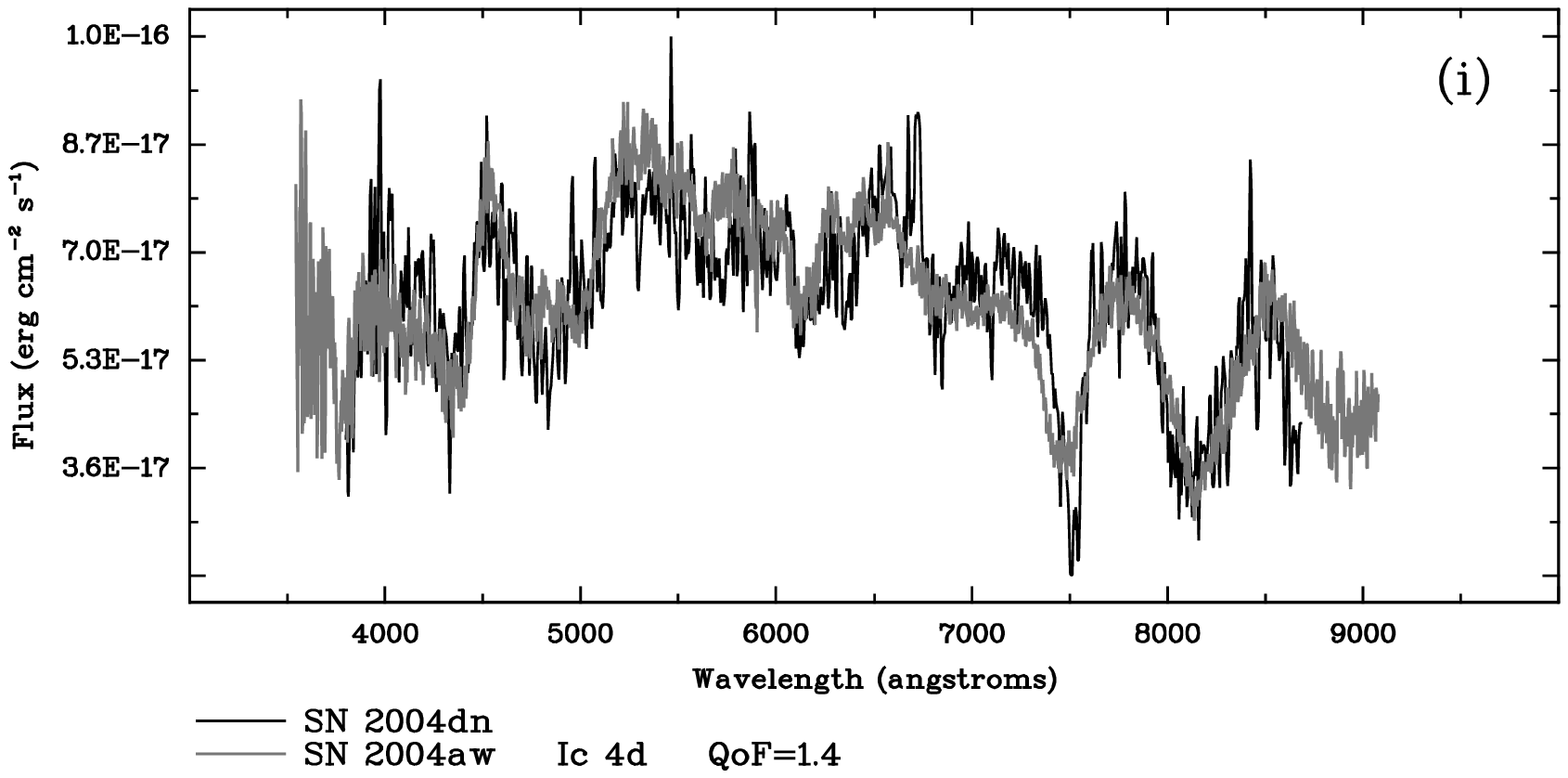}
\includegraphics[height=9cm, angle=270]{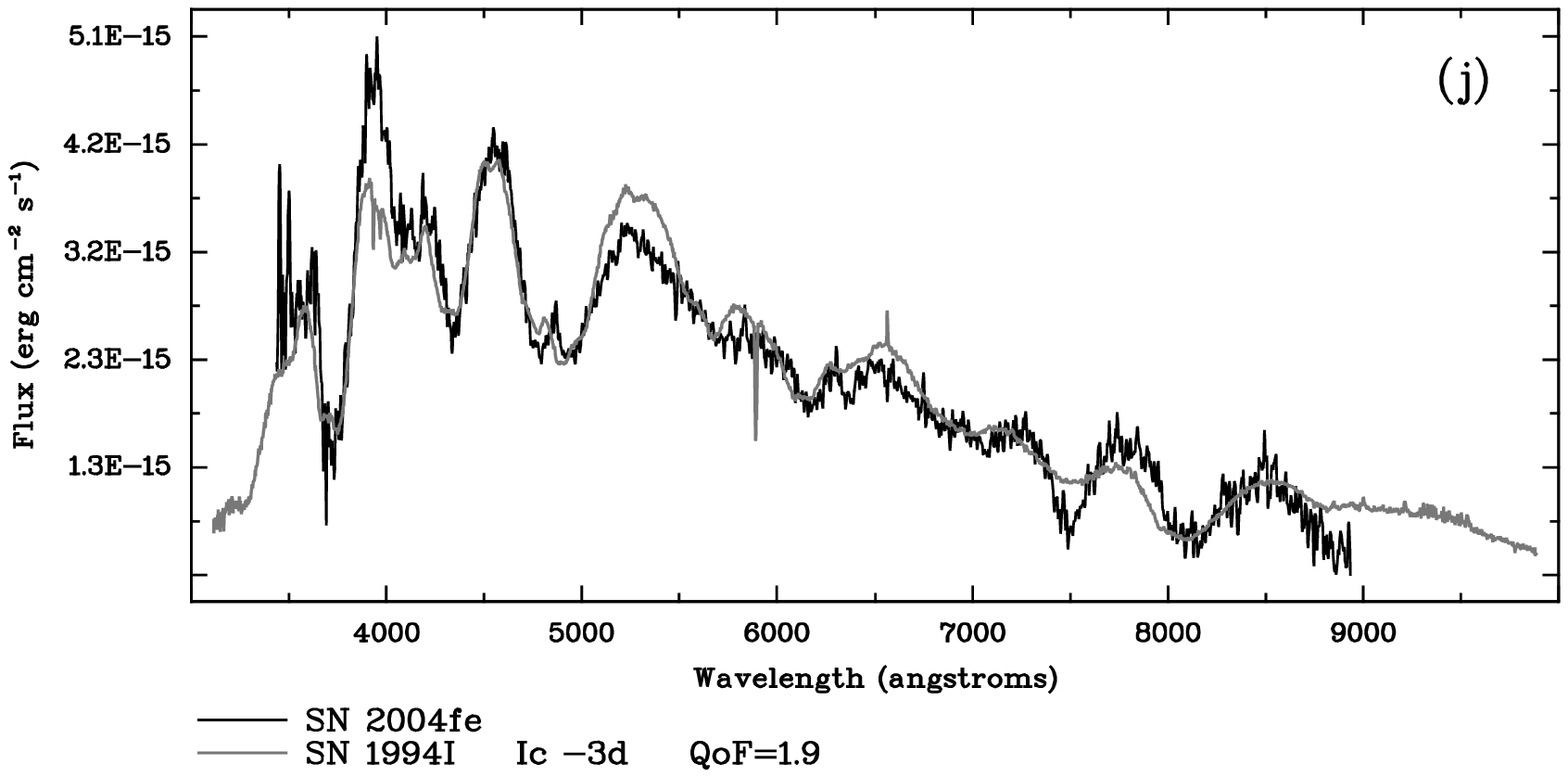}\\
\caption{Same as Fig. \ref{plt1}.}
\label{plt2}
\end{figure*}

\begin{figure*}
\includegraphics[height=9cm, angle=270]{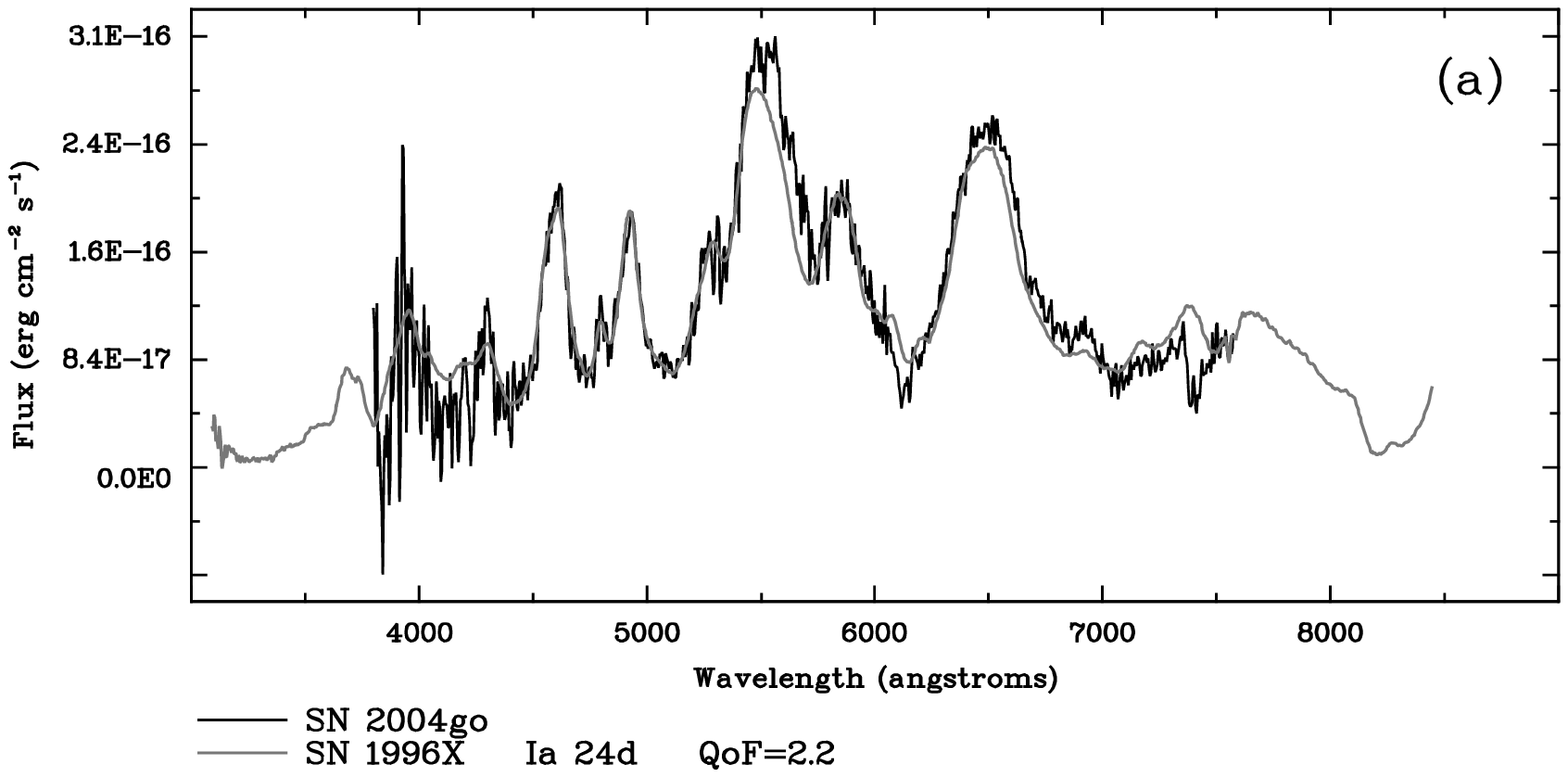}
\includegraphics[height=9cm, angle=270]{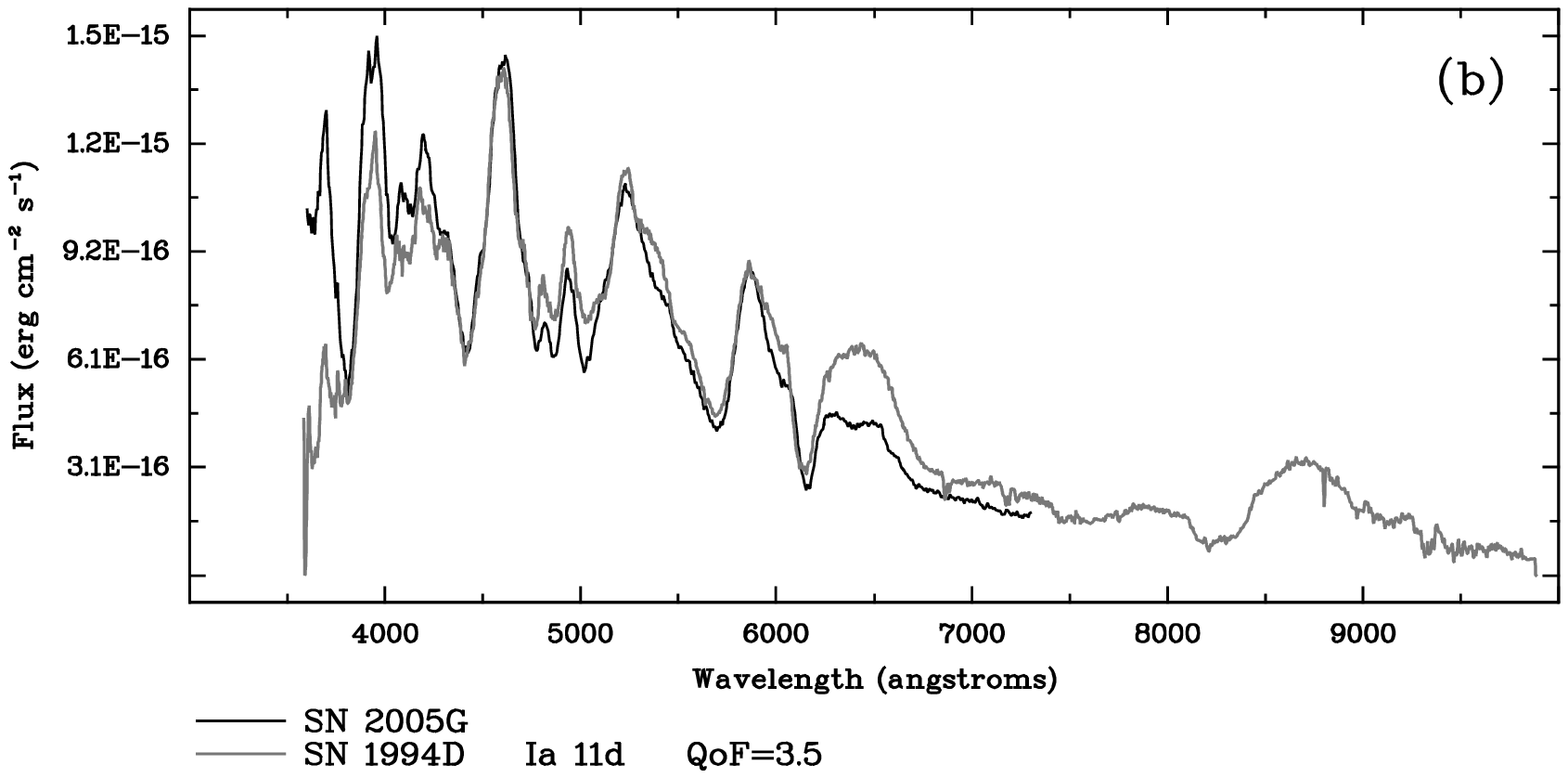}\\
\includegraphics[height=9cm, angle=270]{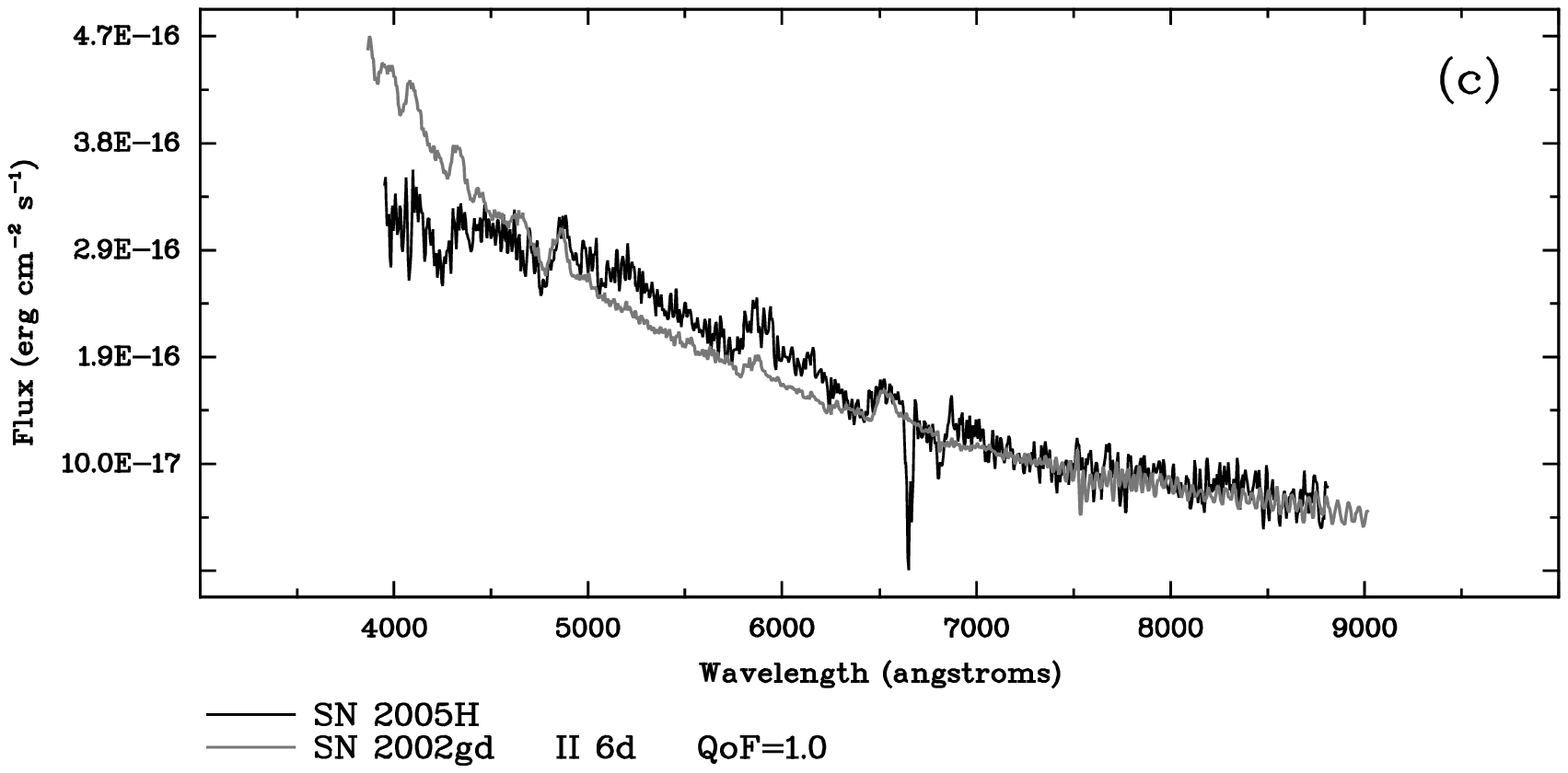}
\includegraphics[height=9cm, angle=270]{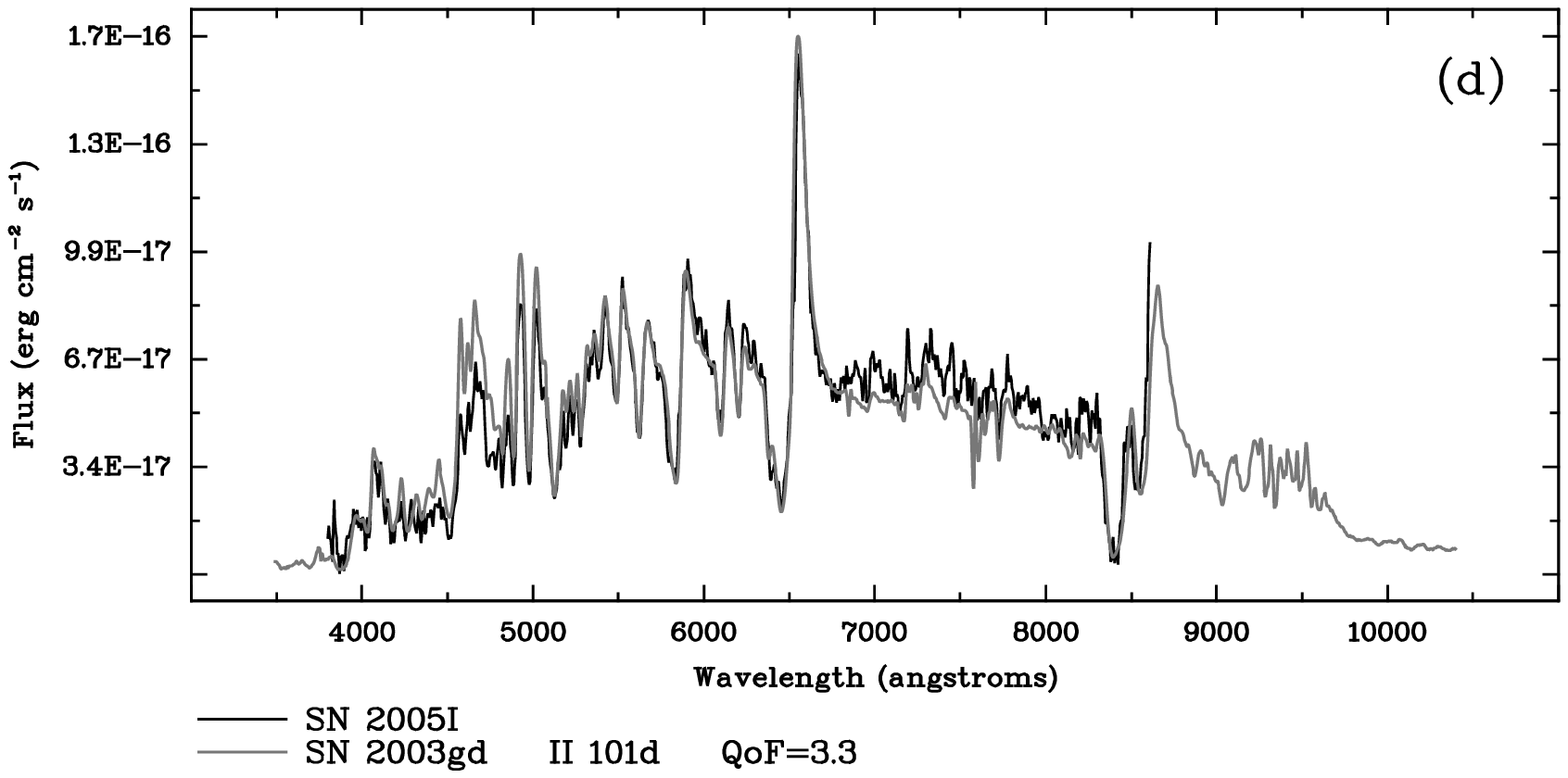}\\
\includegraphics[height=9cm, angle=270]{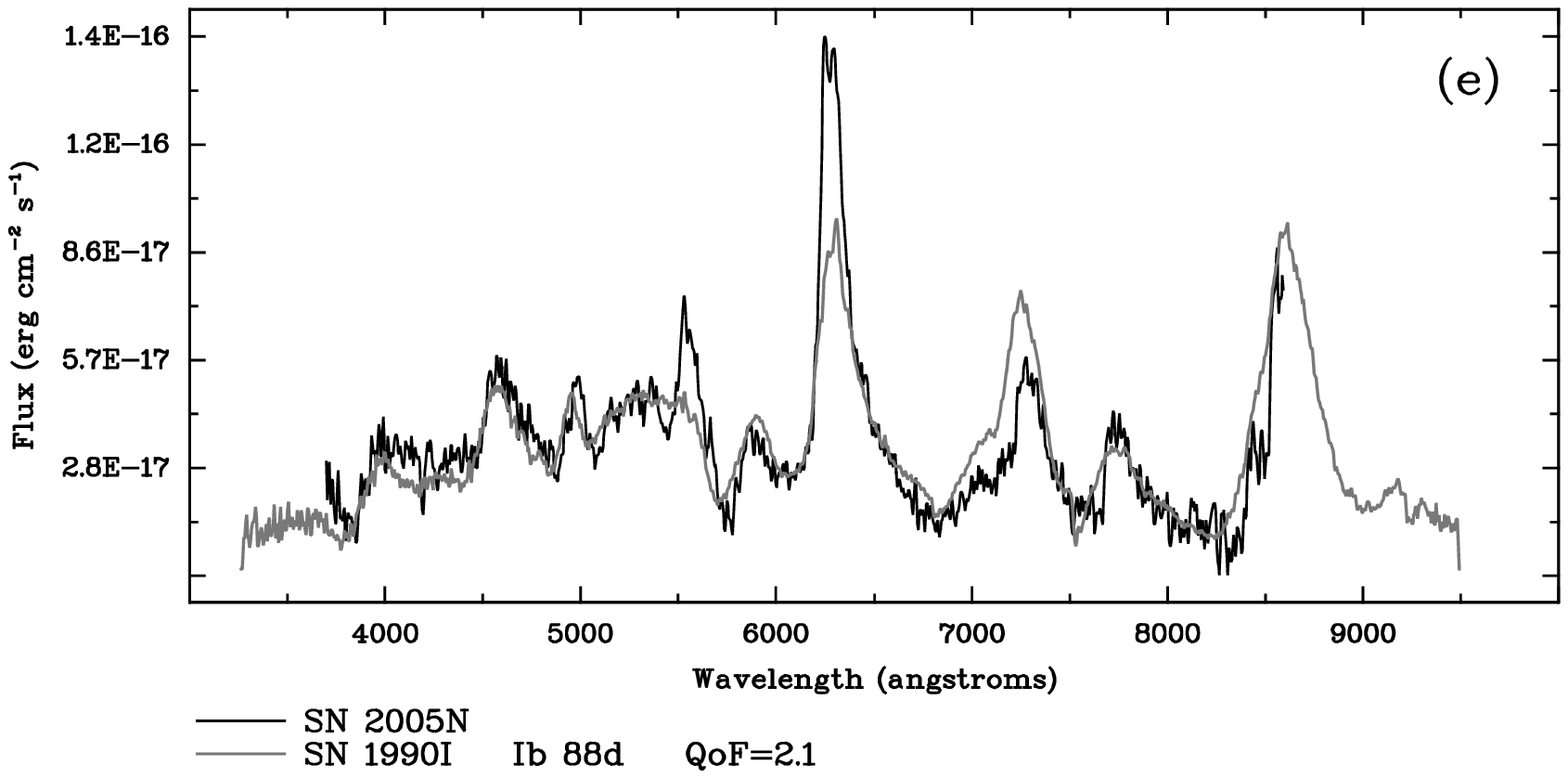}
\includegraphics[height=9cm, angle=270]{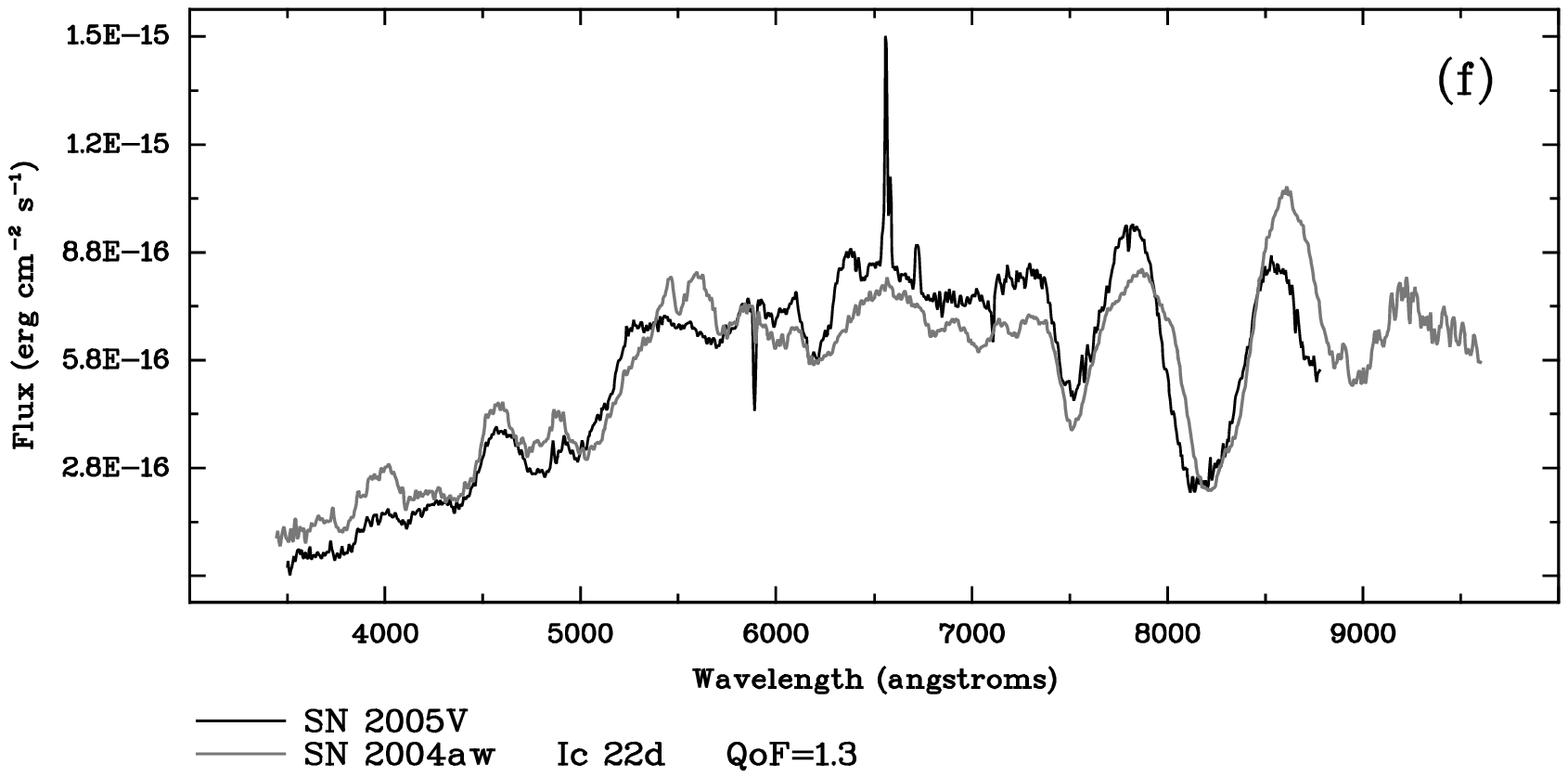}\\
\includegraphics[height=9cm, angle=270]{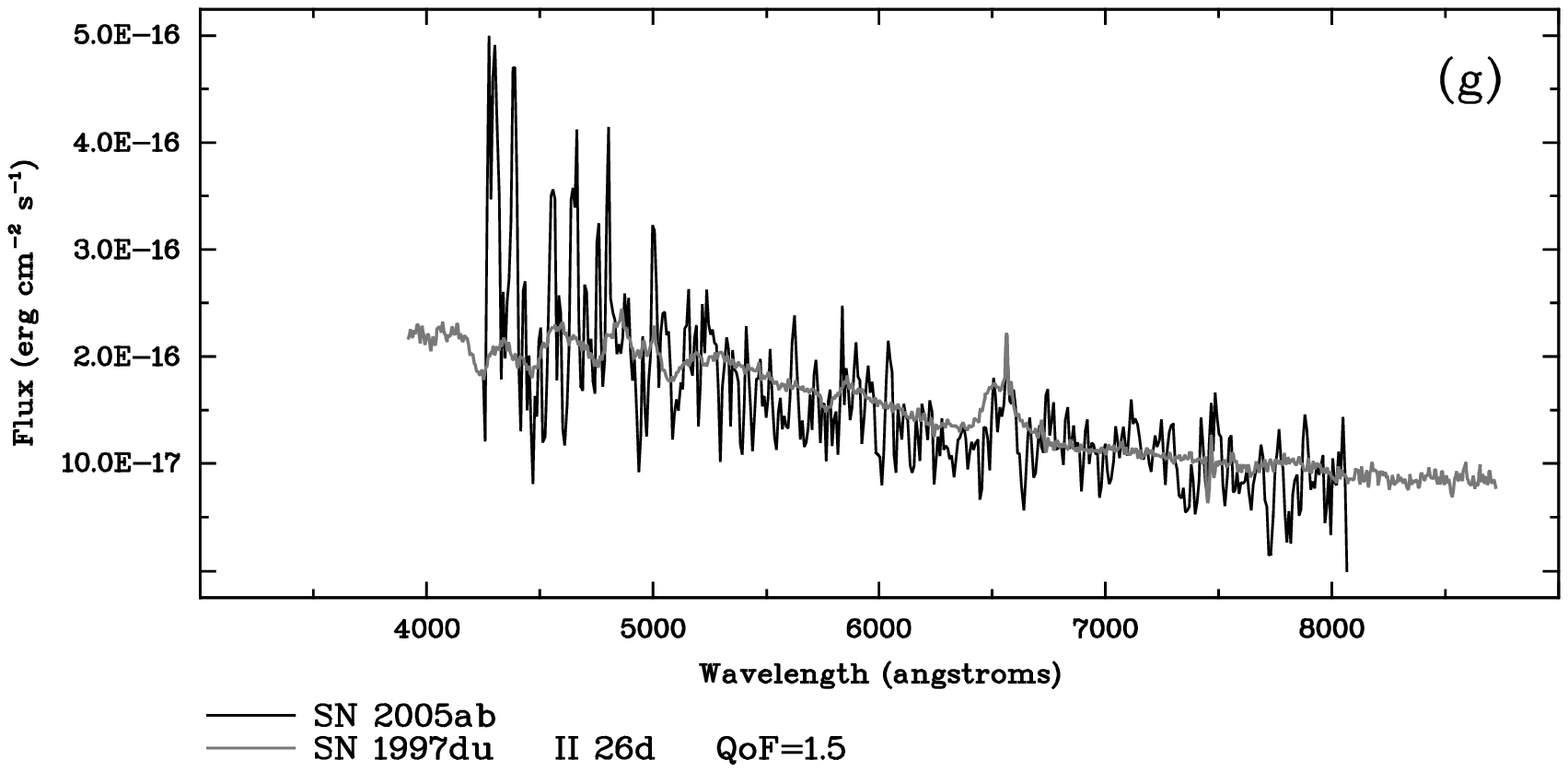}
\includegraphics[height=9cm, angle=270]{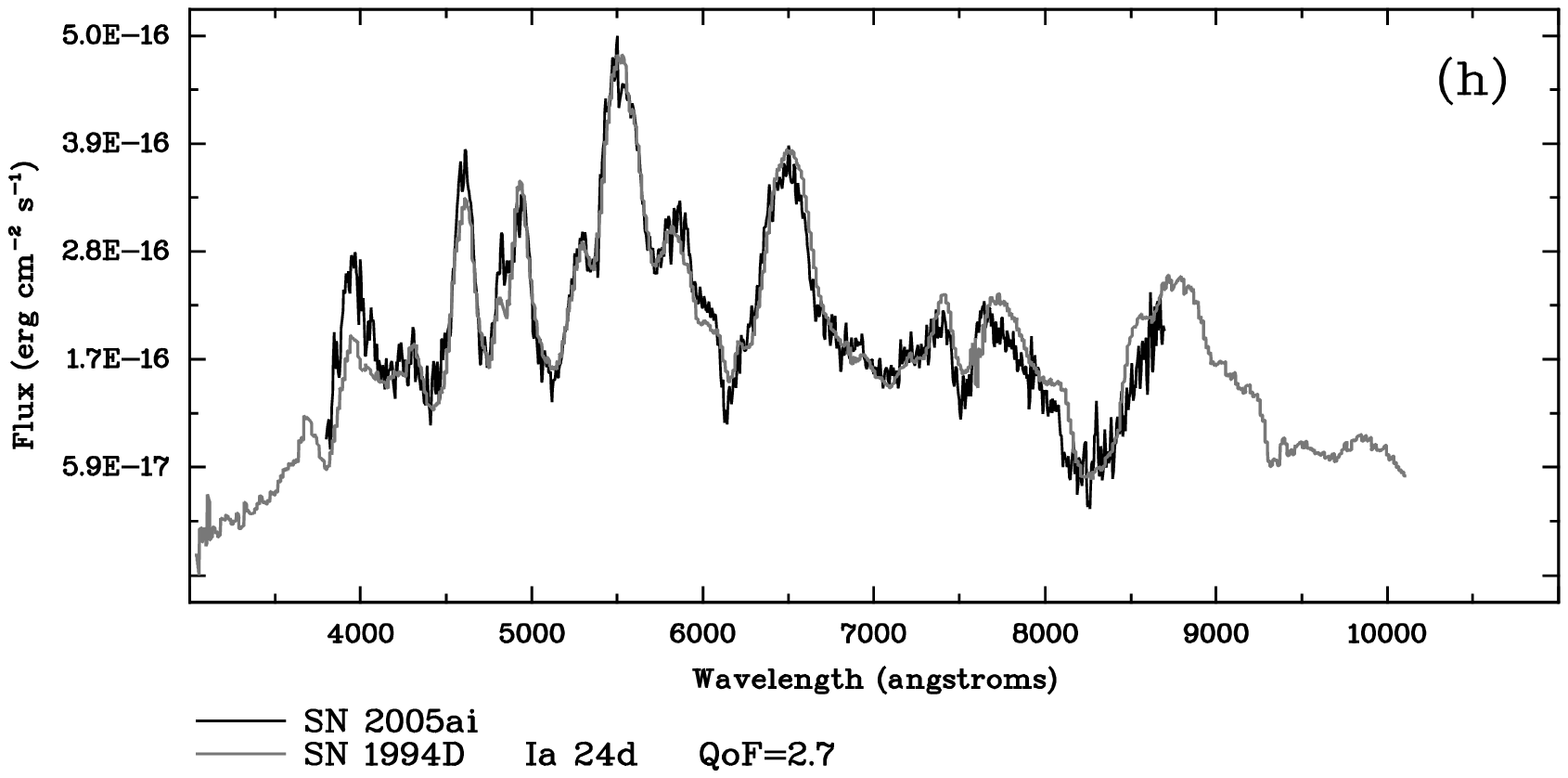}\\
\includegraphics[height=9cm, angle=270]{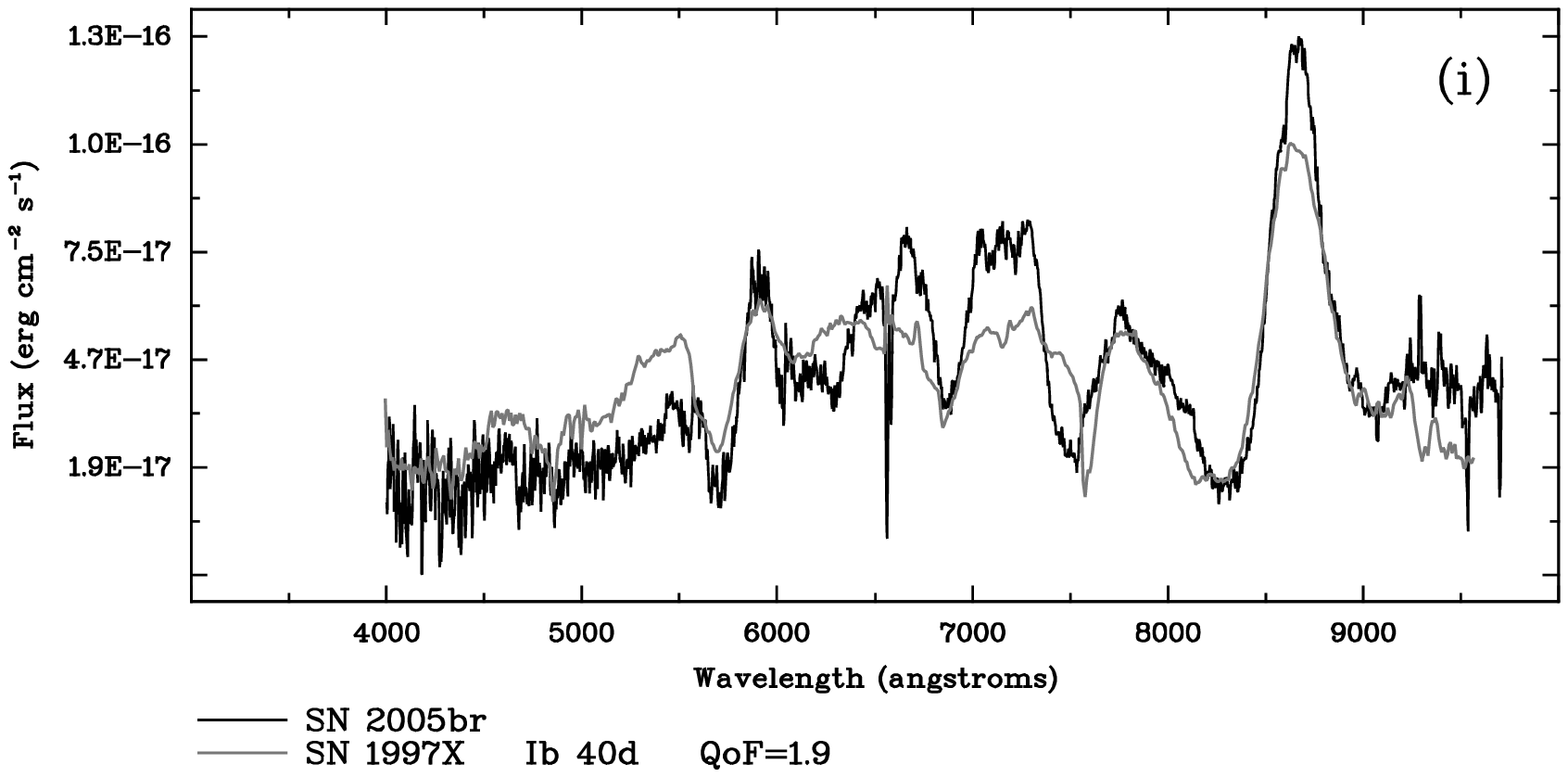}
\includegraphics[height=9cm, angle=270]{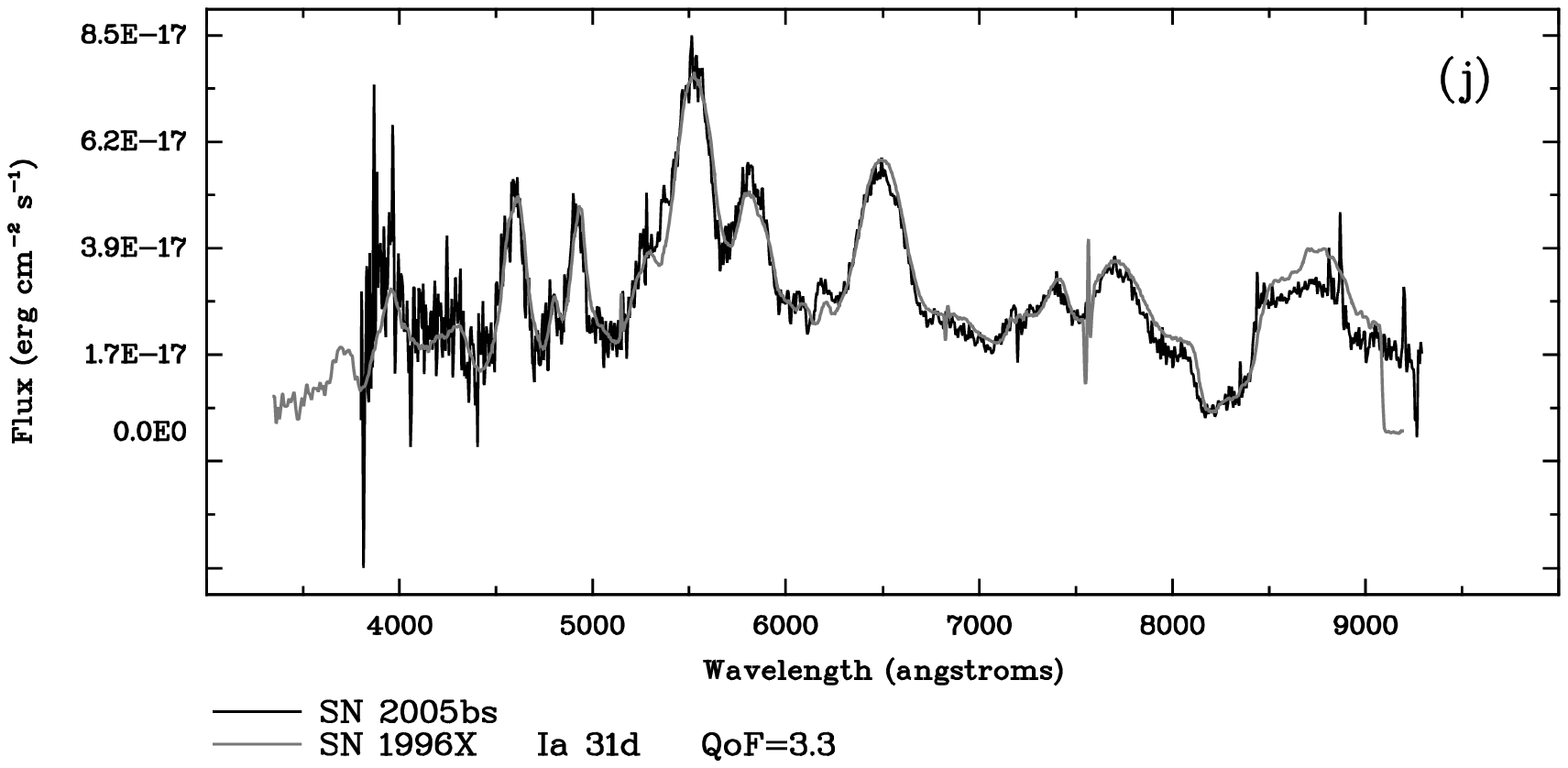}\\
\caption{Same as Fig. \ref{plt1}.}
\label{plt3}
\end{figure*}

\begin{figure*}
\includegraphics[height=9cm, angle=270]{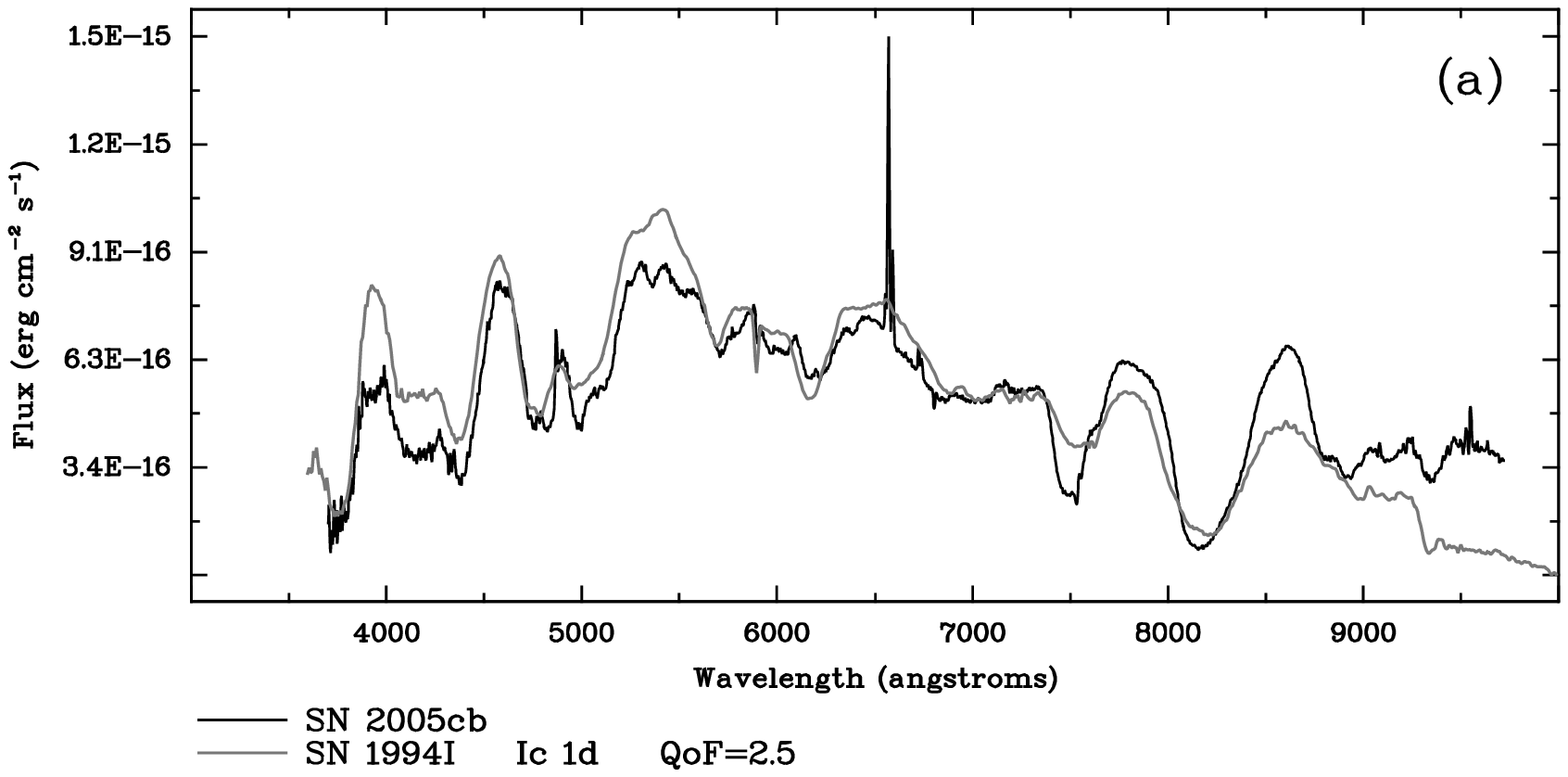}
\includegraphics[height=9cm, angle=270]{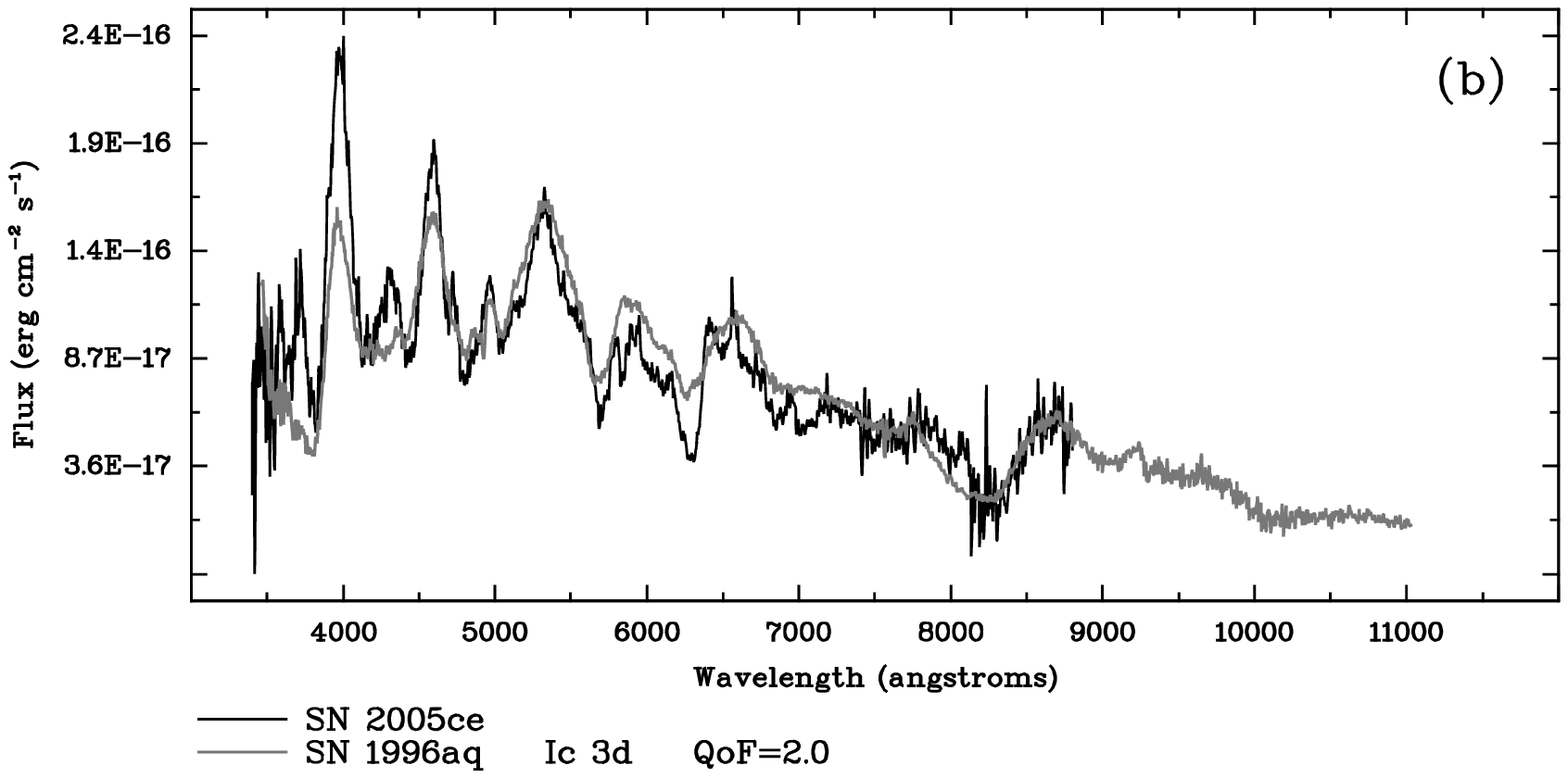}\\
\includegraphics[height=9cm, angle=270]{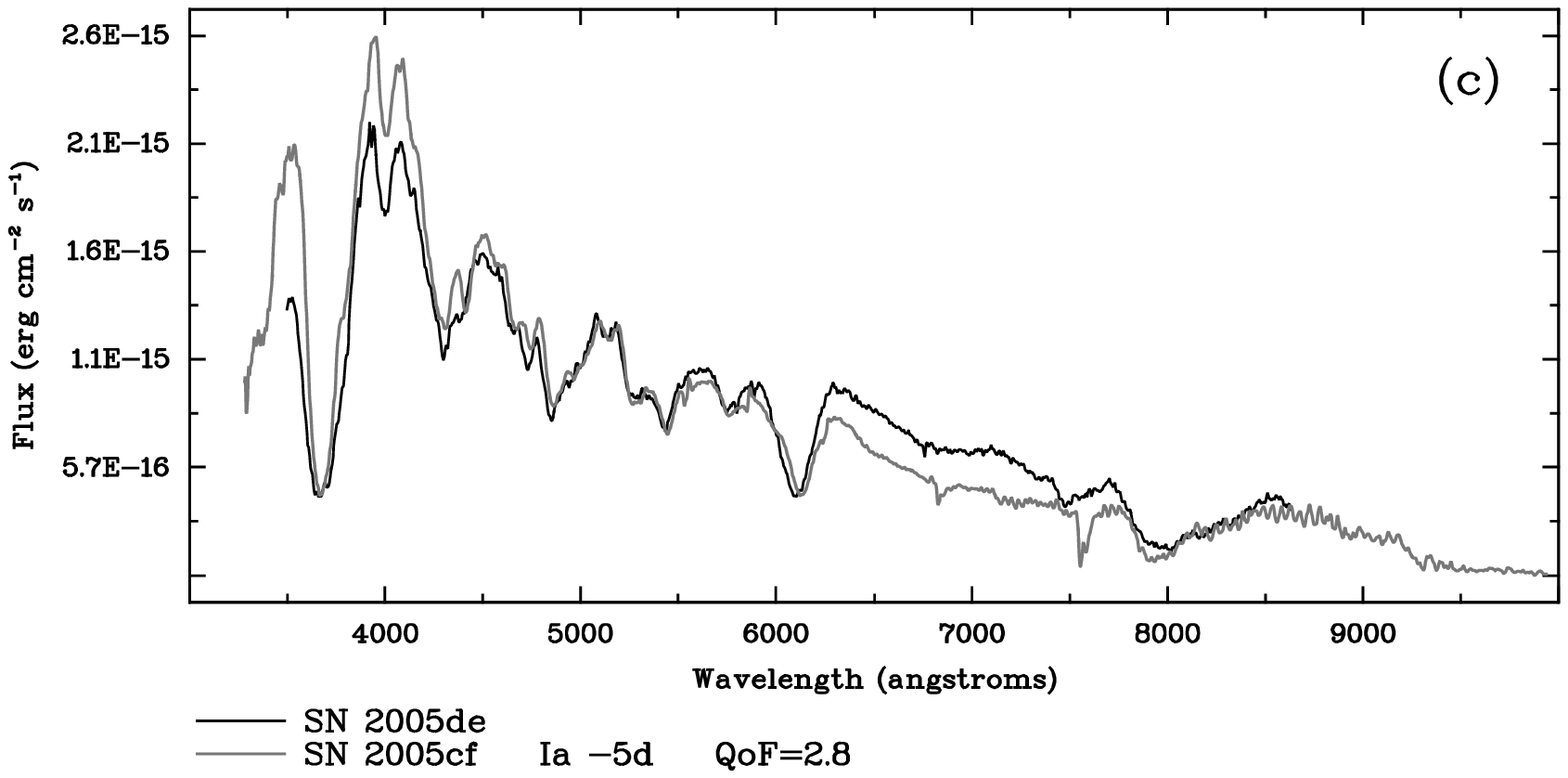}
\includegraphics[height=9cm, angle=270]{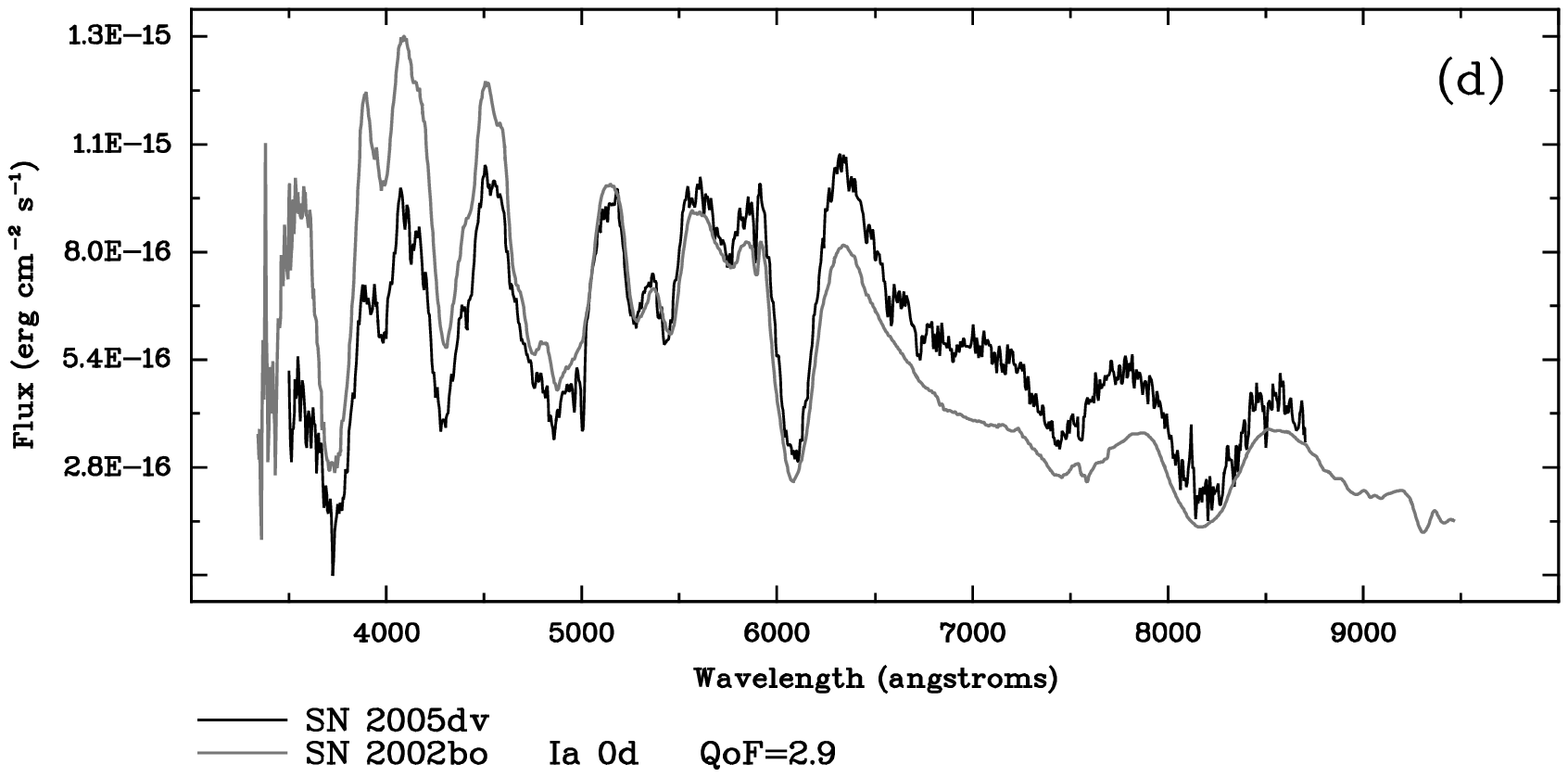}\\
\includegraphics[height=9cm, angle=270]{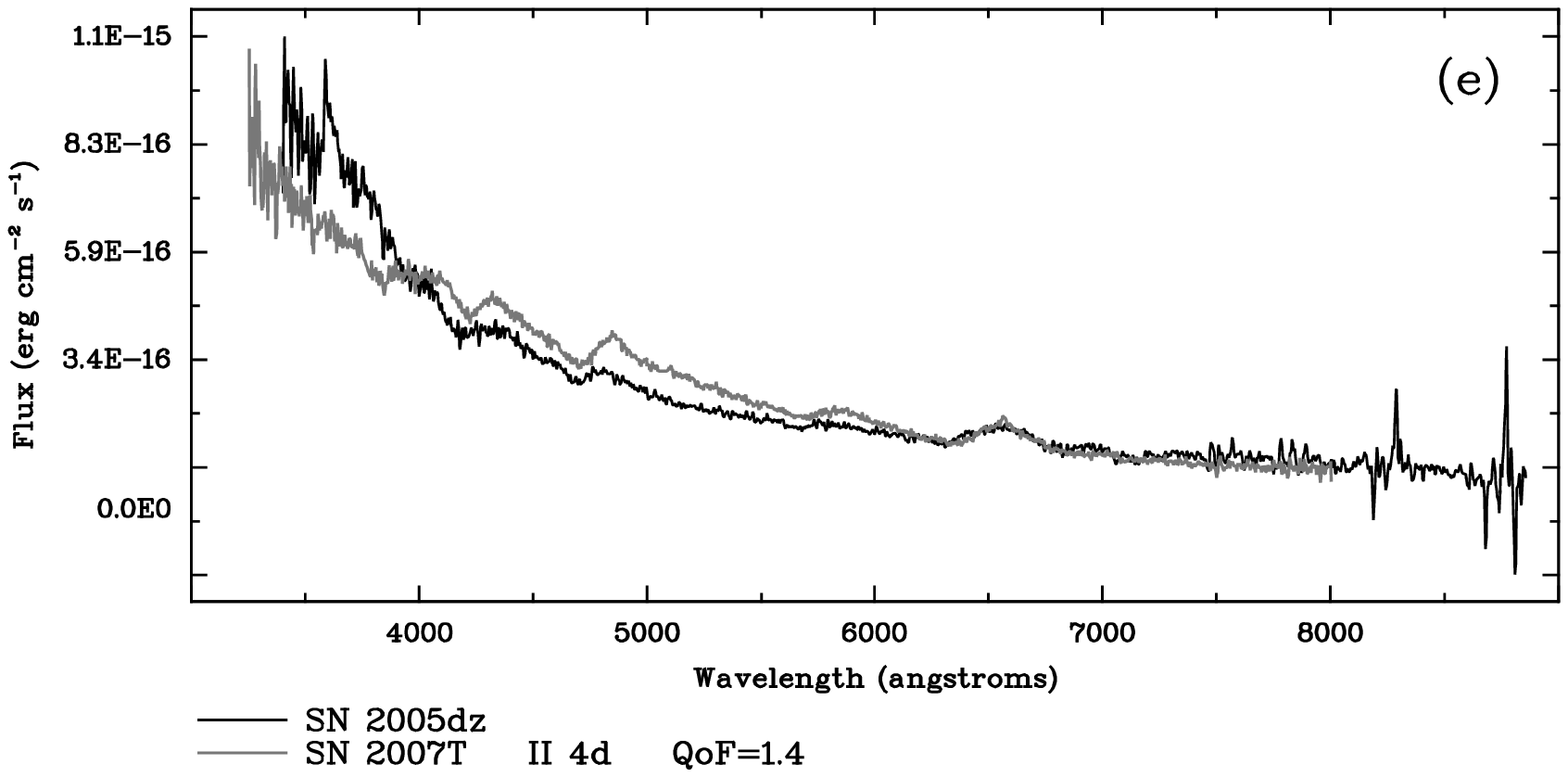}
\includegraphics[height=9cm, angle=270]{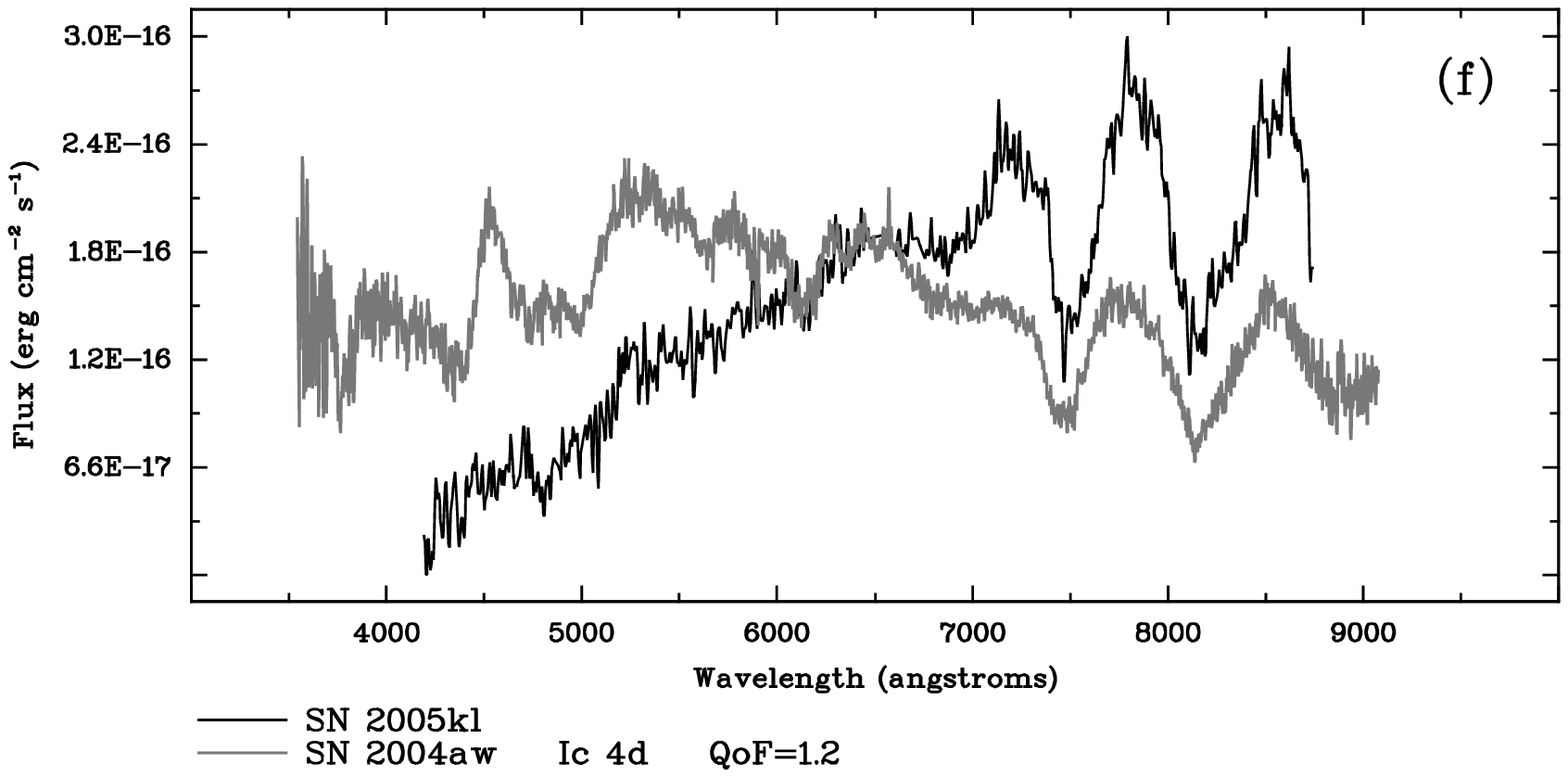}\\
\caption{Same as Fig. \ref{plt1}.}
\label{plt4}
\end{figure*}

The results of the comparison of our set of 36 spectra with 
the templates are summarised in Table \ref{table2}, listing
for each object the best template, the phase and the QoF.
Figures \ref{plt1}, \ref{plt2}, \ref{plt3} and \ref{plt4} show
the plots of the spectra of our objects together with their best fitting
templates. Spectra are shown in the parent
galaxy restframe and are not corrected for extinction. Below we
present a short discussion on individual objects.

\textbf{SN 2002an} was found on Jan. 22.52 by \citet{nakano02}
and classified as a type-II supernova by \citet{benetti02a}.
The spectrum consists of a blue continuum with P-Cyg
profiles of H Balmer and He I $\lambda$5876 lines. H$_\alpha$ is almost
purely in emission (Fig. \ref{plt1}a). The QoF is not high (1.7),
which is due to some discrepancies in the fit. In particular
H$_\alpha$ is not well fitted. The most similar template
spectrum is that of the type-II SN 1996bw obtained 7 days after its
discovery \citep[ASA,][]{benetti96}. For this template the explosion
epoch is not estimated. The spectrum of 
another type-II SN 1995ad 12 days after its explosion
\citep{pastorello03} also provides a satisfactory fit. The expansion velocities
deduced from the minima of He I $\lambda$5876, H$_\beta$ and H$_\gamma$ lines
are 9300, 9690 and 9800 km s$^{-1}$, respectively.

\textbf{SN 2002bh} was found on Feb. 24.3 by \citet{ganeshalingam02}.
\citet{benetti02b} classified it as a type-II supernova. A broad
H$_\alpha$ emission line is the only strong feature in the noisy spectrum. 
Also, broad He I $\lambda$5876,, Fe II and H$_\beta$ features can be identified (Fig. \ref{plt1}b). 
From the minimum of H$_\alpha$ an expansion velocity of about 12000
km s$^{-1}$ is derived. The best fit is with the type-II SN 1995V
spectrum (ASA) 8 days after the explosion \citep{fassia98}.
The template spectrum does not match well the blue slope of the
spectrum, possibly because of some reddening in the host galaxy.

\textbf{SN 2002cs} was discovered on May 5.5 by \citet{ganeshalingam02a}. The
spectrum is that of a type-Ia supernova, and \citet{riello02a} gave
an age estimate of 2 $\pm$ 2 days before maximum light
(Fig. \ref{plt1}c).
The expansion velocity deduced from the Si II $\lambda$6355 minimum
is about 15800 km s$^{-1}$. Both the expansion velocity and the age
estimates are in agreement with those by \citet{matheson02}.
The high expansion velocity may be caused
by contamination by a high-velocity component \citep[see][]{mazzali05}.
High-velocity features are particularly strong and blended with
the photospheric components in the SNe that belong to the high velocity gradient
(HVG) group \citep[see][]{benetti05}.
In fact, the closest match we found
is with the HVG SN 2004dt at a
phase of -7 days \citep{altavilla07}. The shape of the continuum
and interstellar absorption Na I D $\lambda$5876 feature
(in the Galaxy restframe)
of the spectrum suggest the presence of significant extinction.
The value of QoF = 1.6 is mainly due to a
mismatch in the line velocity suggesting either an even more extreme
case of HVG SN than SN 2004dt or an earlier phase.
Because of the first phase determination
(-2 $\pm$ 2 days instead of -7) SN 2002cs was unfortunately not considered
worth of detailed follow-up by the ESC.

\textbf{SN 2002dg} was found on May 31.3 by \citet{woodvasey02} and
classified as a type-Ib supernova 2-3 weeks after maximum by
\citet{riello02b}. The spectrum (Fig. \ref{plt1}d) 
is most similar to that of the type-Ib
SN 1998dt at 15 days from R-band maximum \citep{matheson01}.
For SN 2002dg, adopting a recession velocity of
14000 km s$^{-1}$ derived by \citet{riello02b} from the parent
galaxy emission lines, the expansion
velocities measured from the He I $\lambda$5876 absorption minima of the SN 2002dg
and 1998dt spectra are 9500 and 10900 km s$^{-1}$, respectively.
The largest mismatch to the SN 1998dt spectrum is the absorption
at 6270$\AA$ (rest frame). If this feature is due to H$_\alpha$
\citep[see][]{branch02}, a transitional type-IIb event may be an alternative
classification. Inspection of the fit (see Figure \ref{plt1})
shows that not only the He I features, but also other lines in
the spectrum of SN 2002dg are slightly narrower than in SN 1998dt.

\textbf{SN 2002dm} was found on May 4.76 by \citet{sanders02},
\citet{turatto02} classified it as a type-Ia SN, giving a phase
estimate of about 50 days after maximum light. The high SNR spectrum
(Fig. \ref{plt1}e) is almost identical to that of SN 1994ae (ASA)
at 92 days after maximum light. The large number of features present
in the spectrum and the exceptional resemblance to the 1994ae spectrum
lead to the best match among the SNe of the sample (QoF = 9.06).

\textbf{SN 2002ej} was discovered on Aug. 9.11 by \citet{puckett02} and
classified as a type-II supernova two weeks after maximum by
\citet{desidera02}. On the noisy spectrum P-Cyg profiles of H Balmer,
Fe II and (possibly) Sc II $\lambda$$\lambda$5240, 5527, 5658 lines are present (Fig. \ref{plt1}f).
The best fitting template (QoF = 2.94)
is the one of the type-IIP SN 1995ad at 24 days after explosion epoch
\citep{pastorello03}. The expansion velocity of SN 2002ej deduced from
the minimum of H$_\alpha$ line is about 8070 km s$^{-1}$.

\textbf{SN 2002hg} was found on Oct. 28.22 by \citet{boles02} and
classified as a type-II supernova few weeks past maximum light by
\citet{pignata02a}. The spectrum shows strong P-Cyg profiles of
H Balmer, Fe II, Ca II (H, K plus IR-triplet), Na I D, O I $\lambda$7774,
Sc II , Ba II $\lambda$$\lambda$4934, 6142 lines (Fig. \ref{plt1}g)
and is well-fitted (QoF = 2.4) by a type-II SN 1999em spectrum taken 41 days after
explosion \citep{elmhamdi03}. The expansion velocity of SN 2002hg
deduced from the H$_\alpha$ line profile is about 6900 km s$^{-1}$.

\textbf{SN 2002hm} was detected on Nov. 5.16 by \citet{boles02a} and
classified as a type-II supernova 30 days after maximum light by
\citet{pignata02b}. The rather blue continuum is
overimposed by P-Cyg profiles of H Balmer, Fe II, Ca II lines (Fig. \ref{plt1}h).
The best fit to this spectrum is provided by the type-II SN 1998ce spectrum
at 10 days after discovery \citep[ASA,][]{patat98}.
The SN 2002hm spectrum is also well fitted by
SN 1995ad 12 days after explosion \citep{pastorello03},
in agreement with the fact the Na I D feature, typical of type-II
SNe at more advanced epochs, is not present yet, thus suggesting
an earlier phase than that proposed by \citet{pignata02b}.
The expansion velocities deduced from the H$_\alpha$ and
H$_\beta$ absorption minima in the SN 2002hm spectrum are about
9500 and 8100 km s$^{-1}$, respectively.

\textbf{SN 2002hy} was found on Nov. 12.1 by \citet{monard02} and classified
as a peculiar type-Ib supernova by \citet{benetti02c}. The blue continuum
is overimposed with strong He I lines at $\lambda$$\lambda$3889,
4471, 5015, 5876, 6678, 7065 (Fig. \ref{plt1}i). As \citet{benetti02c}
mentioned, the He I emission
peaks are blueshifted on average by 1800 km s$^{-1}$. Despite the
peculiarity of the object, GELATO finds a matching spectrum that
contains the He I features with similar line velocity,
the type-II SN 2001gh 11
days after discovery \citep[ASA,][]{altavilla01, valenti03}.
The second best fitting template is the type-IIb SN 1993J at
4 days after explosion \citep{barbon95}, which, however, fails to
reproduce most of the He I features.
SN 2001gh is classified as a type-II SN \citep{altavilla01}, but
because of the presence of strong He lines both SN 2002hy and SN
2001gh should be considered IIb/Ib events.

\textbf{SN 2003hg} was discovered on Aug. 18.4 by \citet{moore03}
and classified as a type-II supernova shortly after explosion \citep{eliasrosa03}.
Broad emission lines of H$_\alpha$ (possibly with a boxy profile), and
He I $\lambda$5876 are present together with absorptions of H$_\beta$,
and H$_\gamma$ (Fig. \ref{plt1}j). The best
fit to this spectrum is with that of the type-II SN 1995V 8 days after
explosion (ASA, the same as in SN 2002bh case).
Despite the different SED, most probably due to reddening (in fact,
Na I D is present with an equivalent width (EW) of about 1.3$\AA$),
the QoF is  high (2.09).

\textbf{SN 2003hn} was found on Aug. 25.7 by \citet{evans03} and
classified as a type-II supernova approximately two weeks after
explosion by \citet{salvo03}. The spectrum has a blue continuum
and P-Cyg profiles of the H$_\beta$, H$_\gamma$ and He I $\lambda$5876
lines. H$_\alpha$ is present almost purely in emission (Fig. \ref{plt2}a).
The best match
of this spectrum is the type-II SN 1995ad spectrum 12 days after
explosion date \citep{pastorello03}. The interstellar Na I D
absorption feature present in the SN 2003hn spectrum suggests some
reddening \citep[EW(Na I D) $\approx$ 0.62$\AA$, corresponding to a
lower limit of E(B$-$V)=0.089, see][]{turatto03b}.
The expansion velocities for SN 2003hn deduced from the H$_\beta$ and
H$_\gamma$ lines are of about 8700 and 8200 km s$^{-1}$, respectively.

\textbf{SN 2003ie} was found on Sept. 19.8 by \citet{arbour03} and
classified as a type-II supernova by \citet{benetti03}. 
P-Cyg profiles of H$_\alpha$, H$_\beta$, Fe II, Sc II, Na I D and
Ba II lines are overimposed on red continuum (Fig. \ref{plt2}b).
The expansion velocity deduced
from the H$_\alpha$ absorption minimum is about 5500 km s$^{-1}$. The
best fit to this spectrum is with that of the peculiar type-II SN 1998A 37
days after explosion \citep{pastorello05a}. Like in SN 1998A
the H$_\alpha$ emission shows a significant shift of about
2100 km s$^{-1}$ towards the blue. This effect is seen also in
SN 1987A and in other type-II SNe \citep[see][for discussion]{pastorello05a}.

\textbf{SN 2004G} was found on Jan. 19.8 by \citet{nakano04} and
classified as a type-II supernova about 5 months after explosion by
\citet{eliasrosa04a}. The best fit to the SN 2004G spectrum
(Fig. \ref{plt2}c) is with the type-II SN 1993S
90 days after its discovery (ASA). The spectrum
is well-fitted also by the type-II SN 1995ad spectrum 100 days past
explosion \citep{pastorello03}. The expansion velocities deduced from
the H$_\alpha$ and H$_\beta$ absorption minima 
are about 6700 and 4900 km s$^{-1}$, respectively.

\textbf{SN 2004aq} was discovered on Mar. 2.1 by \citet{armstrong04a} and
classified as a type-II supernova one month after explosion by
\citet{eliasrosa04b}. The spectrum shows P-Cyg profiles of H$_\alpha$,
H$_\beta$, Ca II and Fe II lines overimposed on a rather blue continuum
(Fig. \ref{plt2}d).
The best fit to this spectrum is a type-II SN 1991al spectrum
taken 25 days after discovery (ASA). The minima of the H$_\alpha$ and H$_\beta$
absorption components on the SN 2004aq spectrum are blueshifted by about
7700 and 6600 km s$^{-1}$, respectively, very similar to those of
SN 1991al (7700 and 6500 km s$^{-1}$). The spectrum of SN 2004aq is
also well fitted by that of the type-II SN 2001du 18 days after
explosion \citep{smartt03}.

\textbf{SN 2004bs} was found on May 16.9 by \citet{armstrong04b} and
classified as a type-Ib supernova about 3 weeks past maximum by
\citet{pignata04}. The spectrum is dominated by He I $\lambda$5876,
6678, 7065 lines with velocities of about 10800, 10200
and 10000 km s$^{-1}$, respectively (Fig. \ref{plt2}e).
Lines of Fe II, O I and Ca II are also present in the spectrum.
The best matching template is the type-Ib SN 1998dt (ASA),
17 days after R-band maximum \citep{matheson01}.

\textbf{SN 2004cc} was found on Jun 10.3 by \citet{monard04} and
classified as a type-I supernova by \citet{matheson04} and type-Ic
supernova one week before maximum by \citet{foley04}. P-Cyg profiles
of Fe II, Na I D (possibly blended with He I $\lambda$5876), Si II
and Ca II are present on the highly reddened spectrum
(Fig. \ref{plt2}f). Strong Na I D interstellar absorption is detected
in the host galaxy restframe (EW $\approx$ 4.1$\AA$). The best fit
to this spectrum is provided by the type-Ic SN 1994I 10 days after
B maximum \citep{filippenko95}, although the Na I D (+ He I) absorption
and Fe II features in SN 2004cc are bluer (i.e. with a higher
expansion velocity) than those of SN 1994I. The deep O I
feature in the SN 1994I spectrum in not present in SN 2004cc, making
this object rather peculiar.

\textbf{SN 2004dg} was discovered on Jul 19.8 by \citet{vagnozzi04} and
classified as a type-II supernova by \citet{eliasrosa04c}. The spectrum
shows P-Cyg profiles of H$_\alpha$, H$_\beta$, Ca II H\&K, Fe II and Ti II
lines (Fig. \ref{plt2}g). There is a narrow emission component, probably due
to a nearby H II
region, on the broad H$_\alpha$ emission profile. The recession
velocity deduced from this narrow emission line is of about 1760 km s$^{-1}$.
Adopting this recession velocity, the photospheric velocities deduced from
H$_\alpha$ and H$_\beta$ are about 8420 and 6970 km s$^{-1}$, respectively.
The spectrum of the type-II SN 2001du 18 days from explosion
\citep{smartt03} is the best match. Our algorithm provides a very high
QoF = 4.76, despite a stronger H$_\alpha$ emission in SN 2004dg and
a slight difference in SED, probably caused by reddening. In fact, on the noisy
continuum a clear Na I D interstellar absorption feature with
EW $\approx$ 1.9$\AA$ is detected, corresponding to E(B-V) = 0.29
\citep{turatto03b}.

\textbf{SN 2004dk} was found on Aug 1.2 by \citet{graham04a} and classified
as a type-Ic supernova by \citet{patat04a}. P-Cyg profiles of Ca II, Fe II,
Na I, Si II and O I lines are present in the spectrum (Fig. \ref{plt2}h).
The spectrum is
similar to that of the type-Ic SN 2004aw 4 days after B maximum light
\citep{taubenberger06}, though
the O I and Ca II absorption profiles of the SN 2004aw spectrum are
significantly bluer than those of 2004dk. The absorption
minima of Si II 6355$\AA$ and O I 7774$\AA$ in the SN 2004dk spectrum
suggest expansion velocities of about 9200 and 7300 km s$^{-1}$,
respectively. A narrow H$_\alpha$
emission probably due to a nearby H II region is present.
The second best fit to this spectrum is by that of SN 1994I 3 days before
maximum light \citep{filippenko95}, as mentioned in \citet{patat04a}.

\textbf{SN 2004dn} was discovered on Jul. 29.4 by \citet{graham04b} and
classified as a type-Ic supernova few days before maximum light by
\citet{patat04b}. Spectral features of
Si II, O I and Ca II are clearly visible (Fig. \ref{plt2}i).
The expansion velocities deduced from Si II 6355$\AA$ and O I 7774$\AA$
minima are of about 10500 and 10200 km s$^{-1}$, respectively. The best
fitting template is again that of the type-Ic SN 2004aw 4 days after
maximum \citep{taubenberger06}. Having the same best fitting template
means that the spectra of SN 2004dn and 2004dk are also similar.

\textbf{SN 2004fe} was discovered on Oct. 30.3 by \citet{pugh04} and
classified as a type-Ic supernova a few days before maximum by
\citet{modjaz04}. The spectrum (Fig. \ref{plt2}j) shows P-Cyg profiles
of Ca II, Fe II, Na I D, and O I lines.
The best fitting template spectrum is that of the type-Ic SN 1994I 3
days before B maximum \citep{filippenko95}. The absorption at about
6170$\AA$ is probably due to Si II $\lambda$6355, while the one at
about 6360$\AA$ is possibly either weak H$_\alpha$ or C II $\lambda$6580
\citep[see][]{valenti08}.

\textbf{SN 2004go} was found on Nov. 18.3 by \citet{li04} and classified
as a type-Ia supernova 3-4 weeks past maximum by \citet{navasardyan04}.
The expansion velocity deduced from Si II 6355$\AA$ absorption minimum
is about 10200 km s$^{-1}$. The spectrum (Fig. \ref{plt3}a) is very
similar to that of the type-Ia SN 1996X 24 days after maximum light
\citep{salvo01}. The spectra of SN 1994D \citep{patat96} and 2002bo
\citep{benetti04} 24 and 28 days after their B band maxima also provide
good fits.

\textbf{SN 2005G} was found on Jan. 14.6 by \citet{graham05a} and
classified as a type-Ia supernova about 10 days past maximum by
\citet{navasardyan05}. \citet{ganeshalingam05} report that a
spectrum of SN 2005G taken 2 days before ours shows a narrow
Si II 6355$\AA$ absorption, a blend of two S II absorptions
around 5500$\AA$ and a flux density drop blueward of Ca II H\&K
lines. The Si II and S II features peculiarities are present
in our spectrum as well (Fig. \ref{plt3}b), but we cannot confirm
the blue flux decline, because of the limited spectral range. Nevertheless,
the spectrum closely resembles (QoF = 3.5) that of the type-Ia
SN 1994D 11 days after maximum light \citep{patat96}. The blueshift
of Si II 6355$\AA$ minimum is of 9600 km s$^{-1}$.

\textbf{SN 2005H} was found on Jan. 15.2 by \citet{graham05b} and
classified as a young type-II supernova by \citet{pastorello05b}.
The spectrum is dominated by a blue continuum with overimposed
P-Cyg profiles of H$_\beta$, Fe II and He I (Fig. \ref{plt3}c).
H$_\alpha$ is also present, with broad emission and a shallow
absorption. The H$_\beta$ and He I absorption
minima are blueshifted by about 6600 and 6800 km s$^{-1}$, respectively.
The best fitting template to this spectrum is that of the
SN 2002gd, a low line velocity Ni-poor SN IIP 6 days after explosion
\citep{pastorello04}. The featureless
spectrum the peculiar line profiles lead, however, to a
very low QoF = 1.01.

\textbf{SN 2005I} was discovered on Jan. 15.6 by \citet{graham05b}.
\citet{pastorello05b} classified it as a type-II supernova
about 3 months after the explosion. The spectrum is characterised
by a red continuum and narrow P-Cyg profiles of H$_\alpha$, H$_\beta$,
Ca II and Fe II lines (Fig. \ref{plt3}d). The expansion velocities deduced
from the H$_\alpha$ minimum is about 4900 km s$^{-1}$. The best match is with the type-II 
SN 2003gd spectrum (ASA). Adopting for SN 2003gd the explosion
epoch found by \citet{hendry05}, the template is at an epoch of 101
days from the explosion, in good agreement with the phase estimate
by \citet{pastorello05b}.

\textbf{SN 2005N} was found on Jan.19.6 by \citet{puckett05a} and
classified as a type-Ib/c supernova in the nebular phase by
\citet{taubenberger05a}. The spectrum shows strong emission lines
of Mg I] $\lambda$4571, Na I D, [O I] $\lambda$6300, 6364, [Ca II]
$\lambda$7291, 7323 and Ca II (Fig. \ref{plt3}e). The best fit to
this spectrum is with the type-Ib SN 1990I spectrum 88 days after maximum
light \citep{elmhamdi04}, which, however, shows weaker [O I]
emissions.

\textbf{SN 2005V} was found on Jan. 30.2 by \citet{mattila05} and
classified as a type-Ib/c supernova by \citet{taubenberger05b}. The
spectrum shows P-Cyg profiles of Fe II, Na I D, O I and Ca II lines
overimposed on a very red continuum (Fig. \ref{plt3}f). A narrow
H$_\alpha$ emission line, due to an underlying H II region, is present.
The best match of this spectrum
is with the type-Ic SN 2004aw 22 days after the maximum light
\citep{taubenberger06}. The shape of the continuum and the presence
of a deep Na I D absorption line (EW = 5.4$\AA$) in the host galaxy
restframe, suggest that SN 2005V is affected by heavy extinction
\citep[E(B-V) = 0.9,][]{turatto03b}.
The absorption feature at 6200$\AA$ on the SN 2005V spectrum is most
likely due to Si II (possibly blended with H$_\alpha$). There is
no evidence of He I lines.

\textbf{SN 2005ab} was discovered on Feb. 5.6 by \citet{nakano05}.
\citet{benetti05a} classified it as a type-II SN shortly
after its explosion. On the very noisy spectrum (SNR $\approx$ 6)
a relatively broad ($\sim$ 5000 km s$^{-1}$) H$_\alpha$ emission line
is present (Fig. \ref{plt3}g). The best fitting
template spectrum is that of the type-II SN 1997du 26 days after
its discovery \citep[ASA,][]{patat97}. The minimum of the H$_\beta$ line
absorption component in the SN 1997du spectrum is blueshifted by
about 6900 km s$^{-1}$. 
Despite the low SNR, the absorptions at about 5770$\AA$, 5110$\AA$
and 4950$\AA$ seem to be fitted by the SN 1997du absorptions of
the Na I D 5892$\AA$, Fe II 5169$\AA$ and Fe II 5110$\AA$
lines, therefore the phase of SN 2005ab could be more advanced than
reported by \citet{benetti05a}.

\textbf{SN 2005ai} was found on Feb. 12.23 by \citet{puckett05b} and
classified as a type-Ia supernova about one month past maximum light
by \citet{taubenberger05c}. The spectrum is that of a typical type-Ia
SN (Fig. \ref{plt3}h), and SN 1994D 24 days after the maximum light
\citep{patat96} yields the best match. This spectrum is similar to the
spectrum of SN 2004go of the present sample.

\textbf{SN 2005br} was detected on Mar. 28.1 by \citet{monard05a} and
classified as a type-Ib supernova about 40 days past maximum by
\citet{turatto05}. The reddened spectrum shows features of He I
(probably blended with Na I D), O I and Ca II (Fig. \ref{plt3}i).
Also, a rather strong
(EW $\approx$ 2.6$\AA$) interstellar Na I D absorption line is
detected. The photospheric expansion velocity deduced
from He I 5876$\AA$ is about 9000 km s$^{-1}$. The best
fit to this spectrum is achieved with the type-Ib SN 1997X 40 days
after discovery (ASA), However,
the O I 7774$\AA$ line is shallower (probably contaminated by a
telluric feature and less blueshifted) in 1997X. Also, a SN 1990U
spectrum (ASA) provides a good fit. Adopting the B maximum epoch
given by \citet{piemonte96} for SN 1990U, the template spectrum
phase is 48 days.

\textbf{SN 2005bs} was found on Apr. 19.1 by \citet{monard05b}, and
\citet{turatto05} classified it as a type-Ia supernova. The spectrum
(Fig. \ref{plt3}j)
resembles very well (QoF = 3.33) that of the type-Ia SN 1996X 31
day after the maximum light \citep{salvo01}, in agreement with the
estimate by \citet{turatto05}.

\textbf{SN 2005cb} was found on May 13.22 by \citet{jacques05} and
classified as a type-Ib/c supernova at about 10 days after maximum
by \citet{turatto05}. P-Cyg profiles of Fe II, Na I D, Si II, O I
and Ca II are present in the spectrum (Fig. \ref{plt4}a).
The expansion velocities deduced from the minima of Si II and O I
features are about 9500 and 11000 km s$^{-1}$, respectively.
The best fitting
template is that of the type-Ic SN 1994I 1 day after B
maximum light \citep{filippenko95}. SN 2005cb seems to be slightly
redder than 1994I at maximum, and has a stronger O I with a
higher line velocity.

\textbf{SN 2005ce} was found on May 28.5 by \citet{pugh05} and
classified as a type-Ib/c supernova a few days after explosion
by \citet{stanishev05a}. The spectrum has a blue continuum
with several P-Cyg profiles of Fe II, Ca II lines
and moderately weak absorptions at about 5700$\AA$, 6496$\AA$,
6850$\AA$ and 7060$\AA$ due to He I, with expansion velocities of
about 8980, 8170, 9100 and 9140 km s$^{-1}$, respectively (Fig. \ref{plt4}b).
However, the best fit to this spectrum is provided by the type-Ic SN 1996aq
5 days after discovery (ASA). The spectrum is also similar to that of
SN 1994I 10 days after B maximum, though in the SN 2005ce
spectrum the O I feature is much weaker (if any).
The deep absorption at about 6294$\AA$ is
most likely due to H$_\alpha$ (probably blended with Si II and C II $\lambda$6580).
This is supported by the identification of an absorption at about
4689$\AA$ with H$_\beta$. 
The presence of H$_\alpha$ in various SN Ib/c was suspected
\citep[see][]{branch02, elmhamdi06}. Usually the H$_\alpha$ line optical
depth is small and no apparent H$_\beta$ is detected, while in this case
the H$_\alpha$ feature is strong and there is also some hint of H$_\beta$ presence.
With these identifications the photospheric expansion velocities derived
from H$_\alpha$ and H$_\beta$ lines are of about 12300 and 10600
km s$^{-1}$.
Although the best fits found by our program are with type-Ic SNe,
SN 2005ce is a rare and a very interesting example of an intermediate
case between a Ib (with possibly some contamination of H) and
Ic event.

\textbf{SN 2005de} was discovered on Aug. 2.28 by \citet{lee05} and
classified as a type-Ia supernova one week before maximum by
\citet{wang05}. Our spectrum
of SN 2005de shows P-Cyg lines of Ca II, Fe II, S II and Si II
(Fig. \ref{plt4}c).
The Ca II near-infrared triplet seems to be present with two
components. This was noted by \citet{wang05} who measured a
velocity of about 20000 km s$^{-1}$ for the bluer component,
in agreement with our spectrum where the Ca II absorption minimum
is blueshifted by 20200 km s$^{-1}$, taking 8579$\AA$ as the
wavelength reference for the multiplet. Indeed, this high velocity
feature is also present in the best fitting template spectrum of the
type-Ia SN 2005cf 5 days before the maximum light
\citep{garavini07}. \citet{mazzali05} find that the
presence of high-velocity features is very common, if not
ubiquitous, in the early spectra of SNe Ia.

\textbf{SN 2005dv} was found on Sep. 4.8 by \citet{dimai05} and
classified as a type-Ia supernova probably before maximum light
by \citet{leonard05}. The minimum of the Si II $\lambda$6355 line is
blue-shifted by about 12600 km s$^{-1}$ suggesting that the phase
of the spectrum is near-maximum (Fig. \ref{plt4}d).
The best fitting template is that of the type-Ia SN 2002bo at 
maximum light \citep{benetti04}. The fit reproduces all spectral
features rather well, though in the blue region the continuum of
SN 2005dv spectrum is weaker, probably due to some reddening. This
is confirmed by the presence of a Na I D absorption line
(EW $\approx$ 2.3$\AA$) in the rest frame of the host galaxy,
as noted also by \citet{leonard05}.

\textbf{SN 2005dz} was found on Sep. 10 by \citet{puckett05c} and
classified as a young type-II supernova by \citet{stanishev05b}.
The blue continuum of the spectrum is superimposed by broad and shallow
P-Cyg profiles of H Balmer and He I $\lambda$5876 lines
(Fig. \ref{plt4}e). H$_\alpha$ is present mostly in emission.
The expansion velocities derived from the H$_\alpha$ and H$_\beta$
absorptions are about 12200 and 10600 km s$^{-1}$, respectively.
The best fitting template spectrum is that of the type-II SN 2007T
4 days after discovery \citep{benetti07}, although the expansion
velocities of 2007T are smaller. The SN 2005dz spectrum is also similar
to that of the type-II SN 2002gd 6 days after
explosion \citep{pastorello04}.

\textbf{SN 2005kl} was discovered on Nov. 22 by \citet{dimai05a}
and classified as a type-Ic supernova by \citet{taubenberger05d}.
P-Cyg profiles of O I, Ca II and also absorptions due to Si II and
Fe II are present in the spectrum (Fig. \ref{plt4}f). The red
continuum suggests that the SN is heavily extinguished.
The best fitting template is the type-Ic SN 2004aw 4 days
after B maximum \citep{taubenberger06}, which was also the case for
SN 2004dk and 2004dn. The expansion velocity deduced from the Si II
absorption is of about 10000 km s$^{-1}$.

\section{Summary}
\label{summ}
The European Supernova Collaboration (ESC) was conceived to
perform very detailed studies of nearby SNe Ia and carried
out extensive follow-up programs on 15 objects (plus one SN Ic).
Integral part of the ESC was a prompt classification program
with ToO observations of selected, newly discovered SNe, with
the aim to single out candidates for the following intensive monitoring.
In this context several tens of SN candidates were observed
which did not meet the ESC requirements as to
epoch of discovery and SN type. In this paper we
have presented and discussed the spectra of these objects which
include 8 type-Ia, 13 type-Ib/c (in 2 cases possibly IIb) and 15 type-II SNe.
Each SN spectrum has been analysed by means of
a new software tool that compares it with a vast database of
SN spectra. For each object we have identified the best fit SN template
providing type and a phase estimate. Other information
such as the identification of the most prominent spectral
features, the expansion velocities of the absorbing layers
and possible peculiarities have also been reported.

The comparison with ASA spectra has confirmed
the previous classification of all objects. Nevertheless, in
some cases the new spectral ages differ from the estimates
reported in the original IAU circulars. In the case of SN 2002cs
the new determination of the spectral age (-7 days)
to be compared to the previous one (-2 $\pm$ 2 days), shows the
importance of having a reliable and objective classification
tool and a complete archive.
Had such a tool been available in 2002, SN 2002cs would
have become an ESC target for detailed follow-up observations.

The current version
of the comparison software, which is now routinely used by
the Padova team for the classification of newly discovered
SNe, has been presented. In particular, we have discussed the
general algorithm and the accuracy of the classification and
phase determination. The classification can be considered
``safe'' when the QoF parameter is larger than 1.4,
while for QoF $<$ 1.4 the type determination must be done ``cum
grano salis''. Values of QoF smaller than 1 mean that no
spectral template similar to the input spectrum is
present in the archive.
Typical errors in the epoch determination of $\pm$1.9 days
have been found for SN Ia in
the early, fast evolving phases, while for later epochs the
uncertainty rises to 3.1 days. Much larger (about 10 days)
is the uncertainty on the epoch determination for CC SNe,
because of their heterogeneity and lack of the extensively
observed templates.

Although this tool already provides satisfactory results,
we plan to implement some refinement for what concerns the
quantitative measure of the quality of the fits. The SN
research community will have access to the tool through
a web-based interface that we plan to develop.

\begin{acknowledgements}
AH acknowledges the support by the Padova municipality ``Padova
Citt\`{a} delle Stelle'' prize. MT, EC and SB are supported by the
Italian Ministry of Education via the PRIN 2006 n.022731\_002.
GP acknowledges support by the Proyecto FONDECYT 30700034.
This work is supported in part by the European Community's Human
Potential Programme under contract HPRN-CT-2002-00303, ``The Physics
of Type Ia Supernovae''. 
The observations on which this paper is
based were collected at the European Southern Observatory (Chile),
the Calar Alto Observatory (Spain), the Italian Telescopio Nazionale
Galileo (La Palma), the Nordic Optical Telescope (La Palma), the Asiago
Observatory (Italy). We made use of the NASA/IPAC Extragalactic Database
(NED) which is operated by the Jet Propulsion Laboratory, California
Institute of Technology, under contract with the National Aeronautics
and Space Administration. We thank the anonymous referee for
useful comments and suggestions.
\end{acknowledgements}


\begin{thebibliography}{widest-lable}

\bibitem[Altavilla et al.(2001)]{altavilla01} Altavilla, G., et Tl.\ 2001, \iaucirc, 7762, 1 
\bibitem[Altavilla et al.(2007)]{altavilla07} Altavilla, G., et al.\ 2007, \aap, 475, 585 
\bibitem[Arbour \& Boles(2003)]{arbour03} Arbour, R., \& Boles, T.\ 2003, \iaucirc, 8205, 1
\bibitem[Armstrong \& Buczynski(2004)]{armstrong04a} Armstrong, M., \& Buczynski, D.\ 2004, \iaucirc, 8301, 1
\bibitem[Armstrong(2004)]{armstrong04b} Armstrong, M.\ 2004, Central Bureau Electronic Telegrams, 66, 1
\bibitem[Barbon et al.(1995)]{barbon95} Barbon, R., Benetti, S., Cappellaro, E., Patat, F., Turatto, M., \& Iijima, T.\ 1995, \aaps, 110, 513 
\bibitem[Barbon et al.(1999)]{barbon99} Barbon, R., Buond{\'{\i}}, V., Cappellaro, E., \& Turatto, M.\ 1999, \aaps, 139, 531 
\bibitem[Benetti \& Turatto(1996)]{benetti96} Benetti, S., \& Turatto, M.\ 1996, \iaucirc, 6520, 1
\bibitem[Benetti et al.(2002a)]{benetti02a} Benetti, S., Altavilla, G., Pastorello, A., Turatto, M., Desidera, S., Giro, E., \& Cappellaro, E.\ 2002a, \iaucirc, 7828, 2
\bibitem[Benetti et al.(2002b)]{benetti02b} Benetti, S., Altavilla, G., Pastorello, A., Turatto, M., Desidera, S., \& Cappellaro, E.\ 2002b, \iaucirc, 7844, 2 
\bibitem[Benetti et al.(2002c)]{benetti02c} Benetti, S., Altavilla, G., Pastorello, A., Riello, M., Turatto, M., Zampieri, L., \& Cappellaro, E.\ 2002c, \iaucirc, 8019, 3
\bibitem[Benetti et al.(2003)]{benetti03} Benetti, S., et al.\ 2003, \iaucirc, 8207, 3
\bibitem[Benetti et al.(2004)]{benetti04} Benetti, S., et al.\ 2004, \mnras, 348, 261
\bibitem[Benetti et al.(2005)]{benetti05} Benetti, S., et al.\ 2005, \apj, 623, 1011
\bibitem[Benetti et al.(2007)]{benetti07} Benetti, S., Harutyunyan, A., Turatto, M., Cappellaro, E., \& Magazzu, A.\ 2007, Central Bureau Electronic Telegrams, 837, 1
\bibitem[Benetti \& Di Mille(2005)]{benetti05a} Benetti, S., \& di Mille, F.\ 2005, \iaucirc, 8480, 2 
\bibitem[Blondin \& Tonry(2007)]{blondin07} Blondin, S., \& Tonry, J.~L.\ 2007, \apj, 666, 1024 
\bibitem[Boles \& Schwartz(2002)]{boles02} Boles, T., \& Schwartz, M.\ 2002, \iaucirc, 8004, 2
\bibitem[Boles(2002)]{boles02a} Boles, T.\ 2002, \iaucirc, 8009, 1
\bibitem[Botticella et al.(2007)]{botticella07} Botticella, M.~T., et al. \ 2007, \aap, in press
\bibitem[Branch et al.(2002)]{branch02} Branch, D., et al.\ 2002, \apj, 566, 1005
\bibitem[Branch et al.(2003)]{branch03} Branch, D., Baron, E.~A., \& Jeffery, D.~J.\ 2003, Supernovae and Gamma-Ray Bursters, 598, 47 
\bibitem[Desidera et al.(2002)]{desidera02} Desidera, S., Giro, E., Della Valle, A., Benetti, S., Altavilla, G., Pastorello, A., Turatto, M., \& Cappellaro, E.\ 2002, \iaucirc, 7963, 2
\bibitem[Dimai \& Dainese(2005)]{dimai05} Dimai, A., \& Dainese, P.\ 2005, Central Bureau Electronic Telegrams, 217, 1 
\bibitem[Dimai \& Migliardi(2005)]{dimai05a} Dimai, A., \& Migliardi, M.\ 2005, Central Bureau Electronic Telegrams, 300, 1
\bibitem[Elias-Rosa et al.(2003)]{eliasrosa03} Elias-Rosa, N., et al.\ 2003, \iaucirc, 8187, 2
\bibitem[Elias-Rosa et al.(2004a)]{eliasrosa04a} Elias-Rosa, N., et al.\ 2004a, \iaucirc, 8273, 2
\bibitem[Elias-Rosa et al.(2004b)]{eliasrosa04b} Elias-Rosa, N., Benetti, S., Stanishev, V., Goobar, A., \& Jaervinen, A.\ 2004b, \iaucirc, 8301, 2
\bibitem[Elias et al.(2004)]{eliasrosa04c} Elias, N., et al.\ 2004, \iaucirc, 8376, 2
\bibitem[Elias-Rosa et al.(2006)]{eliasrosa06} Elias-Rosa, N., et al.\ 2006, \mnras, 369, 1880
\bibitem[Elias-Rosa et al.(2008)]{eliasrosa08} Elias-Rosa, N., et al.\ 2008, \mnras, 384, 107 
\bibitem[Elmhamdi et al.(2003)]{elmhamdi03} Elmhamdi, A., et al.\ 2003, \mnras, 338, 939
\bibitem[Elmhamdi et al.(2004)]{elmhamdi04} Elmhamdi, A., Danziger, I.~J., Cappellaro, E., Della Valle, M., Gouiffes, C., Phillips, M.~M., \& Turatto, M.\ 2004, \aap, 426, 963 
\bibitem[Elmhamdi et al.(2006)]{elmhamdi06} Elmhamdi, A., Danziger, I.~J., Branch, D., Leibundgut, B., Baron, E., \& Kirshner, R.~P.\ 2006, \aap, 450, 305 
\bibitem[Evans et al.(2003)]{evans03} Evans, R., Bock, G., Krisciunas, K., \& Espinoza, J.\ 2003, \iaucirc, 8186, 1
\bibitem[Fassia et al.(1998)]{fassia98} Fassia, A., Meikle, W.~P.~S., Geballe, T.~R., Walton, N.~A., Pollacco, D.~L., Rutten, R.~G.~M., 
\& Tinney, C.\ 1998, \mnras, 299, 150
\bibitem[Filippenko et al.(1995)]{filippenko95} Filippenko, A.~V., et al.\ 1995, \apjl, 450, L11
\bibitem[Filippenko(1997)]{filippenko97} Filippenko, A.~V.\ 1997, \araa, 35, 309 
\bibitem[Foley et al.(2004)]{foley04} Foley, R.~J., Wong, D.~S., Moore, M., \& Filippenko, A.~V.\ 2004, \iaucirc, 8353, 3
\bibitem[Ganeshalingam \& Li(2002)]{ganeshalingam02} Ganeshalingam, M., \& Li, W.~D.\ 2002, \iaucirc, 7837, 1
\bibitem[Ganeshalingam et al.(2002)]{ganeshalingam02a} Ganeshalingam, M., Li, W.~D., \& Armstrong, M.\ 2002, \iaucirc, 7891, 1
\bibitem[Ganeshalingam et al.(2005)]{ganeshalingam05} Ganeshalingam, M., Serduke, F.~J.~D., \& Filippenko, A.~V.\ 2005, \iaucirc, 8468, 2
\bibitem[Garavini et al.(2007)]{garavini07} Garavini, G., et al.\ 2007, \aap, 471, 527 
\bibitem[Graham \& Li(2004a)]{graham04a} Graham, J., \& Li, W.\ 2004a, Central Bureau Electronic Telegrams, 75, 1
\bibitem[Graham \& Li(2004b)]{graham04b} Graham, J., \& Li, W.\ 2004b, \iaucirc, 8381, 1 
\bibitem[Graham et al.(2005a)]{graham05a} Graham, J., Li, W., Schwartz, M., \& Trondal, O.\ 2005a, \iaucirc, 8465, 1 
\bibitem[Graham et al.(2005b)]{graham05b} Graham, J., Li, W., Trondal, O., \& Schwartz, M.\ 2005b, \iaucirc, 8467, 1
\bibitem[Hachinger et al.(2006)]{hachinger06} Hachinger, S., Mazzali, P.~A., \& Benetti, S.\ 2006, \mnras, 370, 299
\bibitem[Harutyunyan et al.(2005)]{harutyunyan05} Harutyunyan, A., Benetti, S., Cappellaro, E., \& Turatto, M.\ 2005, 1604-2004: Supernovae as Cosmological Lighthouses, 342, 258
\bibitem[Hendry et al.(2005)]{hendry05} Hendry, M.~A., et al.\ 2005, \mnras, 359, 906
\bibitem[Jacques et al.(2005)]{jacques05} Jacques, C., Colesanti, C., Pimentel, E., \& Napoleao, T.\ 2005, \iaucirc, 8530, 2 
\bibitem[Jeffery \& Branch(1990)]{jeffery90} Jeffery, D.~J., \& Branch, D.\ 1990, Supernovae, Jerusalem Winter School for Theoretical Physics, 149 
\bibitem[Kotak et al.(2005)]{kotak05} Kotak, R., et al.\ 2005, \aap, 436, 1021 
\bibitem[Lee et al.(2005)]{lee05} Lee, E., Ponticello, N.~J., Foley, R.~J., Puckett, T., \& Tigner, J.\ 2005, Central Bureau Electronic Telegrams, 191, 1 
\bibitem[Leonard(2005)]{leonard05} Leonard, D.~C.\ 2005, Central Bureau Electronic Telegrams, 218, 1 
\bibitem[Li et al.(2004)]{li04} Li, W., Yamaoka, H., \& Itagaki, K.\ 2004, \iaucirc, 8448, 2
\bibitem[Matheson et al.(2001)]{matheson01} Matheson, T., Filippenko, A.~V., Li, W., Leonard, D.~C., \& Shields, J.~C.\ 2001, \aj, 121, 1648
\bibitem[Matheson et al.(2002)]{matheson02} Matheson, T., Jha, S., Challis, P., Kirshner, R., Calkins, M., Chornock, R., Li, W.~D., \& Filippenko, A.~V.\ 2002, \iaucirc, 7894, 1
\bibitem[Matheson et al.(2004)]{matheson04} Matheson, T., Challis, P., Kirshner, R., \& Penev, K.\ 2004, \iaucirc, 8353, 2
\bibitem[Mattila et al.(2005)]{mattila05} Mattila, S., Greimel, R., Gerardy, C., \& Meikle, W.~P.~S 2005, \iaucirc, 8474, 1
\bibitem[Mazzali et al.(2000)]{mazzali00} Mazzali, P.~A., Iwamoto, K., \& Nomoto, K.\ 2000, \apj, 545, 407
\bibitem[Mazzali et al.(2005)]{mazzali05} Mazzali, P.~A., et al.\ 2005, \apjl, 623, L37
\bibitem[Mazzali et al.(2007)]{mazzali07} Mazzali, P.~A., R{\"o}pke, F.~K., Benetti, S., \& Hillebrandt, W.\ 2007, Science, 315, 825
\bibitem[Modjaz et al.(2004)]{modjaz04} Modjaz, M., Challis, P., Kirshner, R., Matheson, T., Garg, A., \& Stubbs, C.\ 2004, \iaucirc, 8426, 3 
\bibitem[Monard(2002)]{monard02} Monard, L.~A.~G.\ 2002, \iaucirc, 8016, 1
\bibitem[Monard \& Li(2004)]{monard04} Monard, L.~A.~G., \& Li, W.\ 2004, \iaucirc, 8350, 2
\bibitem[Monard(2005a)]{monard05a} Monard, L.~A.~G.\ 2005a, \iaucirc, 8516, 1
\bibitem[Monard(2005b)]{monard05b} Monard, L.~A.~G.\ 2005b, Central Bureau Electronic Telegrams, 143, 1 
\bibitem[Moore \& Li(2003)]{moore03} Moore, M., \& Li, W.\ 2003, Central Bureau Electronic Telegrams, 40, 1
\bibitem[Nakano et al.(2002)]{nakano02} Nakano, S., Sano, Y., Kushida, R., \& Kushida, Y.\ 2002, \iaucirc, 7805, 1
\bibitem[Nakano et al.(2004)]{nakano04} Nakano, S., Kushida, R., Kushida, Y., \& Itagaki, K.\ 2004, \iaucirc, 8272, 1
\bibitem[Nakano \& Kadota(2005)]{nakano05} Nakano, S., \& Kadota, K.\ 2005, \iaucirc, 8478, 1 
\bibitem[Navasardyan et al.(2004)]{navasardyan04} Navasardyan, H., Turatto, M., Harutunyan, A., Benetti, S., Elias-Rosa, N., Pastorello, A., Viotti, R., \& Rossi, C.\ 2004, \iaucirc, 8454, 3
\bibitem[Navasardyan et al.(2005)]{navasardyan05} Navasardyan, H., Benetti, S., Elias-Rosa, N., Harutunyan, A., \& Pastorello, A.\ 2005, \iaucirc, 8468, 3
\bibitem[Pastorello(2003)]{pastorello03}Pastorello, A.\ 2003, PhD thesis, Univ. Padova
\bibitem[Pastorello et al.(2004)]{pastorello04} Pastorello, A., et al.\ 2004, \mnras, 347, 74 
\bibitem[Pastorello et al.(2005a)]{pastorello05a} Pastorello, A., et al.\ 2005, \mnras, 360, 950
\bibitem[Pastorello et al.(2005b)]{pastorello05b} Pastorello, A., Taubenberger, S., Patat, F., Benetti, S., Harutyunyan, A., Elias-Rosa, N., \& Alises, M.\ 2005b, \iaucirc, 8467, 2
\bibitem[Pastorello et al.(2007a)]{pastorello07a} Pastorello, A., et al.\ 2007a, \mnras, 377, 1531 
\bibitem[Pastorello et al.(2007b)]{pastorello07b} Pastorello, A., et al.\ 2007b, \nat, 447, 829
\bibitem[Pastorello et al.(2007c)]{pastorello07c} Pastorello, A., et al.\ 2007c, \mnras, 376, 1301 
\bibitem[Pastorello et al.(2008)]{pastorello08} Pastorello, A., et al.\ 2008, ArXiv e-prints, 801, arXiv:0801.2277 
\bibitem[Patat et al.(1996)]{patat96} Patat, F., Benetti, S., Cappellaro, E., Danziger, I.~J., della Valle, M., Mazzali, P.~A., \& Turatto, M.\ 1996, \mnras, 278, 111
\bibitem[Patat(1997)]{patat97} Patat, F.\ 1997, \iaucirc, 6776, 1
\bibitem[Patat \& Turatto(1998)]{patat98} Patat, F., \& Turatto, M.\ 1998, \iaucirc, 6922, 1 
\bibitem[Patat et al.(2004a)]{patat04a} Patat, F., Pignata, G., Benetti, S., \& Aceituno, J.\ 2004a, \iaucirc, 8379, 3
\bibitem[Patat et al.(2004b)]{patat04b} Patat, F., Pignata, G., Benetti, S., \& Aceituno, J.\ 2004b, \iaucirc, 8381, 2 
\bibitem[Perlmutter \& Schmidt(2003)]{perlmutterschmidt03} Perlmutter, S., \& Schmidt, B.~P.\ 2003, Supernovae and Gamma-Ray Bursters, 598, 195
\bibitem[Piemonte (1996)]{piemonte96} Piemonte, A.\ 1996, Tesi di Laurea, Univ. Padova
\bibitem[Pignata et al.(2002a)]{pignata02a} Pignata, G., Patat, F., \& Turatto, M.\ 2002a, \iaucirc, 8007, 4
\bibitem[Pignata et al.(2002b)]{pignata02b} Pignata, G., Patat, F., \& Benetti, S.\ 2002b, \iaucirc, 8009, 2 
\bibitem[Pignata et al.(2004a)]{pignata04} Pignata, G., Patat, F., Benetti, S., \& Harutyunyan, A.\ 2004a, \iaucirc, 8344, 2
\bibitem[Pignata et al.(2004b)]{pignata04b} Pignata, G., et al.\ 2004b, \mnras, 355, 178 
\bibitem[Puckett \& Kerns(2002)]{puckett02} Puckett, T., \& Kerns, B.\ 2002, \iaucirc, 7951, 2
\bibitem[Puckett et al.(2005a)]{puckett05a} Puckett, T., George, D., Graham, J., \& Li, W.\ 2005a, \iaucirc, 8470, 1
\bibitem[Puckett et al.(2005b)]{puckett05b} Puckett, T., Orff, T., George, D., \& Crowley, T.\ 2005b, \iaucirc, 8486, 2
\bibitem[Puckett et al.(2005c)]{puckett05c} Puckett, T., Pelloni, A., Ponticello, N., Baek, M., Burket, J., \& Li, W.\ 2005c, Central Bureau Electronic Telegrams, 222, 1
\bibitem[Pugh et al.(2004)]{pugh04} Pugh, H., Park, S., \& Li, W.\ 2004, \iaucirc, 8425, 1
\bibitem[Pugh \& Li(2005)]{pugh05} Pugh, H., \& Li, W.\ 2005, Central Bureau Electronic Telegrams, 158, 1
\bibitem[Riello et al.(2002a)]{riello02a} Riello, M., et al.\ 2002a, \iaucirc, 7894, 4
\bibitem[Riello et al.(2002b)]{riello02b} Riello, M., Benetti, S., Altavilla, G., Pastorello, A., Turatto, M., \& Cappellaro, E.\ 2002b, \iaucirc, 7922, 2
\bibitem[Riess et al.(1997)]{riess97} Riess, A.~G., et al.\ 1997, \aj, 114, 722
\bibitem[Rizzi (1998)]{rizzi98} Rizzi, L.\ 1998, Tesi di Laurea, Univ. Padova
\bibitem[Salvo et al.(2001)]{salvo01} Salvo, M.~E., Cappellaro, E., Mazzali, P.~A., Benetti, S., Danziger, I.~J., Patat, F., \& Turatto, M.\ 2001, \mnras, 321, 254 
\bibitem[Salvo et al.(2003)]{salvo03} Salvo, M., Bessell, M., \& Schmidt, B.\ 2003, \iaucirc, 8187, 1
\bibitem[Sanders(2002)]{sanders02} Sanders, E.\ 2002, \iaucirc, 7921, 1
\bibitem[Schlegel et al.(1998)]{schlegel98} Schlegel, D.~J., Finkbeiner, D.~P., \& Davis, M.\ 1998, \apj, 500, 525
\bibitem[Smartt et al.(2003)]{smartt03} Smartt, S.~J., Maund, J.~R., Gilmore, G.~F., Tout, C.~A., Kilkenny, D., \& Benetti, S.\ 2003, \mnras, 343, 735
\bibitem[Stanishev et al.(2005a)]{stanishev05a} Stanishev, V., Goobar, A., \& Augusteijn, T.\ 2005a, Central Bureau Electronic Telegrams, 159, 1
\bibitem[Stanishev et al.(2005b)]{stanishev05b} Stanishev, V., Goobar, A., \& Naranen, J.\ 2005b, Central Bureau Electronic Telegrams, 225, 1
\bibitem[Stanishev et al.(2007)]{stanishev07} Stanishev, V., et al.\ 2007, \aap, 469, 645
\bibitem[Taubenberger et al.(2005a)]{taubenberger05a} Taubenberger, S., Pastorello, A., \& Alises, M.\ 2005a, \iaucirc, 8472, 3
\bibitem[Taubenberger et al.(2005b)]{taubenberger05b} Taubenberger, S., Pastorello, A., Benetti, S., \& Aceituno, J.\ 2005b, \iaucirc, 8474, 3 
\bibitem[Taubenberger et al.(2005c)]{taubenberger05c} Taubenberger, S., Patat, F., Benetti, S., \& Alises, M.\ 2005c, \iaucirc, 8487, 4
\bibitem[Taubenberger et al.(2005d)]{taubenberger05d} Taubenberger, S., Pastorello, A., Mazzali, P.~A., Witham, A., \& Guijarro, A.\ 2005d, Central Bureau Electronic Telegrams, 305, 1 
\bibitem[Taubenberger et al.(2006)]{taubenberger06} Taubenberger, S., et al.\ 2006, \mnras, 371, 1459
\bibitem[Taubenberger et al.(2008)]{taubenberger08} Taubenberger, S., et al.\ 2008, \mnras, 177 
\bibitem[Turatto et al.(2002)]{turatto02} Turatto, M., Riello, M., Altavilla, G., Benetti, S., Pastorello, A., \& Cappellaro, E.\ 2002, \iaucirc, 7921, 2
\bibitem[Turatto(2003)]{turatto03a} Turatto, M.\ 2003, Supernovae and Gamma-Ray Bursters, 598, 21
\bibitem[Turatto et al.(2003)]{turatto03b} Turatto, M., Benetti, S., \& Cappellaro, E.\ 2003, From Twilight to Highlight: The Physics of Supernovae, 200
\bibitem[Turatto et al.(2005)]{turatto05} Turatto, M., Benetti, S., Harutyunyan, A., Riello, M., Cappellaro, E., Botticella, M.~T., \& Mason, E.\ 2005, Central Bureau Electronic Telegrams, 156, 1
\bibitem[Turatto et al.(2007)]{turatto07} Turatto, M., Benetti, S., \& Pastorello, A.\ 2007, ArXiv e-prints, 706, arXiv:0706.1086 
\bibitem[Vagnozzi et al.(2004)]{vagnozzi04} Vagnozzi, A., et al.\ 2004, \iaucirc, 8375, 1
\bibitem[Valenti (2003)]{valenti03} Valenti, S.\ 2003, Tesi di Laurea, Univ. Napoli
\bibitem[Valenti et al.(2008)]{valenti08} Valenti, S., et al.\ 2008, \apjl, 673, L155 
\bibitem[Wang \& Baade(2005)]{wang05} Wang, L., \& Baade, D.\ 2005, Central Bureau Electronic Telegrams, 193, 1 
\bibitem[Wood-Vasey et al.(2002)]{woodvasey02} Wood-Vasey, W.~M., Aldering, G., \& Nugent, P.\ 2002, \iaucirc, 7915, 2

\end{thebibliography}
\clearpage
\longtab{5}{
\begin{longtable}{c c c c c}
\caption{The archive SNe and spectra.}\\
\label{archive_table}\\
\hline\hline
SN & Type & Redshift & Number & Phase range \\
   &      & (z)        & of files  & (days)        \\
&&&&\\
\hline
\endfirsthead
\caption{continued.}\\
\hline\hline
SN & Type & Redshift & Number & Phase range \\
   &      & (z)        & of files  & (days)        \\
&&&&\\
\endhead
\hline
\endfoot
&&&&\\
1937C & Ia & 0.0011 & 5 & 8 -- 39\\
1969L & IIP & 0.0016 & 7 & 2 -- 61\\
1974G & Ia & 0.0024 & 1 & 9\\
1978K & II & 0.0015 & 2 & 7683,7687\\
1979B & Ia & 0.0032 & 5 & 9 -- 47\\
1979C & IIL & 0.0053 & 24 & 5 -- 223\\
1980K & IIL & 0.0002 & 19 & 2 -- 95\\
1980N & Ia & 0.0060 & 1 & 31\\
1981B & Ia & 0.0060 & 22 & -3 -- 356\\
1982B & Ia & 0.0074 & 2 & 3,28\\
1983G & Ia & 0.0039 & 6 & -1 -- 8\\
1983N & Ib & 0.0017 & 6 & -8 -- 228\\
1983U & Ia & 0.0039 & 1 & 14\\
1983V & Ic & 0.0055 & 3 & -9 -- -4\\
1984A & Ia & -0.0009 & 12 & -7 -- 54\\
1984E & IIL & 0.0041 & 1 & 3959\\
1984L & Ib & 0.0051 & 3 & 10 -- 61\\
1985L & IIL & 0.0029 & 2 & 13,203\\
1986E & IIL & 0.0037 & 5 & 23 -- 5907\\
1986G & Iapec & 0.0018 & 33 & -6 -- 324\\
1987A & IIpec & 0.0011 & 52 & -76 -- 5014\\
1987B & IInL & 0.0085 & 2 & 6,7\\
1987K & IIb & 0.0027 & 7 & 0 -- 208\\
1988A & IIP & 0.0051 & 5 & 3 -- 444\\
1988G & Ia & n.a. & 1 & 29\\
1988H & IIP & 0.0066 & 5 & 21 -- 874\\
1988L & Ib & 0.0062 & 8 & 11 -- 138\\
1988Z & IIn & 0.0222 & 21 & 115 -- 3773\\
1989B & Ia & 0.0024 & 20 & -6 -- 348\\
1989C & IIP & 0.0063 & 3 & 3 -- 58\\
1989M & Ia & 0.0051 & 12 & 20 -- 421\\
1990B & Ic & 0.0075 & 20 & 5 -- 148\\
1990E & IIP & 0.0041 & 21 & 13 -- 537\\
1990H & II & 0.0053 & 8 & 2 -- 27\\
1990I & Ib & 0.0097 & 24 & 0 -- 357\\
1990K & II & 0.0053 & 21 & 5 -- 475\\
1990M & Ia & 0.0088 & 8 & 1 -- 55\\
1990N & Ia & 0.0033 & 6 & -14 -- 246\\
1990Q & II & 0.0064 & 3 & 3 -- 299\\
1990S & IIn & 0.0256 & 1 & 4\\
1990U & Ic & 0.0079 & 25 & -3 -- 367\\
1990W & Ib/c & 0.0049 & 10 & 1 -- 3995\\
1990aa & Ic & 0.0166 & 20 & 7 -- 141\\
1990ah & II & 0.0175 & 1 & -84\\
1990aj & Iac & 0.0053 & 5 & 42 -- 82\\
1991A & Ic & 0.0107 & 8 & 5 -- 96\\
1991D & Ib & 0.0418 & 14 & 5 -- 78\\
1991E & II & 0.0240 & 1 & 4\\
1991H & II & 0.0180 & 1 & 1\\
1991I & II & 0.0360 & 1 & 2\\
1991K & Ia & 0.0170 & 3 & 70 -- 122\\
1991L & Ib/c & 0.0300 & 1 & 13\\
1991M & Ia & 0.0072 & 9 & 1 -- 147\\
1991N & Ic & 0.0033 & 5 & 5 -- 282\\
1991S & Ia & 0.0550 & 2 & 16,19\\
1991T & Iapec & 0.0058 & 35 & -12 -- 1785\\
1991ah & IIn & 0.0370 & 3 & 62 -- 84\\
1991al & II & 0.0100 & 4 & 20 -- 28\\
1991ar & Ib & 0.0152 & 2 & 14,30\\
1991bb & Ia & 0.0266 & 1 & 45\\
1991bc & Ia & 0.0214 & 2 & 19,49\\
1991bd & Ia & 0.0127 & 2 & 16,46\\
1991bg & Iapec & 0.0030 & 21 & 1 -- 203\\
1991bj & Iapec & 0.0182 & 1 & 2\\
1992A & Ia & 0.0062 & 35 & -6 -- 406\\
1992B & Ia & 0.0550 & 1 & 13\\
1992C & II & 0.0101 & 9 & 15 -- 455\\
1992D & IIn & 0.0500 & 2 & 7,33\\
1992E & Ia & 0.0600 & 2 & 9,9\\
1992G & Ia & 0.0053 & 4 & -2 -- 35\\
1992H & II & 0.0060 & 29 & 10 -- 403\\
1992K & Iapec & 0.0104 & 6 & 45 -- 45\\
1992O & Ia & 0.0370 & 4 & 25 -- 25\\
1992al & Ia & 0.0146 & 1 & 61\\
1992ao & II & 0.0122 & 25 & 4 -- 1382\\
1992av & Ia & n.a. & 1 & 9\\
1992ay & IIn & 0.0620 & 1 & 13\\
1992ba & II & 0.0041 & 3 & 3 -- 150\\
1992bb & Ia & n.a. & 2 & 31,31\\
1992bm & II & 0.0500 & 1 & -315\\
1993H & Ia & 0.0241 & 5 & 1 -- 416\\
1993J & IIb & -0.0001 & 39 & 1 -- 1264\\
1993K & II & 0.0091 & 13 & 31 -- 354\\
1993L & Ia & 0.0064 & 15 & 16 -- 378\\
1993M & Ia & 0.0901 & 2 & 18,18\\
1993N & IIn & 0.0098 & 11 & 32 -- 655\\
1993S & II & 0.0320 & 2 & 33,90\\
1993T & Ia & 0.0881 & 1 & 33\\
1993W & II & 0.0180 & 2 & 2,6\\
1993ad & II & 0.0172 & 11 & 3 -- 326\\
1993ae & Ia & 0.0190 & 1 & 10\\
1993af & Ia & 0.0033 & 7 & -304 -- 318\\
1993aj & Ia & 0.0751 & 1 & 14\\
1994D & Ia & 0.0015 & 42 & -11 -- 373\\
1994I & Ic & 0.0015 & 50 & -6 -- 146\\
1994L & II & 0.0068 & 17 & 31 -- 356\\
1994M & Ia & 0.0232 & 3 & 15 -- 15\\
1994N & II & 0.0098 & 8 & 0 -- 265\\
1994R & II & 0.0070 & 1 & 10\\
1994S & Ia & 0.0152 & 1 & -4\\
1994U & Ia & 0.0044 & 1 & 0\\
1994Z & II & 0.0118 & 13 & 2 -- 365\\
1994ae & Ia & 0.0043 & 13 & -6 -- 531\\
1994ai & Ic & 0.0050 & 8 & 4 -- 70\\
1994aj & II & 0.0320 & 30 & 43 -- 540\\
1995D & Ia & 0.0066 & 8 & -6 -- 364\\
1995F & Ic & 0.0051 & 5 & 10 -- 260\\
1995G & IIn & 0.0163 & 21 & 2 -- 942\\
1995H & II & 0.0047 & 6 & -10 -- 247\\
1995J & II & 0.0099 & 1 & 30\\
1995M & Ia & 0.0520 & 1 & 38\\
1995N & IIn & 0.0062 & 63 & 4 -- 3374\\
1995P & Ia & 0.0560 & 1 & 22\\
1995R & Ia & 0.0237 & 1 & 7\\
1995T & Ia & 0.0560 & 1 & 8\\
1995U & Ia & 0.0556 & 1 & 5\\
1995V & II & 0.0051 & 16 & 1 -- 402\\
1995W & II & 0.0113 & 21 & 12 -- 780\\
1995X & II & 0.0052 & 1 & 28\\
1995Z & II & 0.0158 & 2 & 88,91\\
1995aa & IIn & 0.1900 & 1 & 23\\
1995ac & Iapec & 0.0500 & 8 & 0 -- 43\\
1995ad & II & 0.0061 & 23 & 7 -- 508\\
1995ae & Ia & 0.0689 & 1 & 10\\
1995ag & II & 0.0049 & 2 & 33,207\\
1995ak & Ia & 0.0227 & 3 & -1 -- 19\\
1995al & Ia & 0.0051 & 10 & -3 -- 166\\
1995bb & Ic & 0.0058 & 1 & 18\\
1995bd & Iapec & 0.0154 & 3 & 16 -- 19\\
1996A & II & 0.0330 & 7 & 10 -- 39\\
1996D & Ic & 0.0158 & 5 & 9 -- 214\\
1996L & IIL & 0.0330 & 16 & 9 -- 336\\
1996M & II & 0.0200 & 4 & 3 -- 61\\
1996W & II & 0.0055 & 13 & 8 -- 309\\
1996X & Ia & 0.0068 & 17 & -4 -- 298\\
1996Z & Ia & 0.0076 & 3 & 5 -- 269\\
1996ae & IIn & 0.0052 & 5 & 7 -- 23\\
1996al & II & 0.0061 & 48 & 1 -- 2155\\
1996an & II & 0.0047 & 18 & 3 -- 487\\
1996aq & Ic & 0.0053 & 17 & 2 -- 270\\
1996ar & Ia & 0.0600 & 1 & 3\\
1996as & II & 0.0360 & 2 & 3,3\\
1996bl & Ia & 0.0360 & 1 & -3\\
1996bo & Ia & 0.0175 & 2 & -6,49\\
1996bw & II & 0.0181 & 3 & 7 -- 22\\
1996bx & Ic & 0.0600 & 1 & 19\\
1996cb & IIb & 0.0024 & 19 & 12 -- 155\\
1996cc & II & 0.0072 & 6 & 83 -- 140\\
1997B & Ic & 0.0104 & 18 & 1 -- 385\\
1997C & Ia & 0.0227 & 1 & 27\\
1997D & IIpec & 0.0052 & 16 & -1 -- 383\\
1997X & Ic & 0.0037 & 22 & 5 -- 101\\
1997Y & Ia & 0.0160 & 4 & 31 -- 32\\
1997Z & II & 0.0086 & 5 & 2 -- 7\\
1997ab & IIn & 0.0125 & 6 & 357 -- 777\\
1997bp & Iapec & 0.0083 & 12 & -1 -- 414\\
1997bq & Ia & 0.0094 & 1 & 18\\
1997br & Iapec & 0.0069 & 13 & -4 -- 404\\
1997bs & IIn & 0.0024 & 1 & 14\\
1997bt & II & 0.0648 & 1 & 16\\
1997by & Ia & 0.0453 & 1 & 3\\
1997cn & Iapec & 0.0167 & 3 & 3 -- 78\\
1997cr & II & 0.0771 & 2 & 8,8\\
1997cw & Iapec & 0.0176 & 9 & 44 -- 106\\
1997cy & IIn & 0.0642 & 15 & 8 -- 635\\
1997dc & Ib & 0.0115 & 2 & 6,32\\
1997dd & IIb & 0.0152 & 1 & 17\\
1997de & Ia & 0.0129 & 1 & 25\\
1997dh & Ic & 0.0500 & 1 & 4\\
1997dq & Ib/cpec & 0.0032 & 5 & 6 -- 428\\
1997ds & II & 0.0094 & 1 & 8\\
1997du & II & 0.0200 & 2 & 26,35\\
1997ef & Ib/cpec & 0.0118 & 8 & 4 -- 102\\
1997eg & IIn & 0.0089 & 4 & 76 -- 547\\
1997ei & Ic & 0.0107 & 1 & 25\\
1997ej & Ia & 0.0223 & 2 & 32,37\\
1998A & IIpec & 0.0070 & 10 & 18 -- 397\\
1998E & IIn & 0.0080 & 1 & 374\\
1998R & II & 0.0067 & 1 & 36\\
1998S & IIn & 0.0028 & 5 & 7 -- 489\\
1998T & Ib & 0.0101 & 3 & 3 -- 27\\
1998V & Ia & 0.0174 & 1 & 12\\
1998W & II & 0.0119 & 2 & 6,14\\
1998bn & Ia & 0.0061 & 1 & 360\\
1998bp & Ia & 0.0106 & 2 & 27,346\\
1998bu & Ia & 0.0031 & 12 & -7 -- 681\\
1998bw & Ic & 0.0085 & 34 & -9 -- 376\\
1998ce & II & 0.0084 & 1 & 10\\
1998cg & Ia & 0.1190 & 1 & 29\\
1998co & Ia & 0.0182 & 1 & 31\\
1998cv & Ic & 0.0270 & 1 & 28\\
1998cx & Ia & 0.0197 & 1 & 18\\
1998dg & Ia & 0.0082 & 1 & 203\\
1998dh & Ia & 0.0089 & 4 & 8 -- 58\\
1998dj & Ia & 0.0137 & 2 & 4,52\\
1998dk & Ia & 0.0132 & 2 & 29,31\\
1998dl & II & 0.0047 & 2 & 59,116\\
1998dm & Ia & 0.0065 & 2 & 23,27\\
1998dn & II & 0.0013 & 2 & 43,95\\
1998dq & Ia & 0.0108 & 1 & 39\\
1998dt & Ib & 0.0150 & 7 & 19 -- 125\\
1998ee & IIpec & 0.0497 & 1 & 116\\
1998es & Iapec & 0.0105 & 3 & -2 -- 58\\
1998et & IIn & 0.0404 & 1 & 296\\
1998ew & II & 0.0103 & 1 & 179\\
1998fa & IIb & 0.0244 & 5 & 25 -- 77\\
1999E & IIn & 0.0260 & 17 & 8 -- 449\\
1999J & Iapec & 0.0334 & 1 & 21\\
1999P & Ib/c & 0.0600 & 2 & 6,7\\
1999Z & IIn & 0.0504 & 2 & 31,115\\
1999aa & Iapec & 0.0149 & 3 & 13 -- 50\\
1999ac & Iapec & 0.0095 & 7 & -6 -- 391\\
1999as & Icpec & 0.1270 & 1 & 54\\
1999br & IIpec & 0.0034 & 8 & 11 -- 99\\
1999bv & Ia & 0.0184 & 1 & 5\\
1999by & Iapec & 0.0021 & 1 & 183\\
1999cf & Ia & 0.0244 & 1 & 13\\
1999cl & Ia & 0.0071 & 3 & 16 -- 297\\
1999cn & Ic & 0.0226 & 3 & 1 -- 299\\
1999cw & Ia & 0.0124 & 10 & 3 -- 397\\
1999cz & Ic & 0.0072 & 1 & 15\\
1999da & Ia & 0.0123 & 1 & 60\\
1999dh & II & 0.0108 & 1 & 22\\
1999di & Ib & 0.0164 & 3 & 9 -- 40\\
1999dk & Ia & 0.0152 & 6 & -14 -- 67\\
1999dn & Ib & 0.0094 & 12 & 6 -- 379\\
1999dq & Iapec & 0.0145 & 1 & -6\\
1999eb & IIn & 0.0181 & 7 & 5 -- 88\\
1999ec & Iac & 0.0092 & 1 & 6\\
1999ee & Ia & 0.0114 & 14 & -9 -- 41\\
1999el & IIn & 0.0044 & 4 & 18 -- 103\\
1999em & IIP & 0.0024 & 23 & -2 -- 635\\
1999et & II & 0.0163 & 2 & 0,0\\
1999eu & IIpec & 0.0042 & 4 & 5 -- 43\\
1999ex & Ib/c & 0.0114 & 8 & -1 -- 13\\
1999ey & IIn & 0.0931 & 2 & 4,25\\
1999ga & II & 0.0047 & 4 & 40 -- 441\\
1999ge & II & 0.0188 & 1 & 10\\
1999gi & IIP & 0.0020 & 2 & 49,173\\
1999go & II & 0.0148 & 2 & 5,6\\
1999gt & Ia & 0.2740 & 3 & 13 -- 13\\
1999gu & II & 0.1470 & 2 & 13,13\\
2000B & Ia & 0.0191 & 2 & 10,19\\
2000C & Ic & 0.0127 & 5 & 17 -- 34\\
2000D & II & 0.0172 & 2 & 8,19\\
2000E & Ia & 0.0044 & 7 & 14 -- 144\\
2000H & IIb & 0.0130 & 6 & 9 -- 66\\
2000M & II & 0.0103 & 1 & 5\\
2000N & II & 0.0133 & 2 & 5,9\\
2000O & Ia & 0.0235 & 2 & 4,9\\
2000P & IIn & 0.0074 & 6 & 3 -- 506\\
2000bg & IIn & 0.0245 & 1 & 4\\
2000ck & IIpec & 0.0268 & 1 & 5\\
2000cm & Ia & 0.0072 & 1 & 4\\
2000cn & Ia & 0.0235 & 2 & 8,86\\
2000cu & Ia & n.a. & 1 & 2\\
2000cx & Iapec & 0.0081 & 5 & 1 -- 114\\
2000da & II & 0.0244 & 1 & 19\\
2000db & II & 0.0023 & 1 & 16\\
2000de & Ib & 0.0080 & 4 & 13 -- 14\\
2000dg & Ia & 0.0385 & 1 & 6\\
2000dj & II & 0.0158 & 2 & 7,8\\
2000eo & IIn & 0.0108 & 1 & 10\\
2000ev & IIn & 0.0146 & 1 & 1\\
2000ew & Ic & 0.0032 & 2 & 65,109\\
2000fc & Ia & 0.4200 & 1 & 9\\
2000fe & II & 0.0141 & 1 & 11\\
2000fp & II & 0.3000 & 1 & 5\\
2001N & Ia & 0.0210 & 5 & 11 -- 57\\
2001V & Ia & 0.0151 & 3 & 13 -- 54\\
2001X & IIP & 0.0049 & 2 & 129,468\\
2001bb & Ic & 0.0158 & 4 & 18 -- 74\\
2001bc & II & 0.1950 & 1 & 8\\
2001bd & II & 0.0961 & 1 & 10\\
2001be & Ia & 0.2410 & 2 & 9,9\\
2001bg & Ia & 0.0071 & 2 & 3,11\\
2001cz & Ia & 0.0157 & 4 & 1 -- 29\\
2001dc & IIP & 0.0071 & 3 & 41 -- 86\\
2001dk & IIP & 0.0180 & 1 & 164\\
2001dr & II & 0.0239 & 1 & 11\\
2001du & II & 0.0055 & 1 & 16\\
2001ed & Ia & 0.0165 & 1 & 7\\
2001eh & Ia & 0.0371 & 17 & 3 -- 69\\
2001ep & Ia & 0.0130 & 25 & 7 -- 103\\
2001fh & Iapec & 0.0130 & 2 & 3,14\\
2001fv & II & 0.0049 & 2 & 66,68\\
2001fw & Ib & 0.0295 & 1 & 7\\
2001ge & Ia & 0.2200 & 1 & 1\\
2001gf & Ia & 0.1300 & 2 & 1,1\\
2001gg & II & 0.6100 & 1 & 1\\
2001gh & II & 0.1600 & 2 & 1,29\\
2001gi & Ia & 0.2000 & 1 & 1\\
2001gj & II & 0.2700 & 1 & 1\\
2001ie & Ia & 0.0308 & 1 & 4\\
2001ig & IIb & 0.0030 & 1 & 188\\
2001io & Ia & 0.1900 & 3 & 12 -- 12\\
2001ip & Ia & 0.5400 & 3 & 12 -- 1433\\
2001is & Ib & 0.0133 & 2 & 17,18\\
2001it & II & 0.0345 & 1 & 18\\
2002A & IIn & 0.0096 & 1 & 10\\
2002an & II & 0.0129 & 1 & 14\\
2002ap & Icpec & 0.0021 & 38 & -7 -- 250\\
2002bh & II & 0.0173 & 1 & 9\\
2002bo & Ia & 0.0043 & 28 & -14 -- 367\\
2002cl & Ic & 0.0720 & 2 & 14,14\\
2002cm & II & 0.0871 & 2 & 14,14\\
2002cn & Ia & 0.3020 & 2 & 14,14\\
2002co & II & 0.3180 & 1 & 14\\
2002cr & Ia & 0.0094 & 5 & 4 -- 46\\
2002cs & Ia & 0.0157 & 1 & 2\\
2002cv & Ia & 0.0043 & 10 & -5 -- 26\\
2002dg & Ib & 0.0467 & 1 & 15\\
2002dj & Ia & 0.0093 & 9 & 2 -- 287\\
2002dm & Ia & 0.0252 & 1 & 43\\
2002du & II & 0.2100 & 1 & 68\\
2002ej & II & 0.0162 & 1 & 21\\
2002eo & II & 0.0204 & 1 & 11\\
2002er & Ia & 0.0091 & 27 & -11 -- 582\\
2002gd & II & 0.0090 & 13 & 3 -- 111\\
2002hy & Ibpec & 0.0127 & 1 & 3\\
2002ic & Iapec & 0.0660 & 8 & 16 -- 258\\
2002ji & Ib/c & 0.0049 & 1 & 5\\
2003G & IIn & 0.0115 & 3 & 15 -- 16\\
2003J & II & 0.0026 & 1 & 13\\
2003L & Ic & 0.0213 & 1 & 13\\
2003M & Iapec? & 0.0242 & 5 & 12 -- 42\\
2003Z & II & 0.0042 & 12 & 23 -- 149\\
2003bg & Icpec & 0.0044 & 1 & 4\\
2003cg & Ia & 0.0041 & 40 & -8 -- 386\\
2003dt & Ia & 0.0142 & 1 & 2\\
2003du & Ia & 0.0064 & 10 & -11 -- 72\\
2003ei & IIn & 0.0268 & 2 & 61,62\\
2003gd & II & 0.0021 & 7 & 15 -- 73\\
2003gs & Iapec & 0.0047 & 2 & 14,26\\
2003hg & II & 0.0143 & 1 & 4\\
2003hn & II & 0.0039 & 1 & 3\\
2003ie & II & 0.0023 & 1 & 3\\
2003jd & Icpec & 0.0188 & 13 & 0 -- 29\\
2004G & II & 0.0053 & 1 & 2\\
2004aq & II & 0.0140 & 1 & 8\\
2004aw & Ic & 0.0158 & 32 & 1 -- 261\\
2004dg & II & 0.0045 & 1 & 2\\
2004dh & II & 0.0194 & 2 & 93,93\\
2004dj & IIP & 0.0004 & 5 & 133 -- 157\\
2004dt & Ia & 0.0195 & 33 & -10 -- 354\\
2004eo & Ia & 0.0157 & 21 & 2 -- 241\\
2004et & II & 0.0002 & 5 & 58 -- 263\\
2004ex & IIb & 0.0174 & 3 & 36 -- 85\\
2004gd & IIn & 0.0174 & 1 & 39\\
2004go & Ia & 0.0291 & 1 & 19\\
2004gt & Ib/c & 0.0055 & 2 & 24,163\\
2005G & Ia & 0.0231 & 1 & 4\\
2005N & Ib/c & 0.0163 & 1 & 3\\
2005W & Ia & 0.0087 & 2 & 1,15\\
2005ab & II & 0.0154 & 1 & 4\\
2005au & II & 0.0182 & 1 & 15\\
2005aw & Ic & 0.0133 & 1 & 62\\
2005ay & IIP & 0.0027 & 12 & 1 -- 308\\
2005bl & Ia & 0.0241 & 1 & 33\\
2005br & Ib & 0.0103 & 1 & 58\\
2005bs & Ia & 0.0552 & 1 & 36\\
2005cb & Ic & 0.0105 & 1 & 12\\
2005cf & Ia & 0.0065 & 31 & -12 -- 77\\
2005cq & Ia & 0.3100 & 1 & 10\\
2005cs & IIP & 0.0015 & 17 & -1 -- 222\\
2005ip & II & 0.0071 & 6 & 3 -- 95\\
2006G & II/IIb & 0.0168 & 1 & 25\\
2006W & IIL & 0.0159 & 1 & 2\\
2006X & Ia & 0.0053 & 2 & 1,9\\
2006aj & Ib/c & 0.0330 & 16 & -7 -- 10\\
2006ao & II & 0.0299 & 1 & 10\\
2006ca & II & 0.0089 & 1 & 2\\
2006gi & Ib & 0.0094 & 1 & 146\\
2006gy & IIn & 0.0188 & 3 & -31 -- 106\\
2006gz & Ia & 0.0236 & 17 & -13 -- 12\\
2006jc & Ib/cpec & 0.0056 & 28 & 3 -- 78\\
2006ov & IIP & 0.0053 & 2 & 39,79\\
2007C & Ib & 0.0056 & 1 & 12\\
2007F & Ia & 0.0238 & 1 & 2\\
2007I & Ic & 0.0216 & 1 & 28\\
2007R & Ia & 0.0308 & 1 & 5\\
2007T & II & 0.0135 & 1 & 4\\
2007af & Ia & 0.0053 & 1 & 67\\
2007bj & Ia & 0.0166 & 1 & 2\\
2007bm & Ia & 0.0062 & 5 & 2 -- 28\\
2007bt & IIn & 0.0400 & 1 & 21\\
2007bw & IIn & 0.1400 & 1 & 32\\
2007fo & Ib & 0.0094 & 1 & 7\\
\hline
\end{longtable}
}

\end{document}